\documentclass[superscriptaddress,amsmath, nofootinbib]{revtex4}
\usepackage{amsfonts}
\usepackage{graphicx}
\usepackage[latin1]{inputenc}
\usepackage{amsmath}
\usepackage{amssymb}
\usepackage[latin1]{inputenc}
\usepackage{amsmath}
\usepackage{amssymb}
\usepackage{amsfonts}
\usepackage{graphicx}
\usepackage[latin1]{inputenc}
\usepackage{amsmath}
\usepackage{amssymb}

\begin{document}


\title{Recent progress on the description of relativistic spin: vector model of spinning particle and rotating body
with gravimagnetic moment in General Relativity}
\author{Alexei A. Deriglazov }
\email{alexei.deriglazov@ufjf.edu.br} \affiliation{Departamento de Matem\'atica, ICE, Universidade Federal de Juiz de
Fora, MG, Brazil} \affiliation{Laboratory of Mathematical Physics, Tomsk Polytechnic University, 634050 Tomsk, Lenin
Ave. 30, Russian Federation}

\author{Walberto Guzm\'an Ram\'irez }
\email{wguzman@cbpf.br} \affiliation{Departamento de Matem\'atica, ICE, Universidade Federal de Juiz de Fora, MG,
Brazil}

\date{\today}

\begin{abstract}
We review the recent results on development of vector models of spin and apply them to study the influence of
spin-field interaction on the trajectory and precession of a spinning particle in external gravitational and
electromagnetic fields. The formalism is developed starting from the Lagrangian variational problem, which implies both
equations of motion and constraints which should be presented in a model of spinning particle. We present a detailed
analysis of the resulting theory and show that it has reasonable properties on both classical and quantum level. We
describe a number of applications and show how the vector model clarifies some issues presented in theoretical
description of a relativistic spin: A) One-particle relativistic quantum mechanics with positive energies and its
relation with the Dirac equation and with relativistic {\it Zitterbewegung}; B) Spin-induced non commutativity and the
problem of covariant formalism; C) Three-dimensional acceleration consistent with coordinate-independence of the speed
of light in general relativity and rainbow geometry seen by spinning particle; D) Paradoxical behavior of the
Mathisson-Papapetrou-Tulczyjew-Dixon equations of a rotating body in ultra relativistic limit, and equations with
improved behavior.
\end{abstract}

\maketitle 

\section{Introduction}

Basic notions of Special and General Relativity have been formulated before the discovery of spin, so they describe the
properties of space and time as they are seen by spinless test-particle. It is natural to ask, whether these notions
remain the same if the spinless particle is replaced by more realistic spinning test-particle. To analyze this issue,
it is desirable to have a systematic formalism for semiclassical description of spinning degrees of freedom in
relativistic (Poincaré invariant) and generally covariant theories.

Search for the relativistic equations that describe evolution of rotational degrees of freedom and their influence on
the trajectory of a rotating body represents a problem with almost centenary history \cite{Mathisson:1937zz, Fock1939,
Papapetrou:1951pa, Tulc, Dixon1964, Dixon1965, pirani:1956}. The equations are necessary for current applications of
general relativity on various space-time scales: for analysis of Lense-Thirring precession \cite{Adler2015}, for
accounting spin effects in compact binaries and rotating black holes \cite{Pomeranskii1998, Will2014}, and in
discussion of cosmological problems, see \cite{Balakin2015} and references therein. Closely related problem consists in
establishing of classical equations that could mimic quantum mechanics of an elementary particle with spin in a
semiclassical approximation \cite{Frenkel, Frenkel2, Thomas1927, corben:1968, ba1}. While the description of spin
effects of relativistic electron is achieved in QED on the base of Dirac equation, the relationship among classical and
quantum descriptions has an important bearing, providing interpretation of results of quantum-field-theory computations
in usual terms: particles and their interactions. Semiclassical understanding of spin precession of a particle with an
arbitrary magnetic moment is important in the development of experimental technics for measurements of anomalous
magnetic moment \cite{Field:1979, miller2007muon}. In accelerator physics \cite{hoffstaetter2006adiabatic} it is
important to control resonances leading to depolarization of a beam. In the case of vertex electrons carrying arbitrary
angular momentum, semiclassical description can also be useful \cite{karlovets2012}. Basic equations of spintronics are
based on heuristic and essentially semiclassical considerations \cite{dyakonov1971}. It would be very interesting to
obtain them from first principles, that is from equations of motion of a spinning particle.

Hence the further development of classical models of relativistic spinning particles/bodies represents an actual task.
A review of the achievements in this fascinating area before 1968 can be found in the works of Dixon \cite{Dixon1964}
and in the book of Corben \cite{corben:1968}. Contrary to these works, where the problem was discussed on the level of
equations of motion, our emphasis has been placed on the Lagrangian and Hamiltonian variational formulations for the
description of rotational degrees of freedom. Taking a variational problem as the starting point, we avoid the
ambiguities and confusion, otherwise arising in the passage from Lagrangian to Hamiltonian description and vice-versa.
Besides, it essentially fixes the possible form of interaction with external fields. In this review we show that so
called vector model of spin represents a unified conceptual framework, allowing to collect and tie together a lot of
remarkable ideas, observations and results accumulated on the subject after 1968.

The present review article is based mainly on the recent works \cite{deriglazov2014Monster, DW2015.1, DWGR2016Conf,
DWGR2016S, DWGR2017, DPM2, DPM3, DPM2016, deriglazovMPL2015, DPW2}. In \cite{deriglazov2014Monster} we constructed
final Lagrangian for a spinning particle with an arbitrary magnetic moment. In \cite{DW2015.1} we presented the
Lagrangian minimally interacting with gravitational field, while in \cite{DWGR2016Conf, DWGR2016S} it has been extended
to the case of non minimal interaction through the gravimagnetic moment. In all cases, our variational problem leads to
both dynamical equations of motion and appropriate constraints, the latter guarantee the fixed value of spin, as well
as the spin supplementary condition $S^{\mu\nu}p_\nu=0$. The works \cite{DWGR2016S, DWGR2017, DPM2, DPM3, DPM2016,
deriglazovMPL2015, DPW2} are devoted to some applications of the vector model to various classical and quantum
mechanical problems.

We have not tried to establish a variational problem of the most general form \cite{hanson1974, Fuchs1991}. Instead,
the emphasis has been placed on the variational problem leading to the equations which are widely considered the most
promising candidates for description of spinning particles in external fields. For the case of electromagnetic field,
the vector model leads to a generalization of approximate equations of Frenkel and Bargmann, Michel and Telegdi (BMT)
to the case of an arbitrary field. Here the strong restriction on possible form of equations is that the reasonable
model should be in correspondence with the Dirac equation. In this regard, the vector model is of interest because it
yields a relativistic quantum mechanics with positive-energy states, and is closely related to the Dirac equation.

Concerning the equations of a rotating body in general relativity, the widely assumed candidates are the
Mathisson-Papapetrou-Tulczyjew-Dixon (MPTD) equations. While our vector model has been constructed as a semiclassical
model of an elementary spin one-half particle, it turns out to be possible to apply it to the case: the vector model
with minimal spin-gravity interaction and properly chosen parameters (mass and spin, see below), yields Hamiltonian
equations equivalent to the MPTD equations. In the Lagrangian counterpart of MPTD equations emerges the term, which can
be thought as an effective metric generated along the world-line by the minimal coupling. This leads to certain
problems if we assume that MPTD equations remain applicable in the ultra-relativistic limit. In particular,
three-dimensional acceleration of MPTD particle increases with velocity and becomes infinite in the limit. Therefore we
examine the non-minimal interaction, this gives a generalization of MPTD equations to the case of a rotating body with
gravimagnetic moment \cite{Khriplovich1989}. We show that a rotating body with unit gravimagnetic moment has an
improved behavior in the ultra-relativistic regime and is free from the problems detected in MPTD-equations.

{\bf Notation.} Our variables are taken in arbitrary parametrization $\tau$, then $\dot x^\mu=\frac{dx^\mu}{d\tau}$.
The square brackets mean antisymmetrization, $\omega^{[\mu}\pi^{\nu]}=\omega^\mu\pi^\nu-\omega^\nu\pi^\mu$. For the
four-dimensional quantities we suppress the contracted indexes and use the notation  $\dot x^\mu G_{\mu\nu}\dot
x^\nu=\dot xG\dot x$,  $N^\mu{}_\nu\dot x^\nu=(N\dot x)^\mu$, $\omega^2=g_{\mu\nu}\omega^\mu\omega^\nu$, $\mu, \nu=0,
1, 2, 3$.  Notation for the scalar functions constructed from second-rank tensors are $\theta S=
\theta^{\mu\nu}S_{\mu\nu}$, $S^2=S^{\mu\nu}S_{\mu\nu}$. When we work in four-dimensional Minkowski space with
coordinates $x^\mu=(x^0=ct,~  x^i)$, we use the metric $\eta_{\mu\nu}=(-, +, +, +)$, then $\dot x\omega=\dot
x^\mu\omega_\mu=-\dot x^0\omega^0+\dot x^i\omega^i$ and so on. Levi-Civita tensors in four and three dimensions are
defined by $\epsilon^{0123}=-\epsilon_{0123}=1$ and $\epsilon^{ijk}=\epsilon_{ijk}=1$. Suppressing the indexes of
three-dimensional quantities, we use bold letters, $v^i\gamma_{ij}a^j={\bf v}\gamma{\bf a}$, $v^iG_{i\mu}v^\mu={\bf
v}Gv$, $i, j=1, 2, 3$, and so on.

 The covariant derivative is $\nabla
P^\mu=\frac{dP^\mu}{d\tau}+\Gamma^\mu_{\alpha\beta}\dot x^\alpha P^\beta$. The tensor of Riemann curvature is
$R^\sigma{}_{\lambda\mu\nu}=\partial_\mu\Gamma^\sigma{}_{\lambda\nu} -\partial_\nu
\Gamma^\sigma{}_{\lambda\mu}+\Gamma^\sigma{}_{\beta\mu}\Gamma^{\beta}{}_{\lambda\nu}-
\Gamma^\sigma{}_{\beta\nu}\Gamma^{\beta}{}_{\lambda\mu}$.

Electromagnetic field:
\begin{eqnarray}\label{L.0}
F_{\mu\nu}=\partial_\mu A_\nu-\partial_\nu A_\mu=(F_{0i}=-E_i, ~ F_{ij}= \epsilon_{ijk}B_k), \cr
E_i=-\frac{1}{c}\partial_tA_i+\partial_i A_0, \quad B_i=\frac12\epsilon_{ijk}F_{jk}=\epsilon_{ijk}\partial_j A_k.
\end{eqnarray}

\section{Lagrangian form of Mathisson-Papapetrou-Tulczyjew-Dixon equations of a rotating body} \label{ch09:sec9.10}

Equations of motion of a rotating body in curved background formulated usually in the multipole approach to description
of the body, see \cite{Trautman2002} for the review. In this approach, the energy-momentum of the body is modelled by a
set of quantities called multipoles. Then the conservation law for the energy-momentum tensor, $\nabla_\mu
T^{\mu\nu}=0$, implies certain equations for the multipoles. The first results were reported by Mathisson
\cite{Mathisson:1937zz} and Papapetrou \cite{Papapetrou:1951pa}. They have taken the approximation which involves only
first two terms (the pole-dipole approximation).  A manifestly covariant equations were formulated by Tulczyjew
\cite{Tulc} and Dixon \cite{Dixon1964}. In the current literature they usually appear in the form given by Dixon (the
equations (6.31)-(6.33) in \cite{Dixon1964}), we will refer them as  Mathisson-Papapetrou-Tulczyjew-Dixon equations.

We discuss MPTD-equations in the form studied by Dixon\footnote{Our $S$ is twice of that of Dixon.}
\begin{eqnarray}
\nabla P^\mu=-\frac 14 R^\mu{}_{\nu\alpha\beta}S^{\alpha\beta}\dot x^\nu \equiv-\frac 14 (\theta\dot x)^\mu,\label{r1}\\
\nabla S^{\mu\nu}= 2(P^\mu \dot x^\nu - P^\nu \dot x^\mu), \label{r2} \\
S^{\mu\nu}P_\nu  =0. \label{r3}
\end{eqnarray}
In the multipole approach, $x^\mu(\tau)$ is called representative point of the body, we take it in arbitrary
parametrization $\tau$ (contrary to Dixon, we do not assume the proper-time parametrization\footnote{We will be
interested in ultra-relativistic behavior of a body. The proper-time parametrization has no sense when $v\rightarrow
c$.}, that is we do not add the equation $g_{\mu\nu}\dot x^\mu\dot x^\nu=-c^2$ to the system above). $S^{\mu\nu}(\tau)$
is associated with inner angular momentum, and $P^\mu(\tau)$ is called momentum. The first-order equations (\ref{r1})
and (\ref{r2}) appear in the pole-dipole approximation, while the algebraic equation (\ref{r3}) has been added by hand.
In the multipole approach it is called the spin supplementary condition (SSC) and corresponds to the choice of
representative point as the center of mass \cite{Tulc, Dixon1964, pirani:1956}. After adding the equation (\ref{r3}) to
the system, the number of equations coincides with the number of variables.

Since we are interested in the influence of spin on the trajectory of a particle, we eliminate the momenta from MPTD
equations, thus obtaining a second-order equation for the representative point $x^\mu(\tau)$. The most interesting
property of the resulting equation is the emergence of an effective metric $G_{\mu\nu}$ instead of the original metric
$g_{\mu\nu}$.

Let us start from some useful consequences of the system (\ref{r1})-(\ref{r3}). Take derivative of the constraint,
$\nabla(S^{\mu\nu}P_\nu)=0$, and use (\ref{r1}) and (\ref{r2}), this gives the expression
\begin{eqnarray}\label{r4}
(P\dot x)P^\mu=P^2\dot x^\mu+\frac18(S\theta \dot x)^\mu,
\end{eqnarray}
which can be written in the form
\begin{eqnarray}\label{r5}
P^\mu=\frac{P^2}{(P\dot x)}\left(\delta^\mu{}_\nu+\frac{1}{8P^2}(S\theta)^\mu{}_\nu\right)\dot x^\nu
\equiv\frac{P^2}{(P\dot x)}\tilde{T}^\mu{}_\nu\dot x^\nu.
\end{eqnarray}
Contract (\ref{r4}) with $g_{\mu\alpha}\dot x^\alpha$. Taking into account that $(P\dot x)<0$, this gives $(P\dot
x)=-\sqrt{-P^2}\sqrt{-\dot x\tilde{T}\dot x}$. Using this in Eq. (\ref{r5}) we obtain
\begin{eqnarray}\label{r7}
P^\mu=\frac{\sqrt{-P^2}}{\sqrt{-\dot x\tilde{T}\dot x}}\tilde T^\mu{}_\nu\dot x^\nu, \qquad \tilde
{T}^\mu{}_\nu=\delta^\mu{}_\nu+\frac{1}{8P^2}(S\theta)^\mu{}_\nu.
\end{eqnarray}
Contracting  (\ref{r2}) with $S_{\mu\nu}$ and using  (\ref{r3}) we obtain $\frac{d}{d\tau}(S^{\mu\nu}S_{\mu\nu})=0$,
that is, square of spin is a constant of motion. Contraction of  (\ref{r4}) with $P_\mu$  gives $(PS\theta\dot x)=0$.
Contraction of (\ref{r4}) with $(\dot x\theta)_\mu$ gives $(P\theta\dot x)=0$. Contraction of (\ref{r1}) with $P_\mu$,
gives $\frac{d}{d\tau}(P^2)=-\frac12(P\theta\dot x)=0$, that is $P^2$ is one more constant of motion, say $k$,
$\sqrt{-P^2}=k=\mbox{const}$ (in our vector model developed below this is fixed as $k=mc$). Substituting  (\ref{r7})
into the equations (\ref{r1})-(\ref{r3}) we now can exclude $P^\mu$ from these equations, modulo to the constant of
motion $k=\sqrt{-P^2}$.

Thus, square of momentum can not be excluded from the system (\ref{r1})-(\ref{r4}), that is MPTD-equations in this form
do not represent a Hamiltonian system for the pair $x^\mu, P^\mu$. To improve this point, we note that Eq. (\ref{r7})
acquires a conventional form (as the expression for conjugate momenta of $x^\mu$ in the Hamiltonian formalism), if we
add to the system (\ref{r1})-(\ref{r3}) one more equation, which fixes the remaining quantity $P^2$. To see, how the
equation could look, we note that for non-rotating body (pole approximation) we expect equations of motion of spinless
particle, $\nabla p^\mu=0$, $p^\mu=\frac{\tilde mc}{\sqrt{-\dot xg\dot x}}\dot x^\mu$, $p^2+(\tilde mc)^2=0$.
Independent equations of the system (\ref{r1})-(\ref{r4}) in this limit read $\nabla P^\mu=0$,
$P^\mu=\frac{\sqrt{-P^2}}{\sqrt{-\dot xg\dot x}}\dot x^\mu$. Comparing the two systems, we see that the missing
equation is the mass-shell condition $P^2+(\tilde mc)^2=0$. Returning to the pole-dipole approximation, an admissible
equation should be $P^2+(\tilde mc)^2+f(S, \ldots)=0$, where $f$ must be a constant of motion. Since the only constant
of motion in arbitrary background is $S^2$, we write\footnote{We could equally start with $P^2+(\tilde
mc)^2+f(S^2, P^2)=0$. Assuming that this equation can be resolved with respect to $P^2$, we arrive essentially at the
same expression.}
\begin{eqnarray}\label{r9.2}
P^2=-(\tilde mc)^2-f(S^2)\equiv -(mc)^2, \quad\mbox{where}\quad  m\equiv\sqrt{\tilde m^2+\frac{f(S^2)}{c^2}}.
\end{eqnarray}
With this value of $P^2$, we can exclude $P^\mu$ from MPTD-equations, obtaining closed system with second-order
equation for $x^\mu$ (so we refer the resulting equations as Lagrangian form of MPTD-equations). We substitute
(\ref{r7}) into (\ref{r1})-(\ref{r3}), this gives the system
\begin{eqnarray}
\nabla\frac{(\tilde{T}\dot x)^\mu}{\sqrt{-\dot x\tilde{T}\dot x}}=-\frac{1}{4\sqrt{-P^2}}(\theta\dot x)^\mu,
\qquad \qquad \label{r9} \\
\nabla S^{\mu\nu}=
\frac{1}{4\sqrt{-P^2}{\sqrt{-\dot x\tilde{T}\dot x}}}\dot x^{[\mu}(S\theta\dot x)^{\nu]}, \label{r10} \\
(SS\theta\dot x)^\mu=-8P^2(S\dot x)^\mu, \qquad \qquad \qquad \label{r11}
\end{eqnarray}
where (\ref{r9.2}) is implied. They determine evolution of $x^\mu$ and $S^{\mu\nu}$ for each given function $f(S^2)$.

It is convenient to introduce the symmetric matrix $G$ composed from the "tetrad field" $\tilde{T}$ of Eq. (\ref{r7})
\begin{equation}\label{r12}
G_{\mu\nu}=g_{\alpha\beta} \tilde{T}^{\alpha}{}_{\mu} \tilde{T}^{\beta}{}_{\nu}=g_{\mu\nu}+h_{\mu\nu}(S).
\end{equation}
Since this is composed from the original metric $g_{\mu\nu}$ plus (spin and field-dependent) contribution $h_{\mu\nu}$,
we call $G$ the effective metric produced \textit{along the world-line} by interaction of spin with gravity. Eq.
(\ref{r11}) implies the identity
\begin{eqnarray}\label{r13}
\dot x\tilde{T}\dot x=\dot xG\dot x,
\end{eqnarray}
so we can replace $\sqrt{-\dot x\tilde{T}\dot x}$ in (\ref{r9})-(\ref{r11}) by $\sqrt{-\dot xG\dot x}$. In particular,
Eq. (\ref{r9}) reads
\begin{eqnarray}\label{r9.1}
\nabla\frac{(\tilde{T}\dot x)^\mu}{\sqrt{-\dot xG\dot x}}=-\frac{1}{4\sqrt{-P^2}}(\theta\dot x)^\mu.
\end{eqnarray}
Adding the consequences found above to the MPTD equations (\ref{r1})-(\ref{r3}), we have the system
\begin{eqnarray}
P^\mu=\frac{\sqrt{-P^2}}{\sqrt{-\dot xG\dot x}}(\tilde{T}\dot x)^\mu, \qquad  \nabla
P^\mu=-\frac 14 (\theta\dot x)^\mu, \cr \nabla S^{\mu\nu}= 2P^{[\mu} \dot x^{\nu]}, \qquad S^{\mu\nu}P_\nu  =0, \qquad \label{r003} \\
P^2+(mc)^2+f(S^2)=0, \qquad \qquad \label{r004} \\
S^2 , \quad m(S^2) \quad \mbox{are constants of motion}, \quad   \label{r005}
\end{eqnarray}
with $\tilde{T}$ given in (\ref{r7}). In section \ref{ch09:sec9.9.5} we will see that they essentially coincide with
Hamiltonian equations of our spinning particle with vanishing gravimagnetic moment.

Let us finish this section with the following comment. Our discussion in the next two sections will be around the
factor $\dot xG\dot x$, where appeared the effective metric $G_{\mu\nu}$. The equation for trajectory (\ref{r9.1})
became singular for the particle's velocity which annihilates this factor, $\dot xG\dot x=0$. We call this the {\it
critical velocity}. The observer independent scale $c$ of special relativity is called, as usual, the speed of light.
The singularity determines behavior of the particle in ultra-relativistic limit. To clarify this point, consider the
standard equations of a spinless particle interacting with electromagnetic field in the physical-time parametrization
$x^\mu(t)=(ct, {\bf x}(t))$,
$\left(\frac{\dot x^\mu}{\sqrt{c^2-{\bf v}^2}}\right)^.=\frac{e}{mc^2}F^\mu{}_\nu \dot x^\nu$.
Then the factor is just $c^2-{\bf v}^2$, that is critical speed coincides with the speed of light. Rewriting the
equations of motion in the form of second law of Newton, we find an acceleration. For the case, the longitudinal
acceleration reads $a_{||}={\bf v} {\bf a}=\frac{e(c^2-{\bf v}^2)^{\frac{3}{2}}}{mc^3}({\bf E}{\bf v})$, that is the
factor, elevated in some degree, appears on the right hand side of the equation, and thus determines the value of
velocity at which the longitudinal acceleration vanishes, $a_{||}\stackrel{v\rightarrow c}{\longrightarrow}0$. So the
singularity implies\footnote{We point out that the factor can be hidden using the singular parametrization. For
instance, in the proper-time parametrization this would be encoded into the definition of $ds$, $ds=\sqrt{c^2-{\bf
v}^2}dt$.}, that during its evolution in the external field the spinless particle can not exceed the speed of light
$c$.
\par

\section{Three-dimensional acceleration and speed of light in General Relativity}
\label{ch06:sec6.9}

The ultra-relativistic behavior of MPTD particle in an arbitrary gravitational field will be analyzed by estimation of
three-acceleration as $v\rightarrow c$. Let us discuss the necessary notions.

By construction of Lorentz transformations, the speed of light in special relativity is an observer-independent
quantity. In the presence of gravity, we replace the Minkowski space by a four-dimensional pseudo Riemann manifold
\begin{eqnarray}\label{La.1}
{\bf M}^{(1,3)}=\{x^\mu,  ~  g_{\mu\nu}(x^\rho),  ~  g_{00}<0\}.
\end{eqnarray}
To discuss the physics behind this abstract four-dimensional construction, we should establish a correspondence between
the quantities computed in an arbitrary coordinates of the Riemann space and the three-dimensional quantities used by
an observer in his laboratory. In particular, in a curved space we replace the Lorentz transformations on the
general-coordinate ones, so we need to ensure the coordinate-independence of the speed of light for that case. It turns
out that this essentially determines the relationship between the four-dimensional and three-dimensional geometries
\cite{bib16}. We first recall the most simple part of this problem, which consist in determining of basic differential
quantities of three-dimensional geometry: infinitesimal distances, time intervals and velocity \cite{bib16}. Then we
define the three-dimensional acceleration which guarantees that a particle, propagating along a four-dimensional
geodesic, can not exceed the speed of light. This gives us the necessary tool for discussion of a fast moving body.

The behavior of ultra-relativistic particles turns out to be important for analysis of near horizon geometry of
extremal black holes, see \cite{Gal2013}, and for accurate accounting of the corresponding corrections to geodesic
motion near black hole \cite{Shao2016, Yu2016, Wen22016, Wen12016, Fen2016, Mik2016, Mik2012, frob2016, frob2017}.

\subsection{Coordinate independence of speed of light.}
Consider an observer that label events by some coordinates of
pseudo Riemann space (\ref{La.1}) to describe the motion of a point particle in a gravitational field with given metric
$g_{\mu\nu}$. Formal definitions of the three-dimensional quantities  can be obtained representing four-interval in
$1+3$ block-diagonal form
\begin{eqnarray}\label{La.2}
-ds^2=g_{\mu\nu}dx^\mu dx^\nu= \qquad \qquad \qquad  \cr
-c^2\left[\frac{\sqrt{-g_{00}}}{c}(dx^0+\frac{g_{0i}}{g_{00}}dx^i)\right]^2+\left(g_{ij}-\frac{g_{0i}g_{0j}}{g_{00}}\right)dx^idx^j.
\nonumber
\end{eqnarray}
This prompts to introduce infinitesimal time interval and distance as follows:
\begin{eqnarray}\label{La.3.0}
dt=\frac{\sqrt{-g_{00}}}{c}(dx^0+\frac{g_{0i}}{g_{00}}dx^i)\equiv-\frac{g_{0\mu}dx^\mu}{c\sqrt{-g_{00}}}.
\end{eqnarray}
\begin{eqnarray}\label{La.3}
dl^2=(g_{ij}-\frac{g_{0i}g_{0j}}{g_{00}})dx^idx^j\equiv\gamma_{ij}dx^idx^j.
\end{eqnarray}
Therefore the conversion factor between intervals of the coordinate time $\frac{dx^0}{c}$ and the time $dt$ measured by
laboratory clock is
\begin{eqnarray}\label{La.3.1}
\frac{dt}{dx^0}=\frac{\sqrt{-g_{00}}}{c}(1+\frac{g_{0i}}{g_{00}}\frac{dx^i}{dx^0}).
\end{eqnarray}
From Eq. (\ref{La.3.0}) it follows that laboratory time coincides with coordinate time in the synchronous coordinate
systems where metric acquires the form $g_{00}=1$ and $g_{0i}=0$. If metric is not of this form, we can not describe
trajectory ${\bf x}(t)$ using the laboratory time $t$ as a global parameter. But we can describe it by the function
${\bf x}(x^0)$, and then determine various differential characteristics (such as velocity and acceleration) using the
conversion factor (\ref{La.3.1}). For instance, three-velocity of the particle is
\begin{eqnarray}\label{La.5}
v^i=\left(\frac{dt}{dx^0}\right)^{-1}\frac{dx^i}{dx^0},
\end{eqnarray}
so it is convenient to introduce the notation
\begin{eqnarray}\label{La.50.1}
\partial_t\equiv\left(\frac{dt}{dx^0}\right)^{-1}\frac{\partial}{\partial x^0}, \qquad
\frac{d}{dt}\equiv\left(\frac{dt}{dx^0}\right)^{-1}\frac{dx^i}{dx^0}, \quad \mbox{then, symbolically,} \quad
v^i=\frac{dx^i}{dt}, \quad v=\frac{dl}{dt}.
\end{eqnarray}
The definitions of ${\bf v}$ and $v$ are consistent: $v^2=\left(\frac{dl}{dt}\right)^2={\bf v}^2=v^i\gamma_{ij}v^j$.
Three-dimensional geometry is determined by the metric $\gamma_{ij}(x^0, {\bf x})$. In particular, square of length of
a vector is given by ${\bf v}\gamma{\bf v}=v^i\gamma_{ij}v^j$. Using these notation, the infinitesimal interval
acquires the form similar to special relativity
\begin{eqnarray}\label{La.6}
ds^2=c^2dt^2-dl^2=dt^2\left(c^2-{\bf v}\gamma{\bf v}\right).
\end{eqnarray}
This equality holds in any coordinate system $x^\mu$. Hence a particle with the propagation law $ds^2=0$ has the speed
${\bf v}^2=c^2$, and this is a coordinate-independent statement. The value of the constant $c$, introduced by hand, is
fixed from the flat limit: equation (\ref{La.3.0}) implies $dt=cdx^0$ when $g_{\mu\nu}\rightarrow\eta_{\mu\nu}$.

\begin{figure}[t] \centering
\includegraphics[width=200pt, height=100pt]{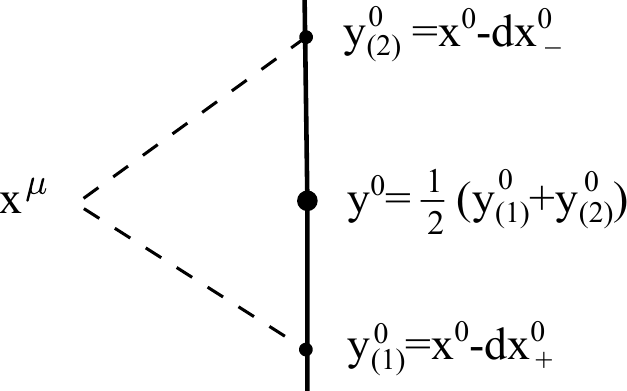}
\caption{Definition of simultaneous events. The vertical line represents a world-line of the laboratory clock. The
points $y^0_{(1)}$ and $y^0_{(2)}$ have null-interval with $x^{\mu}$. Then the middle point $y^0$ represents the event
simultaneous with $x^{\mu}$.}\label{ch06:fig6.7}
\end{figure}

\begin{figure}[t] \centering
\includegraphics[width=200pt, height=130pt]{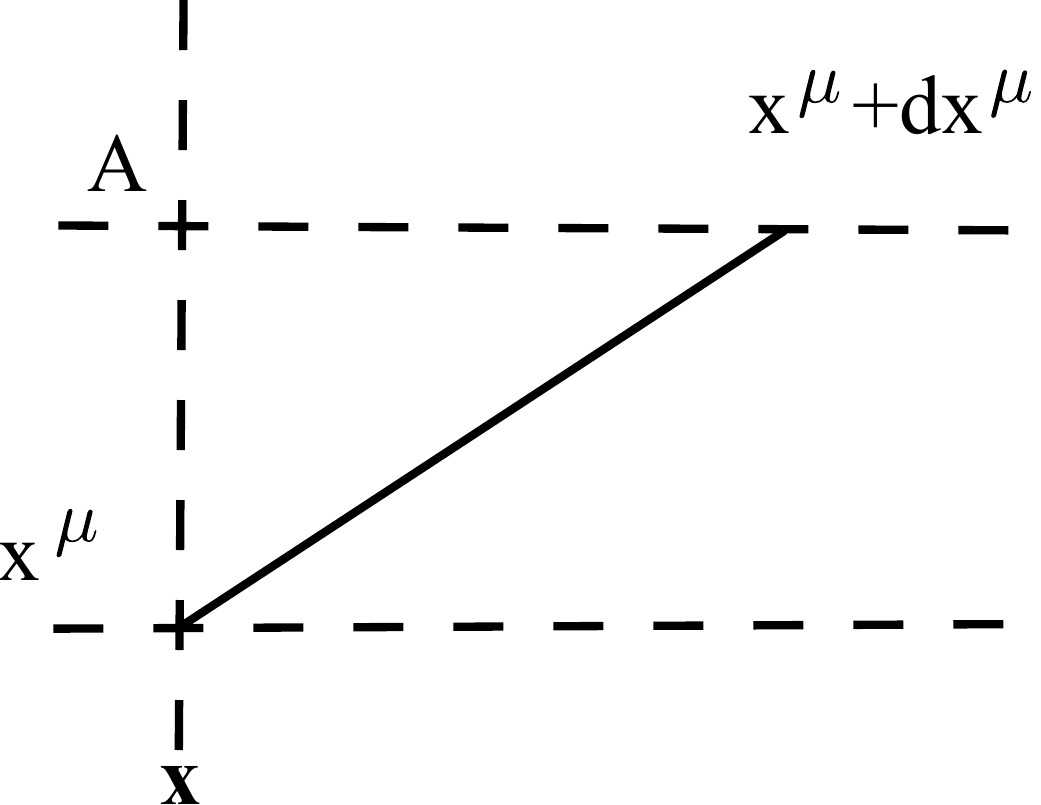}
\caption{Time and distance between the events $x^{\mu}$ and $x^{\mu}+dx^{\mu}$. Equation (\ref{La.6.3}) determines the
event $A$ (at spatial point $\textbf{x}$) simultaneous with $x^{\mu}+dx^{\mu}$. So the time interval between $x^{\mu}$
and $x^{\mu}+dx^{\mu}$ coincide with the interval between $x^{\mu}$ e $A$, and is given by (\ref{La.6.4}). Distance
between $x^{\mu}$ and $x^{\mu}+dx^{\mu}$ coincide with the distance between $x^\mu+ dx^\mu$ and $A$, the latter is
given in (\ref{La.6.5}).}\label{ch06:fig6.8}
\end{figure}

These rather formal tricks are based \cite{bib16} on the notion of simultaneity in general relativity and on the
analysis of flat limit. The four-interval of special relativity has direct physical interpretation in two cases. First,
for two events which occur at the same point, the four-interval is proportional to time interval, $dt=-\frac{ds}{c}$.
Second, for simultaneous events the four-interval coincides with distance, $dl=ds$. Assuming that the same holds in
general relativity, let us analyze infinitesimal time interval and distance between two events with coordinates $x^\mu$
and $x^\mu+dx^\mu$. The world line $y^\mu=(y^0, {\bf y}=\mbox{const})$ is associated with laboratory clock placed at
the spatial point ${\bf y}$. So the time-interval between the events $(y^0, {\bf y})$ and $(y^0+dy^0, {\bf y})$
measured by the clock is
\begin{eqnarray}\label{La.6.020}
dt=-\frac{ds}{c}=\frac{\sqrt{-g_{00}}}{c}dy^0.
\end{eqnarray}
Consider the event $x^\mu$ infinitesimally closed to the world line $(y^0,{\bf y}=\mbox{const})$. To find the event on
the world line which is simultaneous with $x^\mu$, we first look for the events $y^\mu_{(1)}$ and $y^\mu_{(2)}$ which
have null-interval with $x^\mu$, $ds(x^\mu, y^\mu_{(a)})=0$. The equation $g_{\mu\nu}dx^\mu dx^\nu=0$ with
$dx^\mu=x^\mu-y^\mu$ has two solutions $dx^0_{\pm}=\frac{g_{0i}dx^i}{-g_{00}}\pm\frac{\sqrt{d{\bf x}\gamma d{\bf
x}}}{\sqrt{-g_{00}}}$, then $y^0_{(1)}=x^0-dx^0_{+}$ and $y^0_{(2)}=x^0-dx^0_{-}$. Second, we compute the middle point
\begin{eqnarray}\label{La.6.1}
y^0=\frac{1}{2}(y^0_{(1)}+y^0_{(2)})=x^0+\frac{g_{0i}dx^i}{g_{00}}.
\end{eqnarray}
By definition\footnote{In the flat limit the sequence $y^\mu_{(1)}$, $x^\mu$, $y^\mu_{(2)}$ of events can be associated
with emission, reflection and absorbtion of a photon with the propagation law $ds=0$. Then the middle point
(\ref{La.6.1}) should be considered simultaneous with $x^0$.}, the event $(y^0, {\bf y})$ with the null-coordinate
(\ref{La.6.1}) is simultaneous with the event $(x^0, {\bf x})$, see Fig.~\ref{ch06:fig6.7} on page
\pageref{ch06:fig6.7}.
By this way we synchronized clocks at the spatial points ${\bf x}$ and ${\bf y}$. According to (\ref{La.6.1}), the
simultaneous events have different null-coordinates, and the difference $dx^0$ obeys the equation
\begin{eqnarray}\label{La.6.2}
dx^0+\frac{g_{0i}dx^i}{g_{00}}=0.
\end{eqnarray}
Consider a particle which propagated from $x^\mu$ to $x^\mu+dx^\mu$. Let us compute time-interval and distance between
these two events. According to (\ref{La.6.1}),  the event
\begin{eqnarray}\label{La.6.3}
\left(x^0+dx^0+\frac{g_{0i}dx^i}{g_{00}}, ~  {\bf x}\right),
\end{eqnarray}
at the spatial point ${\bf x}$ is simultaneous with $x^\mu+dx^\mu$, see Fig.~\ref{ch06:fig6.8} on page
\pageref{ch06:fig6.8}.
Equation (\ref{La.6.3}) determines the event $A$ (at spatial point $\textbf{x}$) simultaneous with $x^{\mu}+dx^{\mu}$.
So the time interval between $x^{\mu}$ and $x^{\mu}+dx^{\mu}$ coincide with the interval between $x^{\mu}$ e $A$.
Distance between $x^{\mu}$ and $x^{\mu}+dx^{\mu}$ coincide with the distance between $x^\mu+ dx^\mu$ and $A$.

According to (\ref{La.6.020}) and (\ref{La.6.1}), the time interval between the events $x^\mu$ and (\ref{La.6.3}) is
\begin{eqnarray}\label{La.6.4}
dt=\frac{\sqrt{-g_{00}}}{c}(dx^0+\frac{g_{0i}}{g_{00}}dx^i).
\end{eqnarray}
Since the events $x^\mu+dx^\mu$ and (\ref{La.6.3}) are simultaneous, this equation gives also the time interval between
$x^\mu$ and  $x^\mu+dx^\mu$. Further, the difference of coordinates between the events $x^\mu+dx^\mu$ and
(\ref{La.6.3}) is $dz^\mu=(-\frac{g_{0i}dx^i}{g_{00}}, dx^i)$. As they are simultaneous, the distance between them is
\begin{eqnarray}\label{La.6.5}
dl^2=-ds^2=g_{\mu\nu}dz^\mu dz^\nu=(g_{ij}-\frac{g_{0i}g_{0j}}{g_{00}})dx^idx^j\equiv\gamma_{ij}dx^idx^j.
\end{eqnarray}
Since (\ref{La.6.3}) occur at the same spatial point as $x^\mu$, this equation gives also the distance between $x^\mu$
and $x^\mu+dx^\mu$. The equations (\ref{La.6.4}) and (\ref{La.6.5}) coincide with the formal definitions presented
above, Eqs. (\ref{La.3.0}) and (\ref{La.3}).

\subsection{Three-dimensional acceleration and maximum speed of a particle in geodesic motion.}

We now turn to the definition of three-acceleration. Point particle in general relativity follows a geodesic line, and
we expect that during its evolution in gravitational field the particle can not reach the speed of light. This implies
that longitudinal acceleration should vanish when speed of the particle approximates to $c$. To analyze this, we first
use geodesic equation to obtain the derivative $\frac{d v^i}{dt}$ of coordinate of the velocity vector.

If we take the proper time to be the parameter, geodesics obey the system
\begin{eqnarray}\label{L.6.10}
\nabla_s\frac{dx^\mu}{ds}\equiv\frac{d^2x^\mu}{ds^2}+\Gamma^\mu{}_{\alpha\beta}\frac{dx^\alpha}{ds}\frac{dx^\beta}{ds}=0,
\qquad g_{\mu\nu}\frac{dx^\mu}{ds}\frac{dx^\nu}{ds}=-1,
\end{eqnarray}
where
\begin{eqnarray}\label{L.6.10.1}
\Gamma^\mu{}_{\alpha\beta}=\frac12g^{\mu\nu}(\partial_\alpha g_{\nu\beta}+\partial_\beta g_{\alpha\nu}-\partial_\nu
g_{\alpha\beta}).
\end{eqnarray}
Due to this definition, the system (\ref{L.6.10}) obeys the identity
$g_{\mu\nu}\frac{dx^\mu}{ds}\nabla_s\frac{dx^\nu}{ds}=0$. The system in this parametrization has no sense for the case
we are interested in, $ds^2\rightarrow 0$. So we rewrite it in arbitrary parametrization $\tau$
\begin{eqnarray}\label{L.6.11}
\frac{d\tau}{ds}\frac{d}{d\tau}\left(\frac{d\tau}{ds}\frac{d x^\mu}{d\tau}\right)+\left(\frac{d\tau}{ds}\right)^2
\Gamma^\mu{}_{\alpha\beta}(g)\frac{dx^\alpha}{d\tau}\frac{dx^\beta}{d\tau}=0, \qquad
\frac{d\tau}{ds}=\frac{1}{\sqrt{-\dot xg\dot x}}, \qquad \qquad \qquad \nonumber
\end{eqnarray}
this yields the equation of geodesic line in reparametrization-invariant form
\begin{eqnarray}\label{L.6.12}
\frac{1}{\sqrt{-\dot xg\dot x}}\frac{d}{d\tau}\left(\frac{\dot x^\mu}{\sqrt{-\dot xg\dot
x}}\right)=-\Gamma^\mu{}_{\alpha\beta}(g) \frac{\dot x^\alpha}{\sqrt{-\dot xg\dot x}}\frac{\dot x^\beta}{\sqrt{-\dot
xg\dot x}}.
\end{eqnarray}
Using the reparametrization invariance, we set $\tau=x^0$, then Eqs. (\ref{La.2}) and (\ref{La.3.1}) imply $\sqrt{-\dot
xg\dot x}=\frac{dt}{dx^0}\sqrt{c^2-{\bf v}\gamma{\bf v}}$, and spatial part of (\ref{L.6.12}) reads
\begin{eqnarray}\label{La.12}
\left(\frac{dt}{dx^0}\right)^{-1}\frac{d}{dx^0}\frac{v^i}{\sqrt{c^2-{\bf v}\gamma{\bf v}}}=-\frac{1}{\sqrt{c^2-{\bf
v}\gamma{\bf v}}}\Gamma^i{}_{\mu\nu}v^\mu v^\nu,
\end{eqnarray}
where we have denoted
\begin{eqnarray}\label{La.5.1}
v^\mu=\left(\frac{dt}{dx^0}\right)^{-1}\frac{dx^\mu}{dx^0}=(\left(\frac{dt}{dx^0}\right)^{-1}, ~  {\bf v}).
\end{eqnarray}
Direct computation of the derivative on l. h. s. of equation (\ref{La.12}) leads to the desired expression
\begin{eqnarray}\label{La.13}
\left(\frac{dt}{dx^0}\right)^{-1}\frac{dv^i}{dx^0}=-\frac{v^i}{2c^2}({\bf v}\partial_t\gamma{\bf
v})-\frac{v^i}{c^2}({\bf v}\gamma)_p\tilde\Gamma^p{}_{jk}(\gamma)v^jv^k-\tilde M^i{}_j\Gamma^j{}_{\mu\nu}(g)v^\mu
v^\nu,
\end{eqnarray}
where
\begin{eqnarray}\label{La.17.1}
\tilde M^i{}_j=\delta^i{}_j-\frac{v^i({\bf v}\gamma)_j}{c^2}, \quad \mbox{then} \quad \tilde M^i{}_jv^j=\frac{c^2-{\bf
v}\gamma{\bf v}}{c^2}v^i, \quad ({\bf v}\gamma)_i\tilde M^i{}_j=\frac{c^2-{\bf v}\gamma{\bf v}}{c^2}({\bf v}\gamma)_j,
\end{eqnarray}
and three-dimensional Christoffel symbols $\tilde\Gamma^i{}_{jk}(\gamma)$ are constructed with help of
three-dimensional metric $\gamma_{ij}(x^0, x^k)$, where $x^0$ is considered as a parameter
\begin{eqnarray}\label{La.8.1}
\tilde\Gamma^i{}_{jk}(\gamma)=\frac12\gamma^{in}(\partial_j\gamma_{nk}+\partial_k\gamma_{nj}-\partial_n\gamma_{jk}).
\end{eqnarray}
We have $g^{ij}\gamma_{jk}=\delta^{i}{}_k$, so the inverse metric of  $\gamma_{ij}$ turns out to be
$\gamma^{ij}=g^{ij}$. Note that $\tilde M^i{}_jv^j{\stackrel{|{\bf v}|\rightarrow c}\longrightarrow}0$, that is in the
limit the matrix $\tilde M$ turns into the projector on the plane orthogonal to ${\bf v}$.

If we project the derivative (\ref{La.13}) on the direction of motion, we obtain the expression
\begin{eqnarray}\label{La.14}
({\bf v}\gamma\frac{d{\bf v}}{dt})=-\frac{({\bf v}\gamma{\bf v})}{2c^2}({\bf v}\partial_t\gamma{\bf v})-\frac{({\bf
v}\gamma{\bf v})}{c^2}({\bf v}\gamma)_p\tilde\Gamma^p{}_{jk}(\gamma)v^jv^k-\frac{\sqrt{c^2-{\bf v}\gamma{\bf
v}}}{c^2}({\bf v}\gamma)_j\Gamma^j{}_{\mu\nu}(g)v^\mu v^\nu.
\end{eqnarray}
Due to the first and second terms on r. h. s., this expression does not vanish as ${\bf v}\gamma{\bf v}\rightarrow
c^2$. Note that this remains true for stationary metric, $g_{\mu\nu}({\bf x})$, or even for static metric, $g_{00}=1$,
$g_{0i}=0$! The reason is that the derivative $\frac{dv^i}{dt}$ in our three-dimensional geometry consist of three
contributions: variation rate of the vector field ${\bf v}$ itself, variation of basis in the passage from ${\bf x}$ to
${\bf x}+d{\bf x}$, and variation of the metric $\gamma_{ij}$ during the time interval $dt$. Excluding the last two
contributions, we obtain the variation rate of velocity itself, that is an acceleration
\begin{eqnarray}\label{La.8.2}
a^i=\left(\frac{dt}{dx^0}\right)^{-1}\frac{dv^i}{dx^0}+\tilde\Gamma^i{}_{jk}(\gamma)v^jv^k+\frac12({\bf
v}\partial_t\gamma\gamma^{-1})^i\equiv \nabla_tv^i+\frac12({\bf v}\partial_t\gamma\gamma^{-1})^i.
\end{eqnarray}
For the special case of stationary field, $g_{\mu\nu}({\bf x})$, our definition reduces to that of Landau and Lifshitz,
see page 251 in \cite{bib16}. Complementing $\frac{dv^i}{dt}$ in Eq. (\ref{La.13}) up to the acceleration, we obtain
three-dimensional acceleration of the particle moving along the geodesic line (\ref{L.6.12})
\begin{eqnarray}\label{La.11}
a^i=\tilde M^i{}_j\left[\frac12(\gamma^{-1}\partial_t\gamma{\bf v})^j-\Gamma^j_{\mu\nu}(g)v^\mu
v^\nu+\tilde\Gamma^j{}_{kl}(\gamma)v^kv^l\right].
\end{eqnarray}
This is the second Newton law for geodesic motion. Contracting this with $({\bf v}\gamma)_i$, we obtain the
longitudinal acceleration
\begin{eqnarray}\label{La.20}
{\bf v}\gamma{\bf a}=\left(1-\frac{({\bf v}\gamma{\bf v})}{c^2}\right)\left[\frac12({\bf v}\partial_t\gamma{\bf
v})+({\bf v}\gamma)_i[-\Gamma^i_{\mu\nu}(g)v^\mu v^\nu+\tilde\Gamma^i{}_{kl}(\gamma)v^kv^l]\right].
\end{eqnarray}
This implies ${\bf v}\gamma{\bf a}\rightarrow 0$ as ${\bf v}\gamma{\bf v}\rightarrow c^2$.

Let us confirm that $c$ is the only special point of the function (\ref{La.20}). Using Eqs. (\ref{L.6.10.1}),
(\ref{La.3})-(\ref{La.50.1}) and (\ref{La.8.1}) together with the identities
\begin{eqnarray}\label{La.20.1}
\gamma_{ij}g^{jk}=\delta_i{}^k, \qquad \gamma_{ij}g^{j0}=-\frac{g_{0i}}{g_{00}},
\end{eqnarray}
we can present the right hand side of Eq. (\ref{La.20}) in terms of initial metric as follows
\begin{eqnarray}\label{La.20.2}
{\bf v}\gamma{\bf a}=\frac{c^2-{\bf v}\gamma{\bf
v}}{2c\sqrt{-g_{00}}}\left\{\frac{c}{\sqrt{-g_{00}}}[\left(\frac{dt}{dx^0}\right)^{-1}\partial_0
g_{00}+v^k\partial_kg_{00}]-\right.\cr\left. \partial_0 g_{00}\left(\frac{dt}{dx^0}\right)^{-2}-2\partial_0
g_{0k}\left(\frac{dt}{dx^0}\right)^{-1}v^k-\partial_0 g_{kl}v^kv^l\right\}\equiv \cr \frac{c^2-{\bf v}\gamma{\bf
v}}{2c\sqrt{-g_{00}}}\left\{\frac{c}{\sqrt{-g_{00}}}v^\mu\partial_\mu g_{00}-\partial_0g_{\mu\nu}v^\mu v^\nu\right\}.
\qquad
\end{eqnarray}
The quantity $v^\mu$ has been defined in (\ref{La.5.1}). Excluding $v^0$ according to this expression, we obtain
\begin{eqnarray}\label{La.20.3}
{\bf v}\gamma{\bf a}=\frac{c^2-{\bf v}\gamma{\bf v}}{2\sqrt{-g_{00}}}\left\{\frac{v^k\partial_k
g_{00}}{\sqrt{-g_{00}}}-2\partial_0\left(\frac{g_{0i}}{\sqrt{-g_{00}}}\right)v^i-\frac{1}{c}\partial_0\gamma_{ij}v^i
v^j\right\}.
\end{eqnarray}
For the stationary metric, $g_{\mu\nu}(x^k)$, the equation (\ref{La.20.3}) acquires a specially simple form
\begin{eqnarray}\label{La.20.4}
{\bf v}\gamma{\bf a}=-(c^2-{\bf v}\gamma{\bf v})\frac{v^k\partial_kg_{00}}{2g_{00}}.
\end{eqnarray}
This shows that the longitudinal acceleration has only one special point, ${\bf v}\gamma{\bf a}\rightarrow 0$ as ${\bf
v}\gamma{\bf v}\rightarrow c^2$. Hence the spinless particle in the stationary gravitational field can not overcome the
speed of light. Then the same is true in general case (\ref{La.20.2}), at least for the metric which is sufficiently
slowly varied in time.

While we have discussed the geodesic equation, the computation which leads to the formula (\ref{La.20}) can be repeated
for a more general equation. Using the factor $\sqrt{-\dot xg\dot x}$ we construct the reparametrization-invariant
derivative
\begin{eqnarray}\label{La.20.7}
D=\frac{1}{\sqrt{-\dot xg\dot x}}\frac{d}{d\tau}.
\end{eqnarray}
Consider the reparametrization-invariant equation of the form
\begin{eqnarray}\label{La.20.8}
DDx^\mu(\tau)={\mathcal F}^\mu(Dx^\nu, \ldots ).
\end{eqnarray}
and suppose that the three-dimensional geometry is defined by $g_{\mu\nu}$ according to equations
(\ref{La.3.0})-(\ref{La.5}). Then Eq. (\ref{La.20.8}) implies the three-acceleration
\begin{eqnarray}\label{La.20.9}
a^i=\tilde M^i{}_j\left[(c^2-{\bf v}\gamma{\bf v}){\mathcal F}^j+ \tilde\Gamma^j{}_{kl}(\gamma)v^kv^l\right]+ \cr
\frac12\left(\frac{dt}{dx^0}\right)^{-1}\left[({\bf v}\partial_0\gamma\gamma^{-1})^i-\frac{v^i}{c^2}({\bf
v}\partial_0\gamma{\bf v})\right],
\end{eqnarray}
and the longitudinal acceleration
\begin{eqnarray}\label{La.20.10}
{\bf v}\gamma{\bf a}=\frac{(c^2-{\bf v}\gamma{\bf v})^2}{c^2}({\bf v}\gamma{\boldsymbol{\mathcal F}})+ \qquad \qquad
\qquad \cr \frac{c^2-{\bf v}\gamma{\bf v}}{c^2} \left[({\bf
v}\gamma)_i\tilde\Gamma^i{}_{kl}(\gamma)v^kv^l+\frac12\left(\frac{dt}{dx^0}\right)^{-1}({\bf v}\partial_0\gamma{\bf
v})\right].
\end{eqnarray}
The spatial part of the force is ${\mathcal F}^i={\mathcal F}^i(\frac{v^\nu}{\sqrt{c^2-{\bf v}\gamma{\bf v}}})$, where
$v^\mu$ is given by (\ref{La.5.1}), and the connection $\tilde\Gamma^i{}_{kl}(\gamma)$ is constructed with help of the
three-dimensional metric $\gamma_{ij}=(g_{ij}-\frac{g_{0i}g_{0j}}{g_{00}})$ according to (\ref{La.8.1}). For the
geodesic equation in this notation we have ${\mathcal F}^i=-\Gamma^i{}_{\mu\nu}\frac{v^\mu v^\nu}{c^2-{\bf v}\gamma{\bf
v}}$. With this ${\mathcal F}^i$ the equations (\ref{La.20.9}) and (\ref{La.20.10}) coincide with (\ref{La.11}) and
(\ref{La.20}).

Eq. (\ref{La.20.10}) shows that potentially dangerous forces are of degree four or more, ${\mathcal F}^j\sim(Dx)^4$.

\subsection{Parallel transport in three-dimensional geometry.}

Now we consider an arbitrary vector/tensor field in the space with three-dimensional geometry determined by a non
static metric $\gamma_{ij}(x^0, {\bf x})$. Variation rate of the field along a curve ${\bf x}(x^0)$ should be defined
in such a way, that it coincides with (\ref{La.8.2}) for the velocity. Let us show, how this definition follows from a
natural geometric requirements \cite{deriglazovMPL2015}. $1+3$ splitting preserves covariance of the formalism under
the following subgroup of general-coordinate transformations: $x^0=x'{}^0$, $x^i=x^i(x'{}^j)$. Under these
transformations $g_{00}$ is a scalar function, $g_{0i}$ is a vector while $g_{ij}$ and $\gamma_{ij}$ are tensors, then
the conversion factor (\ref{La.3.1}) is a scalar function and the velocity (\ref{La.5}) is a vector. So it is
convenient to introduce the usual covariant derivative $\nabla_k$ of a vector field $\xi^i(x^0, {\bf x})$ in the
direction of $x^k$
\begin{eqnarray}\label{La.8}
\nabla_k\xi^i=\frac{\partial \xi^i}{\partial x^k}+\tilde\Gamma^i{}_{kj}(\gamma)\xi^j,
\end{eqnarray}
with the Christoffel symbols (\ref{La.8.1}). By construction, the metric $\gamma$ is covariantly constant,
$\nabla_k\gamma_{ij}=0$. For the field $\xi^i(x^0, {\bf x}(x^0))$ along the curve ${\bf x}(x^0))$ we have the covariant
derivative in the direction of $x^0$
\begin{eqnarray}\label{La.8}
\nabla_0\xi^i=\frac{d\xi^i}{dx^0}+\tilde\Gamma^i{}_{jk}(\gamma)\frac{dx^j}{dx^0}\xi^k=\frac{\partial\xi^i}{\partial
x^0}+\frac{dx^k}{dx^0}\nabla_k\xi^i.
\end{eqnarray}
To define the variation rate of $\boldsymbol{\xi}$, we need the notion of a constant field (or, equivalently, the
parallel-transport equation). In Euclidean space the scalar product of two constant fields does not depend on the point
where it was computed. In particular, taking the scalar product along a line ${\bf x}(x^0)$, we have
$\frac{d}{dx^0}(\boldsymbol{\xi}, \boldsymbol{\eta})=0$. For the constant fields in our case it is natural to demand
the same (necessary) condition: $\frac{d}{dx^0}[\xi^i(x^0)\gamma_{ij}(x^0, x^i(x^0))\eta^i(x^0)]=0$. Taking into
account that $\nabla_k\gamma_{ij}=0$, this condition can be written as follows
\begin{eqnarray}\label{La.8.3}
(\nabla_0\boldsymbol{\xi}+\frac12\boldsymbol{\xi}\partial_0\gamma\gamma^{-1}, \boldsymbol{\eta})+(\boldsymbol{\xi},
\nabla_0\boldsymbol{\eta}+\frac12\gamma^{-1}\partial_0\gamma\boldsymbol{\eta})=0. \nonumber
\end{eqnarray}
This will be satisfied, if a constant field is defined by the equation
\begin{eqnarray}\label{La.8.4}
\nabla_0\xi^i+\frac12(\xi\partial_0\gamma\gamma^{-1})^i=0.
\end{eqnarray}
This is the parallel-transport equation in our three-dimensional geometry. Deviation from the constant field is the
variation rate. Hence, when the l. h. s. does not vanish, it gives the variation rate, which we write with respect to
physical time:
\begin{eqnarray}\label{La.8.5}
\nabla_t\xi^i\equiv\left(\frac{dt}{dx^0}\right)^{-1}\left[\nabla_0\xi^i+\frac12({\bf
v}\partial_0\gamma\gamma^{-1})^i\right].
\end{eqnarray}
This result is in correspondence with the definition (\ref{La.8.2}) for acceleration.

\section{Behavior of ultra-relativistic MPTD-particle and the rainbow geometry induced by spin}
\label{ch09:sec9.11}

As we saw above, point particle in a gravitational field propagates along a geodesic line with the speed less then
speed of light. Let us study the influence of rotational degrees of freedom on the trajectory of a fast spinning
particle.

Using (\ref{r9.2}) and (\ref{r13}), we present MPTD-equations (\ref{r9})-(\ref{r11}) in the following form
\begin{eqnarray}\label{condition1a}
S^\mu{}_\nu\dot x^\nu-\frac{1}{8(mc)^2}(SS\theta\dot x)^\mu=0.
\end{eqnarray}
\begin{eqnarray}
\nabla\left[ \frac{ \tilde T^\mu_{\ \ \nu} \dot x^\nu}{\sqrt{-\dot xG\dot x}} \right] =- \frac{1}{4mc} R^\mu{}_{
\nu\alpha\beta}S^{\alpha\beta}\dot x^\nu, \label{byM13} \\
\nabla S^{\mu\nu}= \frac{1}{4mc\sqrt{-\dot xG\dot x}}\dot x^{[\mu} S^{\nu]\sigma} \theta_{\sigma\alpha}\dot x^\alpha.
\label{motionJ-5}
\end{eqnarray}
The equations for trajectory and for precession of spin become singular at the critical velocity which obeys the equation
\begin{eqnarray}\label{pe1.1}
\dot xG\dot x=0.
\end{eqnarray}
The singularity determines behavior of the particle in ultra-relativistic limit. The effective metric is composed from
the original one plus (spin and field-dependent) contribution, $G=g+h(S)$. So we need to decide, which of them the
particle probes as the space-time metric. Let us consider separately the two possibilities.

Let us use $g$ to define the three-dimensional geometry  (\ref{La.3.0})-(\ref{La.5}). This leads to two problems. The
first problem is that the critical speed turns out to be slightly more than the speed of light.  To see this, we use
the supplementary spin condition (\ref{condition1a}) to write (\ref{pe1.1}) in the form
\begin{equation}\label{critical1}
-\left(\frac{dt}{dx^0}\right)^2\dot xG\dot x  = \left(c^2-{\bf v}\gamma{\bf v}\right)+\frac{1}{(2m^2c^2)^2}
\left(v\theta SS\theta v\right) =0,
\end{equation}
with $v^\mu$ defined in (\ref{La.5.1}). Using $S^{\mu\nu}=2\omega^{[\mu}\pi^{\nu]}$, we rewrite the last term as
follows:
\begin{equation}\label{critical2}
\left(c^2-{\bf v}\gamma{\bf v}\right)+ \frac{1}{(m^2c^2)^2}\left( \pi^2 (v\theta\omega)^2 + \omega^2(v\theta\pi)^2
\right) =0.
\end{equation}
As $\pi$ and $\omega$ are space-like vectors, the last term is non-negative, this implies $|{\bf v}_{cr}|\ge c$. Let us
confirm that generally this term is nonvanishing function of velocity, then $|{\bf v}_{cr}|> c$. Assume the contrary,
that this term vanishes at some velocity, then
\begin{eqnarray}\label{pe2}
v \theta \omega =\theta_{0i} \omega^i + \theta_{i0} v^i \omega^0 =0,\\
v \theta \pi = \theta_{0i} \pi^i +\theta_{i0} v^i \pi^0 =0.
\end{eqnarray}
We analyze these equations in the following special case.  Consider a space with covariantly-constant curvature
$\nabla_\mu R_{\mu\nu\alpha\beta} =0$. Then $\frac{d}{d\tau}(\theta_{\mu\nu}S^{\mu\nu})=2\theta_{\mu\nu}\nabla
S^{\mu\nu}$, and using (\ref{motionJ-5}) we conclude that $\theta_{\mu\nu}S^{\mu\nu}$ is an integral of motion. We
further assume that the only non vanishing part is the electric part of the curvature, $R_{0i0j}=K_{ij}$. Then the
integral of motion acquires the form
\begin{eqnarray}\label{pe3}
\theta_{\mu\nu} S^{\mu\nu} = 2 K_{ij}S^{0i}S^{0j}.
\end{eqnarray}
Let us take the initial conditions for spin such that $K_{ij}S^{0i}S^{0j}\ne 0$, then this holds at any future instant.
Contrary to this,  the system (\ref{pe2}) implies $K_{ij}S^{0i}S^{0j}=0$. Thus, the critical speed does not always
coincide with the speed of light and, in general case, we expect that ${\bf v}_{cr}$ is both field and spin-dependent
quantity.

The second problem is that acceleration of MPTD-particle grows up in the ultra-relativistic limit. In the spinless
limit the equation (\ref{byM13}) turn into the geodesic equation. Spin causes deviations from the geodesic equation due
to right hand side of this equation, as well as due to the presence of the tetrad field $\tilde T$ and of the effective
metric $G$ in the left hand side. Due to the dependence of the tetrad field on the spin-tensor $S$, the singularity
presented in (\ref{motionJ-5}) causes the appearance of the term proportional to $\frac{1}{\sqrt{\dot xG\dot x}}$ in
the expression for longitudinal acceleration. In the result, the acceleration grows up to infinity as the particle's
speed approximates to the critical speed. To see this, we separate derivative of $\tilde T$ in  Eq. (\ref{byM13})
\begin{eqnarray}\label{pe4}
\nabla\left[ \frac{\dot x^\mu}{\sqrt{-\dot xG\dot x}} \right]=- T^\mu_{\ \alpha} \left( \nabla \tilde T^\alpha_{\
\beta} \right) \frac{\dot x^\beta}{\sqrt{-\dot xG\dot x}} -\frac{1}{4mc} T^\mu_{\ \nu} (\theta \dot x)^\nu, \label{WD1}
\end{eqnarray}
where $T$ is the inverse for $\tilde T$. Using (\ref{motionJ-5}) we obtain
\begin{equation}\label{pe5}
\left[ \nabla \tilde T^\mu{}_\nu\right] \dot x^\nu
=-\frac{S^{\mu\alpha}}{8m^2c^2}\left[\frac{R_{\alpha\nu\beta\sigma}\dot x^\beta (S\theta \dot x)^{\sigma}}{2mc
\sqrt{-\dot x G \dot x}}   +S^{\beta\sigma} (\nabla R_{\alpha\nu\beta\sigma}) \right] \dot x^\nu.
\end{equation}
Using this expression, the equation (\ref{pe4}) reads\footnote{Three-dimensional geometry is defined now by $g$, in
this case we can not use Eqs. (\ref{La.20.9}) and (\ref{La.20.10}) to estimate the acceleration.}
\begin{equation}\label{pe6}
\frac{d}{d\tau}\left[ \frac{\dot x^\mu}{\sqrt{-\dot xG\dot x}} \right] = \frac{f^\mu}{\sqrt{-\dot xG\dot x}},
\end{equation}
where we denoted
\begin{eqnarray}\label{pe7}
f^\mu \equiv \frac{1}{8(mc)^2}(TS)^{\mu\alpha}\left[\frac{R_{\alpha\nu\beta\sigma}\dot x^\beta (S\theta \dot
x)^{\sigma}}{2mc \sqrt{-\dot x G \dot x}} +S^{\beta\sigma} (\nabla R_{\alpha\nu\beta\sigma}) \right] \dot x^\nu -\cr
\left( \Gamma  \dot x \dot x \right)^\mu - \frac{\sqrt{-\dot xG\dot x}}{4mc} \left( T \theta \dot x \right)^\mu. \qquad
\qquad
\end{eqnarray}
It will be sufficient to consider static metric $g_{\mu\nu}({\bf x})$ with $g_{0i}=0$. Then three-dimensional metric
and velocity are
\begin{eqnarray}\label{pe7.1}
\gamma_{ij} = g_{ij}  , \quad v^i = \frac{c}{\sqrt{-g_{00}}} \frac{dx^i}{dx^0},
\end{eqnarray}
Taking $\tau=x^0$, the spatial part of equation (\ref{pe6}) with this metric reads
\begin{equation} \label{pe8}
\left(\frac{dt}{dx^0}\right)^{-1} \frac{d}{dx^0} \left[ \frac{v^i}{\sqrt{-vGv}} \right] = \frac{f^i(v)}{\sqrt{-vGv}}.
\end{equation}
with $v^\mu$ defined in (\ref{La.5.1}), for the case
\begin{equation} \label{pe8.1}
v^\mu=(\frac{c}{\sqrt{-g_{00}}}, ~ {\bf v}),
\end{equation}
and
\begin{equation} \label{pe8.2}
-vGv=-v\tilde Tv=c^2-{\bf v}g{\bf v}+\frac{(vS\theta v)}{8m^2c^2}.
\end{equation}
In the result, we have presented the equation for trajectory in the form convenient for analysis of acceleration, see
(\ref{La.12}). Using  the definition of three-dimensional covariant derivative (\ref{La.8.2}), we present the
derivative on the l.h.s. of (\ref{pe8}) as follows
\begin{equation}\label{pe9}
\frac{d}{dx^0} \left[ \frac{v^i}{\sqrt{-vGv}} \right] = \frac{1}{\sqrt{-vGv}}\left[ {\mathcal M}^i_{\ k} \nabla_0 v^k -
\tilde\Gamma(\gamma) ^i_{jk}v^jv^k \frac{dt}{dx^0} +\frac{Kv^i}{2(-vGv)} \right],
\end{equation}
We have denoted
\begin{eqnarray}\label{pe9.1}
K=(\nabla_0 G_{\mu\nu}) v^\mu v^\nu - v^\mu G_{\mu 0}v^k \partial_k \ln{ (-g_{00}}), \cr {\mathcal M}^i{}_{k} =
\delta^i{}_k -\frac{v^iv^\mu G_{\mu k}}{vGv}. \qquad \qquad
\end{eqnarray}
The matrix ${\mathcal M}^i{}_{k}$ has the inverse
\begin{equation}\label{pe10}
\tilde{\mathcal M}^i{}_{k} = \delta^i{}_k + \frac{ v^iv^\mu G_{\mu k}}{v^\sigma G_{\sigma 0} v^0}, \quad \mbox{then}
\quad \tilde{\mathcal M}^i{}_{k}v^k=v^i\frac{vGv}{v^\sigma G_{\sigma 0} v^0}.
\end{equation}
Combining these equations, we obtain the three-acceleration of our spinning particle
\begin{eqnarray}\label{pe11}
a^i=\left(\frac{dt}{dx^0}\right)^{-1} \nabla_0 v^i = \tilde{\mathcal M}^i_{\ k}\left[ f^k + (\tilde \Gamma v v)^k
\right]+ \frac{ K v^i}{2v^\sigma G_{\sigma 0} }.
\end{eqnarray}
Finally, using manifest form of $f^i$ from (\ref{pe7}) we have
\begin{eqnarray}\label{pe11.1}
a^i= \frac{ \tilde{\mathcal M}^i{}_{k} \hat S^k }{16(mc)^3\sqrt{-vGv}} -  c^2\tilde{\mathcal M}^i_{\
k}\frac{\gamma^{kj}\partial_j  g_{00}}{2g_{00}} -\frac{\sqrt{-vGv}}{4mc} \tilde{\mathcal M}^i_{\ k} (T\theta v)^k  +
\cr \frac{Kv^i}{2v^\sigma G_{\sigma 0}} + \frac{1}{8(mc)^2}\tilde{\mathcal M}^i_{\ k} (TS)^{k\alpha}
R_{\alpha\nu\beta\sigma;\lambda} S^{\beta\sigma} v^\nu v^\lambda. \qquad \qquad
\end{eqnarray}
The longitudinal acceleration is obtained by projecting $a^i$ on the direction of velocity, that is
\begin{eqnarray}\label{pe12}
({\bf v}\gamma {\bf a}) =  \frac{({\bf v}\gamma \tilde{\mathcal M})_k  \hat S^k}{16(mc)^3\sqrt{-vGv}} - c^2 ({\bf
v}\gamma \tilde{\mathcal M})_k\frac{\gamma^{kj}\partial_j g_{00}}{2g_{00}} -\frac{\sqrt{-vGv}}{4mc}({\bf v}\gamma
\tilde{\mathcal M})_k (T\theta v)^k  +\cr \frac{K }{2v^\sigma G_{\sigma 0}} ({\bf v}\gamma{\bf v})+
\frac{1}{8(mc)^2}({\bf v}\gamma \tilde{\mathcal M})_k (TS)^{k\alpha}  R_{\alpha\nu\beta\sigma;\lambda} S^{\beta\sigma}
v^\nu v^\lambda. \qquad \qquad
\end{eqnarray}
where  $\hat S^k = (TS)^{k\mu}R_{\mu\nu\alpha\beta}v^\nu v^\alpha (S\theta v)^\beta$. As the speed of the particle
closes to the critical velocity, the longitudinal acceleration diverges due to the first term in (\ref{pe12}). In
resume, assuming that MPTD-particle sees the original geometry $g_{\mu\nu}$, we have a theory with unsatisfactory
behavior in the ultra-relativistic limit.

Let us consider the second possibility, that is we take $G_{\mu\nu}$ to construct the three-dimensional geometry
(\ref{La.3.0})-(\ref{La.5}).  With these definitions we have, by construction, $-\dot x G \dot x=
(\frac{dt}{dx^0})^2(c^2 - ({\bf v}\gamma{\bf v}))$, so the critical speed coincides with the speed of light. In the
present case, the expression for three-acceleration can be obtained in closed form for an arbitrary curved background.
Taking $\tau = x^0$ the spatial part of (\ref{pe6}) implies
\begin{eqnarray}\label{La.12-a}
\left(\frac{dt}{dx^0}\right)^{-1}\frac{d}{dx^0} \left[ \frac{v^i}{\sqrt{c^2-{\bf v}\gamma{\bf v}}}\right]
=\frac{f^i(v)}{\sqrt{c^2-{\bf v}\gamma{\bf v}}}.
\end{eqnarray}
where, from (\ref{pe7}), $f^i$ is given by
\begin{eqnarray}\label{pe7.0}
f^i \equiv \frac{1}{8(mc)^2}(TS)^{i\alpha}\left[ \frac{R_{\alpha\nu\beta\sigma}v^\beta(S\theta
v)^{\sigma}}{2mc\sqrt{c^2- {\bf v}\gamma{\bf v}}} +S^{\beta\sigma} (\nabla R_{\alpha\nu\beta\sigma}) \right] v^\nu   -
\cr \Gamma^i{}_{\mu\nu}(G) v^\mu v^l\nu - \frac{\sqrt{c^2- {\bf v}\gamma{\bf v}}}{4mc} \left( T \theta v \right)^i.
\qquad
\end{eqnarray}
Equation (\ref{La.12-a}) is of the form (\ref{La.12}), so the acceleration is given by (\ref{La.11}) and (\ref{La.20})
where, for the present case, $\gamma_{i j } = G_{ij} -\frac{G_{0i} G_{0j}}{G_{00}}$
\begin{eqnarray}\label{La.11-a}
a^i=\tilde M^i{}_j[f^j+\tilde\Gamma^j{}_{kl}(\gamma)v^kv^l]+\frac12\left(\frac{dt}{dx^0}\right)^{-1}\left[({\bf
v}\partial_0\gamma\gamma^{-1})^i-\frac{({\bf v}\partial_0\gamma{\bf v})}{c^2}v^i\right], \qquad
\end{eqnarray}
\begin{eqnarray}\label{La.20-a}
{\bf v}\gamma{\bf a}=\left(1-\frac{{\bf v}\gamma{\bf v}}{c^2}\right)\left[ ({\bf
v}\gamma)_i[f^i(v)+\tilde\Gamma^i{}_{kl}(\gamma)v^kv^l]+\frac12\left(\frac{dt}{dx^0}\right)^{-1}({\bf
v}\partial_0\gamma{\bf v})\right]. \qquad
\end{eqnarray}
With $f^i$ given in (\ref{pe7.0}), the longitudinal acceleration vanishes as $v\rightarrow c$.

Let us resume the results of this section. Assuming that spinning particle probes the three-dimensional space-time
geometry determined by the original metric $g$, we have a theory with unsatisfactory ultra-relativistic limit. First,
the critical speed, which the particle can not overcome during its evolution in gravitational field, can be more then
the speed of light. The same observation has been made from analysis of MPTD-particle in specific metrics
\cite{Koch2015, Koch2016, Bal2017, Wang12017, Wang22017}. Second, the longitudinal acceleration grows up to infinity in
the ultra-relativistic limit. Assuming that the the particle sees the effective metric G(S) as the space-time metric,
we avoided the two problems. But the resulting theory still possess the problem.  The acceleration (\ref{La.11-a})
contains the singularity due to $f^i\sim\frac{1}{\sqrt{c^2-({\bf v}\gamma{\bf v})}}$, that is at $v=c$ the acceleration
becomes orthogonal to the velocity, but remains divergent. Besides, due to dependence of effective metric on spin, we
arrive at rather unusual picture of the Universe with rainbow geometry\footnote{Some models of doubly special
relativity predict rainbow geometry at Planck scale \cite{Mag_2003, Gorji16, Gorji216, Feng16}.}: there is no unique
space-time manifold for the Universe of spinning particles: each particle will probe his own three-dimensional
geometry. We conclude that MPTD equations do not seem promising candidate for the description of a relativistic
rotating body. It would be interesting to find their generalization with improved behavior in ultra-relativistic
regime. This will be achieved within the framework of vector model of spinning particle, which we shall describe in the
subsequent sections.

\section{Vector model of non relativistic spinning particle}
\label{ch09:sec9.01}

The data of some experiments with elementary particles and atoms (Stern--Gerlach experiment, fine structure of hydrogen
atom, Zeeman effect) shows that the Schr\"{o}dinger equation for a one-component wave function is not adequate to
describe the behavior of these systems in the presence of an electromagnetic field. This implies a radical modification
of the formalism. Besides the position and the momentum, the state of an electron is specified by some discrete
numbers, which are eigenvalues of suitably defined operators, called the operators of spin. The mathematical theory of
these operators is similar to the formalism of angular momentum. So, intuitively, an elementary particle carries an
intrinsic angular momentum called spin.

To describe a particle with spin $s = \frac{1}{2}$ we introduce the two-component wave function $\varPsi _\alpha $,
$\alpha = 1,2$. The spin operators\index{Operator of spin! non relativistic} $\hat {S}_i $ act on $\varPsi _\alpha $ as
$2\times 2$-matrices, and are defined by
\begin{align}
\hat {S}_i = \frac{\hbar }{2}\sigma _i , \label{ch08:eqn8.156}
\end{align}
where $\sigma_i$ stands for the \textit{Pauli matrices}\index{Pauli matrices}, they form a basis of the vector space of
traceless and Hermitian $2\times 2$-matrices,
\begin{align}
 {\sigma _1 = \left( {{\begin{array}{cc}
 0 & 1 \\
 1 & 0 \\
\end{array} }} \right),\quad \sigma _2 = \left( {{\begin{array}{cc}
 0  & { - i} \\
 i & 0 \\
\end{array} }} \right),\quad \sigma _3 = \left( {{\begin{array}{cc}
 1 & 0 \\
 0 & { - 1} \\
\end{array} }} \right).}
\label{ch08:eqn8.157}
\end{align}
Their basic algebraical properties are
\begin{gather}
\sigma _i \sigma _j + \sigma _j \sigma _i = 2\times {\rm {\bf
1}}\delta _{ij} , \label{ch08:eqn8.159}\\
\sigma _i \sigma _j - \sigma _j \sigma _i \equiv [\sigma _i ,\sigma _j ] = 2i\epsilon_{ijk} \sigma _k , \label{ch08:eqn8.160}\\
\sum\limits_i {\sigma _i } \sigma _i = 3\times {\rm {\bf 1}}.
\label{ch08:eqn8.161}
\end{gather}
Note that the commutators (\ref{ch08:eqn8.160}) of $\sigma$-matrices
are the same as for the angular-momentum vector. The spin
operators, being proportional to the Pauli matrices, have similar
properties, in particular
\begin{gather}
[\hat {S}_i ,\hat {S}_j ] = i\hbar \epsilon_{ijk} \hat {S}_k ,
\label{ch08:eqn8.162}\\
\hat{{\rm {\bf S}}}^2 = \hbar ^2s(s + 1)\times {\rm {\bf 1}} =
\frac{3\hbar ^2}{4}{\rm {\bf 1}}. \label{ch08:eqn8.163}
\end{gather}
Consider Coulomb electric and a constant magnetic fields. The electromagnetic potential can be taken in the form
$A_0=\frac{\alpha}{r}$ and ${\bf A}=\frac12[{\bf B}\times{\bf r}]$. Then evolution of an electron immersed in this
fields described by the equation
\begin{align}
i\hbar \frac{\partial \varPsi }{\partial t} = \left(
{\frac{1}{2m}(\hat{{\rm {\bf p}}} - \frac{e}{c}{\rm {\bf A}})^2 -
eA_0 + \frac{e(g-1)}{2m^2c^2}\hat{{\rm {\bf S}}}[\hat{\bf p}\times{\bf E}]- \frac{eg}{2mc}{\rm {\bf B}}\hat{{\rm {\bf S}}}} \right)\varPsi.
\label{ch08:eqn8.164}
\end{align}
The first and second terms in the Hamiltonian correspond to the minimal interaction of a point particle with an
electromagnetic potential, whereas the last two terms represent interaction of spin with electric and  magnetic fields.
A numeric factor $g$ is called gyromagnetic ratio of the electron\footnote{Quantum
electrodynamics gives $g=2.002322\ldots$ due to radiative corrections.}. The vector $\frac{eg}{2mc}\hat{{\rm {\bf S}}}$
is known as magnetic moment of the particle.

The equation is written in the Schr\"{o}dinger picture, that is we ascribe time-dependence to the wave function,
whereas in semiclassical models we deal with dynamical variables.  We recall that the time-dependence can be ascribed
to operators using the Heisenberg picture. Passing to the Heisenberg picture, we could write dynamical equations for
basic operators of the theory. According to Ehrenfest theorem, expectation values of the operators approximately obey
the classical Hamiltonian equations \cite{bib51}.

The equation (\ref{ch08:eqn8.164}) gives the structure and properties of the energy levels of hydrogen atom in a good
agreement with experiment. The fine structure of hydrogen atom fixes the factor $g-1$ in the third term, while Zeeman
effect requires the factor $g$ in the last term.

To formulate the problem that we wish to discuss, we
recall that quantum mechanics of a spinless particle can be
obtained applying the canonical quantization procedure to a
classical-mechanics system with the Lagrangian $L = \frac{1}{2}mx^2
- U(x)$. To achieve this, we construct a
Hamiltonian formulation for the system, then associate with the
phase-space variables the operators with commutators resembling the
Poisson brackets, and write on this base the Schr\"{o}dinger
equation $i\hbar \dot {\varPsi } = \hat {H}\varPsi $.

It is natural to ask whether this ideology can be realized for the spinning particle. Since the quantum-mechanical
description of a spin implies the use of three extra operators $\hat {S}_i $, the problem can be formulated as follows.
We look for a classical-mechanics system which, besides the position variables $x_{i}$, contains additional degrees of
freedom, suitable for the description of a spin: in the Hamiltonian formulation the spin should be described, \textit
{in the end}, by three variables with fixed square (\ref{ch08:eqn8.163}) and with the classical brackets $\{S_i , S_j
\} = \epsilon_{ijk} S_k $. Then canonical quantization of these variables will yield spin operators with the desired
properties properties (\ref{ch08:eqn8.162}) and (\ref{ch08:eqn8.163}). According to this, typical spinning-particle
model consist of a point on a world-line and some set of variables describing the spin degrees of freedom, which form
an inner space attached to that point\footnote{There is an elegant formalism developed by Berezin and Marinov
\cite{bib42} based on using of anticommuting (Grassmann) variables for the description of spin. We present here another
formulation based on commuting variables, without appealing to a rather formal methods of the Grassmann mechanics.}. In
fact, different spinning particles discussed in the literature differ by the choice of the inner space. An exceptional
case is the rigid particle \cite{86} which consist of only position variables, but with the action containing higher
derivatives. The model yields the Dirac equation \cite{Ner14}, hence it also can be used for description of spin.

It should be noted that equation (\ref{ch08:eqn8.164}) is written in the laboratory system, so we do not
state that our classical variable $S_i$ is a quantity defined in the instantaneous rest frame of the particle.

We intend to construct the spinning particle starting from a suitable variational problem. This is the first task we
need to solve, as the formulation of a variational problem in closed form is known only for the case of a phase space
equipped with canonical Poisson bracket, say $\{\omega_i, \pi_j\}=\delta_{ij}$. The number of variables and their
algebra are different from the number of spin operators and their commutators, (\ref{ch08:eqn8.162}). May be the most
natural way to arrive at the operator algebra (\ref{ch08:eqn8.162}) is to consider spin as a composite quantity,
\begin{eqnarray}\label{intr.9.1}
S_i=\epsilon_{ijk}\omega_j\pi_k, \quad \mbox{or}\quad {\bf S}=\boldsymbol\omega\times\boldsymbol\pi,
\end{eqnarray}
where ${\boldsymbol\omega}, \boldsymbol\pi$ are coordinates of a phase space equipped with canonical Poisson bracket.
This immediately induces $SO(3)$\,-algebra for $S_i$,
$\{S_i(\omega, \pi), S_j(\omega, \pi)\}_{PB}=\epsilon_{ijk}S_k$.
Unfortunately, this is not the whole story. First, we need some mechanism which explains why ${\bf S}$, not
$\boldsymbol{\omega}$ and $\boldsymbol{\pi}$ must be taken for the description of spin degrees of freedom. Second, the
basic space is six-dimensional, while the spin manifold is two-dimensional (we remind that the square of spin operator
has fixed value, Eq. (\ref{ch08:eqn8.163})). To improve this, we look for a variational problem which, besides dynamical
equations, implies the constraints
\begin{eqnarray}\label{intr12.1}
\boldsymbol\omega\boldsymbol\pi=0, \qquad  \boldsymbol\pi^2-\frac{\alpha}{\boldsymbol\omega^2}=0, \quad \mbox{where} \quad
\alpha=\frac{3\hbar^2}{4}.
\end{eqnarray}
According to Dirac's terminology \cite{bib08, bib10, deriglazov2010classical, AAD05, Ban2016}, they form the
first-class set, so in the model with these constraints the spin sector contains $6-2\times 2=2$ physical degrees of
freedom. Geometrically, the constraints determine four-dimensional $SO(3)$\,-invariant surface of the six-dimensional
phase space. The constraints imply the fixed value of spin
\begin{eqnarray}\label{intr.12}
{\bf S}^2={\boldsymbol{\omega}}^2{\boldsymbol{\pi}}^2-({\boldsymbol{\omega}}{\boldsymbol{\pi}})^2=\frac{3\hbar^2}{4}.
\end{eqnarray}
The same square of spin follows from the constraints
\begin{eqnarray}\label{intr.11}
{\boldsymbol{\omega}}^2=\alpha^2, \quad {\boldsymbol{\pi}}^2=\beta^2, \quad {\boldsymbol{\omega}}{\boldsymbol{\pi}}=0,
\end{eqnarray}
if we put $\beta^2=\frac{3\hbar^2}{4\alpha^2}$, any $\alpha$. The combination
$\boldsymbol{\pi}^2-\beta^2+\frac{\beta^2}{\alpha^2}(\boldsymbol{\omega}^2-\alpha^2)$ represents the first-class
constraint of the set (\ref{intr.11}). Hence the model with these constraints also has the desired number of degrees of
freedom, $6-2-1\times 2=2$.  The equalities (\ref{intr.11}) determine essentially unique $SO(3)$\,-invariant
three-dimensional surface of the phase space. The set (\ref{intr12.1}) turns out to be more convenient for
generalization to the case of a relativistic spin.

While ${\bf S}$ in (\ref{intr.9.1}) looks like an angular momentum, the crucial difference is due to the presence of
first-class constraints, and hence of a local symmetry which we refer as spin-plane symmetry. The latter acts on the
basic variables $\boldsymbol{\omega}$, $\boldsymbol{\pi}$, while leaves invariant the spin variable ${\bf{S}}$. Using
analogy with classical electrodynamics, $\boldsymbol{\omega}$ and $\boldsymbol{\pi}$ are similar to four-potential
$A^\mu$ while ${\bf{S}}$ plays the role of $F^{\mu\nu}$. The coordinates $\boldsymbol{\omega}$ of the "inner-space
particle" are not physical (observable) quantities. The only observable quantities are the gauge-invariant variables
${\bf{S}}$. So our construction realizes, in a systematic form, the oldest idea about spin as the "hidden angular
momentum".

\subsection{Lagrangian and Hamiltonian for the spin-sector}
\label{ch09:sec9.1}

As the Lagrangian which implies the constraints (\ref{intr12.1}), we take the expression
\begin{eqnarray}\label{nr2}
L_{spin}=\frac{\sqrt{\alpha}}{\sqrt{\boldsymbol{\omega}^2}}\sqrt{\dot{\boldsymbol{\omega}}N\dot{\boldsymbol{\omega}}},
\quad \alpha=\frac{3\hbar^2}{4},
\end{eqnarray}
where $N_{ij}=\delta_{ij}-\frac{\omega_i\omega_j}{\boldsymbol{\omega}^2}$ is the projector on the plane orthogonal to
$\boldsymbol{\omega}$: $N_{ij}\omega^j=0$, $N^2=N$.  The equivalent forms of the Lagrangian are
\begin{eqnarray}\label{nr2.1}
L_{spin}= \frac{\sqrt{\alpha}\sqrt{\boldsymbol{\omega}^2(\dot{\boldsymbol{\omega}})^2
-(\boldsymbol{\omega}\dot{\boldsymbol{\omega}})^2}}{\boldsymbol{\omega}^2}= \frac{\sqrt{\alpha}\sqrt{{\bf
S}^2}}{\boldsymbol{\omega}^2},
\end{eqnarray}
where $S_i=\epsilon_{ijk}\omega_j\dot\omega_k$. The model is manifestly invariant under global rotations,
$\omega'_i=R_{ij}\omega_j$, where $R^T=R^{-1}$.  There are also two (finite) local symmetries: reparametrizations
$t\rightarrow t'=\sigma(t)\equiv t+\epsilon(t)$, and the scale transformations
$\boldsymbol{\omega}\rightarrow\chi(t)\boldsymbol{\omega}$.

Let us construct the Hamiltonian formulation of the model. Equation for the conjugated momentum reads
$\boldsymbol{\pi}=\frac{\partial
L}{\partial\dot{\boldsymbol{\omega}}}=\frac{\sqrt{\alpha}}{\sqrt{\boldsymbol{\omega}^2}}
\frac{N\dot{\boldsymbol{\omega}}}{\sqrt{\dot{\boldsymbol{\omega}}N\dot{\boldsymbol{\omega}}}}$. This expression
immediately implies (\ref{intr12.1}) as the primary constraints. We also note the equality
$\boldsymbol{\pi}\dot{\boldsymbol{\omega}}=L$, that is $H_0=\boldsymbol{\pi}\dot{\boldsymbol{\omega}}-L=0$. So the
complete Hamiltonian is composed from the primary constraints,
$H=v(\boldsymbol{\omega}\boldsymbol{\pi})+v_1\left(\boldsymbol{\pi}^2-\frac{\alpha}{\boldsymbol{\omega}^2}\right)$, and
the Hamiltonian action reads
\begin{eqnarray}\label{nnr6}
S_H=\int dt ~  \boldsymbol{\pi}\dot{\boldsymbol{\omega}}-
v(\boldsymbol{\omega}\boldsymbol{\pi})-v_1\left(\boldsymbol{\pi}^2-\frac{\alpha}{\boldsymbol{\omega}^2}\right).
\end{eqnarray}
There are no of higher-stage constraints in the problem.

Let us write Hamiltonian counterparts of the Lagrangian local symmetries. \par
\noindent 1. Reparametrizations in extended phase space are
\begin{eqnarray}\label{nnr7}
t'=\sigma(t), \qquad \qquad   \boldsymbol{\omega}'=\boldsymbol{\omega}, \qquad \quad
\boldsymbol{\pi}'=\boldsymbol{\pi}, \qquad \cr  v'=(\dot\sigma)^{-1}v, \quad \quad v'_1=(\dot\sigma)^{-1}v_1.
\end{eqnarray}
They induce the transformations of dynamical variables
\begin{eqnarray}\label{nnr8}
\boldsymbol{\omega}'(\tau)=\boldsymbol{\omega}(\tilde\sigma(\tau)), \qquad \qquad
\boldsymbol{\pi}'(\tau)=\boldsymbol{\pi}(\tilde\sigma(\tau)), \cr \qquad
v'(\tau)=(\dot\sigma)^{-1}v(\tilde\sigma(\tau)), \quad \quad v'_1(\tau)=(\dot\sigma)^{-1}v_1(\tilde\sigma(\tau)).
\end{eqnarray}
Their infinitesimal form read
\begin{eqnarray}\label{nnr9}
\delta\boldsymbol{\omega}=-\epsilon\dot{\boldsymbol{\omega}}, \quad \delta\boldsymbol{\pi}=-\epsilon\dot{\boldsymbol{\pi}}, \quad
\delta v=-(\epsilon v)\dot{}, \quad \delta v_1=-(\epsilon v_1)\dot{}.
\end{eqnarray}
2. Scale transformations of coordinates are
\begin{eqnarray}\label{nnr10}
\tau'=\tau, \quad \boldsymbol{\omega}'=\chi\boldsymbol{\omega}, \quad \boldsymbol{\pi}'=\frac{1}{\chi}\boldsymbol{\pi},
\quad v'=v+\frac{\dot\chi}{\chi}, \quad v'_1=\chi^2v_1.
\end{eqnarray}
Since $\tau$ is not involved, the induced transformations of dynamical variables are the same, for instance
$\boldsymbol{\omega}'(\tau)=\chi\boldsymbol{\omega}(\tau)$. Presenting $\chi=1+\gamma$, infinitesimal transformations
of dynamical variables read
\begin{eqnarray}\label{nnr11}
\delta\boldsymbol{\omega}=\gamma\boldsymbol{\omega}, \quad \delta\boldsymbol{\pi}=-\gamma\boldsymbol{\pi} \quad
\delta v=\dot\gamma, \quad \delta v_1=2\gamma v_1.
\end{eqnarray}
Besides the constraints (\ref{intr12.1}), the variational problem (\ref{nnr6}) implies the Hamiltonian equations
\begin{eqnarray}\label{nnr12}
\dot{\boldsymbol{\omega}}=\rho{\boldsymbol{\omega}}^2{\boldsymbol{\pi}}+ v{\boldsymbol{\omega}}, \quad
\dot{\boldsymbol{\pi}}=-\rho{\boldsymbol{\pi}}^2{\boldsymbol{\omega}}-v{\boldsymbol{\pi}}.
\end{eqnarray}
To make the system more symmetric, we have introduced the variable $\rho=\frac{2v_1}{{\boldsymbol{\omega}}^2}$ instead of $v_1$.

According to general formalism of constrained systems \cite{bib08, bib10, deriglazov2010classical}, neither the
dynamical equations nor the constraints determine the variables $v$ and $\rho$. They enter as arbitrary functions of
time into general solution for the variables $\boldsymbol{\omega}$ and $\boldsymbol{\pi}$, making completely
undetermined their dynamics.  Indeed, for any given functions $v(t)$ and $\rho(t)$, the equations represent a normal
system for determining $\boldsymbol{\omega}$ and $\boldsymbol{\pi}$. Its general solution is
\begin{eqnarray}\label{nnr13}
{\boldsymbol{\omega}}=e^{\int_0^t vd\tau}\left[{\bf b}\cos\left(\sqrt{\alpha}\int_0^t \rho d\tau\right)+{\bf
c}\sin\left(\sqrt{\alpha}\int_0^t \rho d\tau\right)\right], \cr {\boldsymbol{\pi}}=e^{-\int_0^t vd\tau}\left[-{\bf
b}\sin\left(\sqrt{\alpha}\int_0^t \rho d\tau\right)+ {\bf c}\cos\left(\sqrt{\alpha}\int_0^t \rho d\tau\right)\right],
\end{eqnarray}
where the integration constants ${\bf b}$ and ${\bf c}$ are subject to the conditions
\begin{eqnarray}\label{nnr13.1}
{\bf b}{\bf c}=0, \qquad {\bf b}^2={\bf c}^2=\sqrt{\alpha}.
\end{eqnarray}
This implies ${\boldsymbol{\omega}}^2=\sqrt{\alpha}e^{2\int_0^t vd\tau}$ and
${\boldsymbol{\pi}}^2=\sqrt{\alpha}e^{-2\int_0^t vd\tau}$. According to these expressions, the pair of orthogonal
vectors  $\boldsymbol{\omega}$ and $\boldsymbol{\pi}$ rotates in their own plane (or, equivalently, in the plane
determined by ${\bf b}$ and ${\bf c}$) with the variable angular velocity prescribed by the function $\rho(t)$. The
function $v(\tau)$ determines the variation of their magnitudes.  Choosing the functions $v$ and $\rho$ suitably,  we
can make the point with radius-vector $\boldsymbol{\omega}$ move along any prescribed line!

We point out that the two-parametric ambiguity is in correspondence with the invariance of the action (\ref{nnr6})
under the two local symmetries described above. Summing up, all the basic variables of our model are unobservable
quantities.

The spin-vector\footnote{Note that this coincides with ${\bf S}$ appeared in (\ref{nr2.1}).} ${\bf
S}=\boldsymbol{\omega}\times\boldsymbol{\pi}$ has unambiguous evolution
\begin{eqnarray}\label{nnr14}
\dot{\bf S}=0.
\end{eqnarray}
Note also that it is invariant of the local symmetries.  Hence the spin-vector is a candidate for an observable quantity.
In interacting theory ${\bf S}$ will precess under the torque exercised by a magnetic field, see below. Due to Eqs.
(\ref{intr12.1}), the coordinates $S_i$ obey (\ref{intr.12}).

\subsection{Spin fiber bundle and spin-plane local symmetry}

The passage from initial variables $\boldsymbol{\omega}$ and $\boldsymbol{\pi}$ to the observables ${\bf{S}}$ is not a
change of variables, and  acquires a natural interpretation in the geometric terms. It should be noted that basic
notions of the theory of constrained systems have their analogs in differential geometry. Second-class constraints
imply that all true trajectories lie on a submanifold of the initial phase-space. The Dirac bracket, constructed on the
base of second-class constraints, induces canonical symplectic structure on the submanifold. If the first-class
constraints (equivalently, the local symmetries) are presented in the model, a part of variables have ambiguous
evolution. This also can be translated into the geometric language: due to the ambiguity, the submanifold should be
endowed with a natural structure of a fiber bundle. Physical variables are (functions of) the coordinates which
parameterize the base of the fiber bundle. Let us describe, how all this look like in our model.

Consider six-dimensional phase space equipped with canonical Poisson bracket
\begin{eqnarray}\label{ss.1}
\mathbb{R}^{6}=\{~ \omega_i, \pi_j; ~ \{\omega_i, \pi_j\}_{PB}=\delta_{ij} ~ \},
\end{eqnarray}
and three-dimensional spin space $\mathbb{R}^3=\{S_i\}$ with the coordinates $S_i$. Define the map
\begin{eqnarray}\label{ss.2}
f: ~ \mathbb{R}^{6} ~ \rightarrow ~ \mathbb{R}^{3}, \quad f: (\omega_i, \pi_j) ~ \rightarrow ~
S_i=\epsilon_{ijk}\omega_j\pi_k,  \cr \mbox{or} \quad   {\bf S}=\boldsymbol{\omega}\times\boldsymbol{\pi}, \quad
\mbox{rank}\frac{\partial(S_i)}{\partial(\omega_j, \pi_k)}=3.
\end{eqnarray}
Poisson bracket on $\mathbb{R}^{6}$ together with the map induce $SO(3)$ Lie-Poisson bracket on $\mathbb{R}^3$
\begin{eqnarray}\label{ss.3}
\{S_i, S_j\}\equiv\{S_i(\omega, \pi), S_j(\omega, \pi)\}_{PB}, \quad \mbox{then} \quad \{S_i, S_j\}=\epsilon_{ijk}S_k.
\end{eqnarray}
As we saw above, all the trajectories $\boldsymbol{\omega}(t), \boldsymbol{\pi}(t)$ lie on
$SO(3)$\,-invariant surface of $\mathbb {R}^6$ determined by the constraints
\begin{eqnarray}\label{uf4.15}
\mathbb{T}^4=\{~ \boldsymbol\omega\boldsymbol\pi=0, \quad \boldsymbol\pi^2-\frac{\alpha}{\boldsymbol\omega^2}=0 ~ \},
\end{eqnarray}
that is $\boldsymbol\omega$ and $\boldsymbol\pi$ represent a pair of orthogonal vectors with their ends attached to the hyperbole
$y=\frac{\alpha}{x}$.

When $(\boldsymbol{\omega}, \boldsymbol{\pi})\in\mathbb{T}^4$, we have ${\bf
S}^2=\boldsymbol{\omega}^2\boldsymbol{\pi}^2-(\boldsymbol{\omega}\boldsymbol{\pi})^2=\alpha$. So, $f$ maps the manifold
$\mathbb{T}^4$ onto two-dimensional sphere (spin surface) of the radius $\sqrt{\alpha}$, $f(\mathbb{T}^3)=\mathbb{S}^2$.

Denote $\mathbb{F}^2_S\in\mathbb{T}^4$ preimage of a point ${\bf{S}}\in\mathbb{S}^2$,
$\mathbb{F}^2_S=f^{-1}({\bf{S}})$. Let $(\boldsymbol{\omega}, \boldsymbol{\pi})\in\mathbb{F}^2_S$. Then the
two-dimensional manifold $\mathbb{F}^2_S$ contains all pairs $(\chi\boldsymbol\omega, \frac{1}{\chi}\boldsymbol\pi)$,
$\chi\in\mathbb{R^{+}}$, as well as the pairs obtained by rotation of these $(\chi\boldsymbol\omega,
\frac{1}{\chi}\boldsymbol\pi)$ in the plane of vectors $(\boldsymbol{\omega}, \boldsymbol{\pi})$. So elements of
$\mathbb{F}^2_S$ are related by two-parametric transformations
\begin{eqnarray}\label{uf4.18}
\begin{array}{c}
\boldsymbol{\omega}'=\chi\boldsymbol{\omega}, \qquad
\boldsymbol{\pi}'=-\frac{1}{\chi}\boldsymbol{\pi},
\end{array}
\end{eqnarray}
\begin{eqnarray}\label{uf4.18.1}
\begin{array}{c}
\boldsymbol{\omega}'=\boldsymbol{\omega}\cos\phi+\boldsymbol{\pi}\frac{|\boldsymbol\omega|}{|\boldsymbol\pi|}\sin\phi, \quad
\boldsymbol{\pi}'=-\boldsymbol{\omega}\frac{|\boldsymbol\pi|}{|\boldsymbol\omega|}\sin\phi+\boldsymbol{\pi}\cos\phi.
\end{array}
\end{eqnarray}
In the result, the manifold $\mathbb{T}^4$ acquires a natural structure of fiber bundle
\begin{eqnarray}\label{uf4.19}
\mathbb{T}^4=(\mathbb{S}^2, \mathbb{F}^2, f),
\end{eqnarray}
with base $\mathbb{S}^2$, standard fiber $\mathbb{F}^2$, projection map $f$ and structure group given by the
transformations (\ref{uf4.18}) and (\ref{uf4.18.1}). The adjusted with the structure of the fiber bundle local coordinates are
$\chi$, $\phi$, and two coordinates of the vector ${\bf S}$. By construction, the structure-group transformations
leave inert points of the base, $\delta S_i=0$.

Let us discuss the relationship between the structure group and local symmetries of the Hamiltonian action
(\ref{nnr6}). The structure transformation (\ref{uf4.18}) can be identified with the scale transformation
(\ref{nnr10}). Concerning the transformation (\ref{uf4.18.1}), let us apply it to the action (\ref{nnr6}). Inserting
$\boldsymbol{\omega}'$ and $\boldsymbol{\pi}'$ into the action and disregarding the total derivative, we obtain the
expression
\begin{eqnarray}
S_H[q']=\int d\tau ~ {\boldsymbol{\pi}}\dot{\boldsymbol{\omega}}-{\boldsymbol{\omega}}{\boldsymbol{\pi}}\left[v'\cos
2\phi+B-v'_1A\right]-\left({\boldsymbol{\pi}}^2-\frac{\alpha}{{\boldsymbol{\omega}}^2}\right)\left[v'_1-C\right],
\nonumber
\end{eqnarray}
where
\begin{eqnarray}
A=\frac{|\boldsymbol\pi|}{|\boldsymbol\omega|}\left(1-
\frac{\alpha}{{\boldsymbol{\omega}}^2{\boldsymbol{\pi}}^2+({\boldsymbol{\omega}}{\boldsymbol{\pi}})|\boldsymbol\omega||\boldsymbol\pi|\sin
2\phi}\right)\sin 2\phi, \cr  \nonumber
B=\left(\frac{|\boldsymbol\omega|\dot{}}{|\boldsymbol\omega|}-\frac{|\boldsymbol\pi|\dot{}}{|\boldsymbol\pi|}\right)\sin^2\phi,
\qquad C=\frac{\dot\phi{\boldsymbol{\omega}}^2}{|\boldsymbol\omega||\boldsymbol\pi|+\sqrt{\alpha}}, \nonumber
\end{eqnarray}
The action does not change, $S_H[q']=S_H[q]$, if we adopt the following transformation law for $v$ and $v_1$
\begin{eqnarray}\label{uf4.19.1}
v'=\frac{v-B+A(v_1+C)}{\cos 2\phi}, \qquad v'_1=v_1+C.
\end{eqnarray}
Hence we have found one more local symmetry of the action. Its infinitesimal form reads
\begin{eqnarray}\label{uf4.19.2}
\delta{\boldsymbol{\omega}}=\phi\frac{|\boldsymbol\omega|}{|\boldsymbol\pi|}{\boldsymbol{\pi}}, \qquad \qquad
\delta{\boldsymbol{\pi}}=-\phi\frac{|\boldsymbol\pi|}{|\boldsymbol\omega|}{\boldsymbol{\omega}}, \qquad \qquad \cr
\delta v=\frac{2\phi v_1}{|\boldsymbol\omega||\boldsymbol\pi|}\left({\boldsymbol{\pi}}^2-\frac{\alpha}{{\boldsymbol{\omega}}^2}\right) , \quad
\delta v_1=\frac{\dot\phi{\boldsymbol{\omega}}^2}{|\boldsymbol\omega||\boldsymbol\pi|+\sqrt{\alpha}}.
\end{eqnarray}
The three infinitesimal symmetries (\ref{nnr9}), (\ref{nnr11}) and (\ref{uf4.19.2}) are not independent on the subspace
of solutions to equations of motion. To see this, we note that the following infinitesimal transformation:
\begin{eqnarray}\label{uf4.19.3}
\delta_\epsilon+\delta_{\gamma(\epsilon)}+\delta_{\phi(\epsilon)}, \quad \mbox{where} \quad \gamma(\epsilon)=\epsilon
v, \quad \phi(\epsilon)=2v_1\frac{|\boldsymbol\omega|}{|\boldsymbol\pi|}\epsilon,
\end{eqnarray}
being applied to any variable, turns out to be proportional to equations of motion.  For instance,
$[\delta_\epsilon+\delta_{\gamma(\epsilon)}+\delta_{\phi(\epsilon)}]\boldsymbol\pi=-\epsilon\frac{\delta
S_H}{\delta\boldsymbol\pi}- \frac{2\epsilon v_1\boldsymbol{\omega}}{{\boldsymbol{\omega}}^2}\frac{\delta S_H}{\delta
v_1}$. The on-shell symmetries are considered as trivial symmetries, see \cite{bib43}. Hence on the subspace of
solutions the infinitesimal reparametrization can be identified with a special transformation of the structure group
\begin{eqnarray}\label{uf4.19.4}
\delta_\epsilon=-\delta_{\gamma(\epsilon)}-\delta_{\phi(\epsilon)}.
\end{eqnarray}
In the result, the number of infinitesimal symmetries coincides with the number of primary first-class constraints.
Summing up, in the passage from geometric to dynamical realization, the transformations of structure group  of the spin
fiber bundle acts independently at each instance of time and turn into the local symmetries of Hamiltonian action.

{\bf Equivalent formulations.} Let us consider a slightly different Lagrangian
\begin{eqnarray}\label{nnr15}
L_{spin}=\frac12\dot{\boldsymbol{\omega}}N\dot{\boldsymbol{\omega}}+\frac{\alpha}{2\boldsymbol{\omega}^2}.
\end{eqnarray}
The conjugated momentum $\boldsymbol{\pi}=N\dot{\boldsymbol{\omega}}$ implies only one primary constraint
$\boldsymbol{\omega}\boldsymbol{\pi}=0$, then the complete Hamiltonian reads
 \begin{eqnarray}\label{nnr16}
H=\frac12\left(\boldsymbol{\pi}^2-\frac{\alpha}{\boldsymbol{\omega}^2}\right)+v(\boldsymbol{\omega}\boldsymbol{\pi}).
\end{eqnarray}
Computing $\frac{d}{dt}(\boldsymbol{\omega}\boldsymbol{\pi})=\{\boldsymbol{\omega}\boldsymbol{\pi}, H\}$, we obtain
$\boldsymbol\pi^2-\frac{\alpha}{\boldsymbol{\omega}^2}=0$ as the secondary constraint. Hence the Lagrangian implies an
equivalent formulation.

As the Lagrangian which implies the constraints (\ref{intr.11}), we could take the expression
$L_{spin}=\frac{1}{2g}{\dot{\boldsymbol{\omega}}^2}+\frac12g\beta^2-\frac12\lambda(\boldsymbol{\omega}^2-\alpha^2)$. We
remind that in this model $\beta^2=\frac{3\hbar^2}{4\alpha^2}$, while $\alpha$ is any given number. Variation with
respect to auxiliary variables $g(t)$ and $\lambda(t)$ gives the equations $\dot{\boldsymbol{\omega}}^2=g^2\beta^2$ and
$\boldsymbol{\omega}^2=\alpha^2$, the latter implies $\dot{\boldsymbol{\omega}}\boldsymbol{\omega}=0$. In the
Hamiltonian formulation these equations turn into the desired constraints.  We can integrate out the variable $g$,
presenting the Lagrangian in a more compact form
\begin{eqnarray}\label{3nr2}
L_{spin}=\beta\sqrt{{\dot{\boldsymbol{\omega}}}^2}-\frac12\lambda(\boldsymbol{\omega}^2-\alpha^2).
\end{eqnarray}
This also gives the desired constraints. The last term represents kinematic (velocity-independent) constraint. So, we
might follow the known classical-mechanics prescription and exclude $\lambda$ as well. But this would lead to loss of
the manifest rotational invariance of the formalism. The spin fiber bundle corresponding to this formulation turns out
to be the group manifold $SO(3)$, see \cite{DPM1} for details.

\subsection{Canonical quantization and Pauli equation} \label{ch09:sec9.3}

To test our formulation, we show that our spinning particle yields the Pauli equation in a stationary magnetic field
with the vector potential ${\bf A}$. Consider the action
\begin{eqnarray}\label{nr3}
S=\int dt\left[\frac{m}{2}\dot{\bf{x}}^2+\frac{e}{c}{\bf{A}}\dot{\bf{x}}
+\frac{\sqrt{\alpha}}{\sqrt{\boldsymbol{\omega}^2}}\sqrt{D{\boldsymbol{\omega}}ND{\boldsymbol{\omega}}}\right],
\end{eqnarray}
\begin{eqnarray}\label{nr3.0}
D\omega_i=\dot\omega_i-\frac{ge}{2mc}\epsilon_{ijk}\omega_jB_k.
\end{eqnarray}
The configuration-space variables are $x_i(t)$, and $\omega_i(t)$. Here $x_i$ represents the spatial coordinates of the
particle with the mass $m$, charge $e$ and gyromagnetic ratio $g$. In our classical model $g$ appeared as a coupling
constant of $\boldsymbol{\omega}$ with the magnetic field ${\bf{B}}=\mbox{\boldmath$\nabla$}\times{\bf{A}}$ in the last
term of Eq. (\ref{nr3.0}). At the end, it produces the Pauli term in the quantum-mechanical Hamiltonian.

Let us construct Hamiltonian formulation for the model. Equations for the conjugated momenta $p_i$ and $\pi_i$ reads
\begin{eqnarray}\label{nr4}
{\bf p}=m\dot{\bf x}+\frac{e}{c}{\bf A},\ \qquad \Rightarrow \ \dot{\bf x}=\frac{1}{m}({\bf p}-\frac{e}{c}{\bf A}),
\end{eqnarray}
\begin{eqnarray}\label{nr5}
\boldsymbol{\pi}=\frac{\sqrt{\alpha}}{\sqrt{\boldsymbol{\omega}^2}}
\frac{ND{\boldsymbol{\omega}}}{\sqrt{D{\boldsymbol{\omega}}ND{\boldsymbol{\omega}}}}.
\end{eqnarray}
Eq. (\ref{nr5}) implies the primary constraints $\boldsymbol{\omega}\boldsymbol{\pi}=0$ and
$\boldsymbol{\pi}^2-\frac{\alpha}{\boldsymbol{\omega}^2}=0$. The complete Hamiltonian, $H=P\dot Q-L+v_a\Phi_a,
Q=({\bf{x}}, \boldsymbol{\omega}), P=({\bf{p}}, \boldsymbol{\pi})$, reads
\begin{eqnarray}\label{nr6}
H=\frac{1}{2m}({\bf p}-\frac{e}{c}{\bf A})^2-eA_0-\frac{ge}{2mc}({\bf B}{\bf S})+
v(\boldsymbol{\omega}\boldsymbol{\pi})+v_1\left(\boldsymbol{\pi}^2-\frac{\alpha}{\boldsymbol{\omega}^2}\right).
\end{eqnarray}
There are no of higher-stage constraints in the formulation. Besides the constraints, the Hamiltonian (\ref{nr6})
implies the dynamical equations
\begin{eqnarray}\label{nr9}
\dot x_i=\frac{1}{m}(p_i-\frac{e}{c}A_i), \qquad \dot p_i=\frac{e}{c}\dot x_j\partial_iA_j+\frac{g
e}{2mc}S_j\partial_iB_j,
\end{eqnarray}
\begin{eqnarray}\label{nr10}
\dot\omega_i=v\omega_i+2v_1\pi_i+\frac{ge}{2mc}\epsilon_{ijk}\omega_jB_k, \qquad \cr
\dot\pi_i=-v\pi_i-2v_1\frac{\boldsymbol{\pi}^2}{\boldsymbol{\omega}^2}\omega_i+\frac{ge}{2mc}\epsilon_{ijk}\pi_jB_k.
\end{eqnarray}
As a consequence of these equations, the spin-vector $S_i=\epsilon_{ijk}\omega_j\pi_k$ have unambiguous evolution
\begin{eqnarray}\label{nr11}
\dot S_i=\frac{ge}{2mc}\epsilon_{ijk}S_jB_k.
\end{eqnarray}
This is the classical equation for precession of spin in an external magnetic field. Due to Eqs. (\ref{intr.11}), the
coordinates $S_i$ obey (\ref{intr.12}). Equations
(\ref{nr9}) imply the second-order equation for $x_i$
\begin{eqnarray}\label{nr13}
m\ddot x_i=\frac{e}{c}\epsilon_{ijk}\dot x_jB_k+\frac{ge}{2mc}S_k\partial_iB_k.
\end{eqnarray}
Note that in the absence of interaction, the spinning particle does not experience a self-acceleration. The last term
gives non vanishing contribution into the trajectory in unhomogeneous field and can be used for semiclassical
description of Stern-Gerlach experiment. Since ${\bf{S}}^2\sim\hbar^2$, the $S$-term disappears from Eq. (\ref{nr13})
at the classical limit $\hbar\rightarrow 0$. Then Eq. (\ref{nr13}) reproduces the classical motion of a charged
particle subject to the Lorentz force.

{\bf Precession of spin.} Let us denote $-\frac{ge}{2mc}{\bf B}=\boldsymbol{\omega}_p$, then Eq. (\ref{nr11})
reads
\begin{eqnarray}\label{nr11.1}
\dot{\bf S}=\boldsymbol{\omega}_p\times{\bf S}.
\end{eqnarray}
The vector $\dot{\bf S}$ is orthogonal to the plane of $\boldsymbol{\omega}_p$ and ${\bf S}$ at any instant. Besides,
contracting Eq. (\ref{nr11.1}) with ${\bf S}$ we see that magnitude of spin does not change, ${\bf S}^2=\mbox{const}$.
In the result, the end point of ${\bf S}$ rotates around the axis $\boldsymbol{\omega}_p$. Let ${\bf S}(0)={\bf S}_0$
is the initial position of spin.
We present this vector as a sum of longitudinal and transversal parts with respect to $\boldsymbol{\omega}_p$, ${\bf
S}_0={\bf S}_{0\perp}+{\bf S}_{0||}$. Then for the constant vector $\boldsymbol{\omega}_p$, the general solution to Eq.
(\ref{nr11.1}) is
\begin{eqnarray}\label{nr11.2}
{\bf S}={\bf S}_{0||}+|{\bf S}_{0\perp}|({\bf e_1}\cos|\boldsymbol{\omega}_p|t+{\bf e_2}\sin|\boldsymbol{\omega}_p|t).
\end{eqnarray}
Hence the magnitude of vector $\boldsymbol{\omega}_p$ from Eq. (\ref{nr11.1}) is just the frequency of precession.
Equation of trajectory (\ref{nr13}) in the constant magnetic field is $\dot{\bf
v}=\frac{2}{g}[\boldsymbol{\omega}_p\times{\bf v}]$, that is particle's velocity precesses with the frequency
$\frac{2}{g}\boldsymbol{\omega}_p$. For a particle with classical gyromagnetic ratio $g=2$, the two frequences coincide
and the angle between velocity and spin preserves during the evolution. For the anomalous magnetic moment, $g\ne2$, the
frequences are different. The spin precession relative to the velocity is used in a cyclotron experiments for
measurement of anomalous magnetic moment \cite{Field:1979, miller2007muon}.

{\bf Canonical quantization.} We quantize only the physical variables $x_i, p_i, S_i$.  Their classical brackets are
\begin{eqnarray}\label{qq7}
\{x_i, p_j\}=\delta_{ij}, \qquad \{S_i, S_j\}=\epsilon_{ijk}S_k.
\end{eqnarray}
As the last
two terms in (\ref{nr6}) does not contributes into equations of motion for the physical variables, we omit them. This
gives the physical Hamiltonian
\begin{eqnarray}\label{qq6}
H=\frac{1}{2m}({\bf p}-\frac{e}{c}{\bf A})^2-\frac{ge}{2mc}{\bf B}{\bf S}.
\end{eqnarray}
The first equation from (\ref{qq7}) implies the standard quantization of the variables $x$ and $p$, we take $\hat
x_i=x_i$, $\hat p_i=-i\hbar\partial_i$. According to the second equation from (\ref{qq7}), we look for the
wave-function space which is a representation of the group $SO(3)$. Finite-dimensional irreducible representations of
the group are numbered by spin $s$, which is related with the values of Casimir operator as follows: ${\bf{S}}^2\sim
s(s+1)$. Then Eq. (\ref{intr.12}) fixes the spin $s=\frac12$, and $S_i$ must be quantized by $\hat
S_i=\frac{\hbar}{2}\sigma_i$. The operators act on the spinor space of two-component complex columns $\Psi$. Quantum
Hamiltonian is obtained from Eq. (\ref{qq6}) replacing classical variables by the operators. This immediately yields
the Pauli equation, that is Eq. (\ref{ch08:eqn8.164}) with ${\bf E}=0$.

\section{Why we need a semiclassical model of relativistic spin?}
\label{ch09:sec9.0}

{\bf Dirac equation.} We expect that a semiclassical relativistic model of spin should be closely related to the
Dirac equation normally used to describe the relativistic spin in quantum theory. The consistent description of
relativistic spin is achieved in quantum electrodynamics, where the Dirac equation is considered as a quantum field
theory equation.  But it also admits a quantum-mechanical interpretation and thus represents an example of relativistic
quantum mechanics. This is of interest on various reasons. In particular, namely being considered as a
quantum-mechanical equation, the Dirac equation gives the correct energy levels of hydrogen atom. As we saw in
Sect.~\ref{ch09:sec9.1}, dynamical equations for expectation values of operators in quantum mechanics should resemble
the Hamiltonian equations of the corresponding classical system. Let us discuss these equations in the Dirac theory.

Under the infinitesimal Lorentz transformation $\delta x^\mu=\omega^\mu{}_\nu x^\nu$, the Dirac spinor $\Psi=(\Psi_1,
\Psi_2, \Psi_3, \Psi_4)$ transforms as follows:
\begin{eqnarray}\label{de.1}
\delta\Psi=-\frac{i}{4}\omega_{\mu\nu}\gamma^{\mu\nu}\Psi,
\end{eqnarray}
where
\begin{eqnarray}\label{de.2}
\gamma^{\mu\nu}\equiv[\gamma^\mu, \gamma^\nu]=\frac{i}{2}(\gamma^\mu\gamma^\nu-\gamma^\nu\gamma^\mu),
\end{eqnarray}
and the $4\times 4$ $\gamma$\,-matrices can be composed from $\sigma$\,-matrices of Pauli
\begin{eqnarray}\label{de.3}
\gamma^0=
\left(
\begin{array}{cc}
1& 0\\
0& -1
\end{array}
\right), \quad
\gamma^i=
\left(
\begin{array}{cc}
0& \sigma^i\\
-\sigma^i& 0
\end{array}
\right).
\end{eqnarray}
We use the representation with hermitian $\gamma^0$ and antihermitian $\gamma^i$. The matrices do not commute with each
other, and the basic formula for their permutation is as follows
\begin{eqnarray}\label{de.3.1}
[\gamma^\mu, \gamma^\nu]_{+}\equiv\gamma^\mu\gamma^\nu+\gamma^\nu\gamma^\mu=-2\eta^{\mu\nu}.
\end{eqnarray}
The Dirac equation in an external four-potential $A_\mu$
\begin{eqnarray}\label{de.4}
\left[\gamma^\mu(\hat p_\mu-\frac{e}{c}A_\mu)+mc\right]\Psi=0, \quad \mbox{where} \quad \hat p_\mu=-i\hbar\partial_\mu,
\end{eqnarray}
turns out to be covariant under the transformation (\ref{de.1}). Applying the operator $\gamma^\mu(\hat
p_\mu-\frac{e}{c}A_\mu)-mc$ to (\ref{de.4}), we see that the Dirac equation implies the Klein-Gordon equation with
non-minimal interaction
\begin{eqnarray}\label{1.1_1}
\left[(\hat p^\mu-\frac{e}{c}A^\mu)^2-\frac{e\hbar}{2c}F_{\mu\nu}\gamma^{\mu\nu}+m^2c^2\right]\Psi=0,
\end{eqnarray}
where $F_{\mu\nu}=\partial_\mu A_\nu-\partial_\nu A_\mu$.

For the latter use, let us analyze commutators of the matrices involved. The commutators of $\gamma$\,-matrices can not
be presented through themselves, but produce $\gamma^{\mu\nu}$\,-matrices as they are written in (\ref{de.2}). The set
$\gamma^\mu$,  $\gamma^{\mu\nu}$ forms a closed algebra
\begin{eqnarray}\label{1.3}
[\gamma^\mu, \gamma^\nu]=-2i\gamma^{\mu\nu}, \qquad
[\gamma^{\mu\nu},
\gamma^\alpha]=2i(\eta^{\mu\alpha}\gamma^\nu-\eta^{\nu\alpha}\gamma^\mu), ~ \cr
[\gamma^{\mu\nu}, \gamma^{\alpha\beta}]=
2i(\eta^{\mu\alpha}\gamma^{\nu\beta}-
\eta^{\mu\beta}\gamma^{\nu\alpha}-
\eta^{\nu\alpha}\gamma^{\mu\beta}+\eta^{\nu\beta}\gamma^{\mu\alpha}).
\end{eqnarray}
As it was tacitly implied in Eq. (\ref{de.1}), $\gamma^{\mu\nu}$\,-matrices obey $SO(1, 3)$\,-algebra of Lorentz
generators.  The complete algebra (\ref{1.3}) can be identified with the five-dimensional Lorentz algebra $SO(2, 3)$
with generators $J^{AB}$, $A, B=(\mu, 5)=(0, 1, 2, 3, 5)$, and with the metric $\eta^{AB}=(-,+,+,+,-)$
\begin{eqnarray}\label{1.4}
[J^{AB}, J^{CD}]=2i(\eta^{AC}J^{BD}-\eta^{AD} J^{BC}-
\eta^{BC} J^{AD}+\eta^{BD} J^{AC}),
\end{eqnarray}
assuming $\gamma^\mu\equiv  J^{5\mu}$, $\gamma^{\mu\nu}\equiv  J^{\mu\nu}$. Vector model of spinning particle with
$SO(2, 3)$ covariant spin-space has been constructed in \cite{AAD:Ann}.

{\bf Observer-independent probability.} $\Psi$ can be used to construct the adjoint spinor
$\bar\Psi=\Psi^{\dagger}\gamma^0$ with the transformation law
$\delta\bar\Psi=\frac{i}{4}\bar\Psi\gamma^{\mu\nu}\omega_{\mu\nu}$. Then $\bar\Psi\Psi$ is a scalar,
$\bar\Psi\gamma^\mu\Psi$ ia a vector\footnote{With the factor $-\frac{i}{4}$ in (\ref{de.1}) and with the standard
transformation law for a vector, $\delta v_\mu=\epsilon_\mu{}^\nu v_\nu$, the function $v_\mu\bar\Psi\gamma^\mu\Psi$ is
a scalar function.}  and so on. The vector turns out to be a conserved current, that is
\begin{eqnarray}\label{1.5}
\partial_\mu(\bar\Psi\gamma^\mu\Psi)=0,
\end{eqnarray}
on solutions to the Dirac equation. The time-component of the vector is $\Psi^{\dagger}\Psi$. Assuming that symbols
$x^i$ represent the position of a particle, the quantity
\begin{eqnarray}\label{1.6}
P(t)=\Psi^{\dagger}\Psi d^3x.
\end{eqnarray}
is identified with relativistic-invariant probability to find a
particle in the infinitesimal volume $d^3x$ at the instant $t=\frac{x^0}{c}$. To confirm this interpretation, we first
note that the probability density $\Psi^{\dagger}\Psi$ is a positive function. Second, due to the continuity
equation (\ref{1.5}), integral of the density over all space does not depend on time:
$\frac{d}{dx^0}\int_{V}d^3x\Psi^{\dagger}\Psi=\int_{V}d^3x\partial_0(\bar\Psi\gamma^0\Psi)=
-\int_{V}d^3x\partial_i(\bar\Psi\gamma^i\Psi)=-\int_{\partial V}d\Omega_i(\bar\Psi\gamma^i\Psi)=0$ for the solutions
$\Psi$ that vanish on spatial infinity. Third, $P$ coincides with the manifestly Lorentz-invariant quantity
\begin{eqnarray}\label{1.7}
-\frac{1}{6}\epsilon_{\mu\nu\alpha\beta}(\bar\Psi\gamma^\mu\Psi) dx^\nu dx^\alpha dx^\beta,
\end{eqnarray}
when it computed over equal-time surface $x^0=\mbox{const}$ of Minkowski space. This implies an observer-independence
of the probability $P$: all inertial observers, when they compute $P$ using their coordinates,  will compute the same
number (\ref{1.7}).

However, it is well known that adopting the quantum-mechanical interpretation, we arrive at a rather strange and
controversial picture. We outline here the results of analysis on the applicability of quantum-mechanical treatment to
the free Dirac equation made by Schr\"{o}dinger\footnote{For an electron interacting with electromagnetic field this
analysis has been repeated by Feynman in \cite{Feyn61}.} in \cite{Schr30}.  We multiply the Dirac equation on
$\gamma^0$, representing it in the Schr\"odinger-like form
\begin{eqnarray}\label{4.1}
i\hbar\partial_t\Psi=\hat H\Psi, \qquad \hat H= c\alpha^i\hat
p_i+mc^2\beta,
\end{eqnarray}
where $\alpha^i=\gamma^0\gamma^i$ and $\beta=\gamma^0$ are Dirac matrices.
Then $\hat H$ may be interpreted as the Hamiltonian. Passing from the Schr\"odinger to Heisenberg picture, the time
derivative of an operator $a$ is $i\hbar\dot a=[a, H]$, and for the expectation values of basic operators of the Dirac
theory we obtain the equations
\begin{eqnarray}\label{4.2}
\dot x_i=c\alpha_i, \qquad \quad \qquad \dot p_i=0,  \qquad \qquad \cr
i\hbar\dot\alpha_i=2(cp_i-H\alpha_i), \quad  i\hbar\dot\beta=-2c\alpha_ip_i\beta+mc^2.
\end{eqnarray}
Some properties of the equations are in order. \par \noindent 1.  \textit{The wrong balance of the number of degrees of
freedom}. The first equation in (\ref{4.2}) implies that the operator $c\alpha^i$ represents the velocity of the
particle. Then physical meaning of the operator $p^i$ becomes rather obscure in the classical limit. \par \noindent 2.
\textit{Zitterbewegung}. The equations (\ref{4.2}) can be solved, with the result for $x^i(t)$ being
$x^i=a^i+dp^it+c^i\mbox{exp}(-\frac{2iH}{\hbar}t)$. The first and second terms are expected and describe a motion along
the straight line. The last term on the r.h.s. of this equation states that the free electron experiences rapid
oscillations with higher frequency $\frac{2H}{\hbar}\sim\frac{2mc^2}{\hbar}$. \par \noindent 3. \textit{Velocity of an
electron}. Since the velocity operator $c\alpha^i$ has eigenvalues $\pm c$, we conclude that a measurement of a
component of the velocity of a free electron is certain to lead to the result $\pm c$. \par \noindent 4.
\textit{Operator of relativistic spin.} We expect that in the Dirac theory can be constructed the relativistic
generalization of the spin operator (\ref{ch08:eqn8.156}). The question on the definition of a conventional spin
operator has been raised a long time ago \cite{pryce1948mass, foldy:1978} and is under discussion up to date \cite
{Keitel14, Keitel15, Terno2016}.

Many people noticed that in the Dirac theory it is possible to construct another operators that obey to a reasonable
equations, see \cite{Fleming65, corben:1968}. Presenting these equations, Feynman accompanied them with the following
comment (see p. 48 in \cite{Feyn61}): ``The following relations may be verified as true but their meaning is not yet
completely understood, if at all: ...".

In view of all this, it seem desirable to construct a semiclassical model of spin that will be as close as possible to
the Dirac equation. By this we mean the model which, being quantized, yields the Dirac equation. In the following
sections, we will see how the vector model clarifies the issues discussed above. In a few words, this can be resumed as
follows. As we already saw above, the vector model is necessarily invariant under the spin-plane local symmetry which
determines its physical sector formed by observables. We show that observables of the vector model have an expected
behavior on both classical and quantum level. Comparing quantum mechanics of the vector model with that of Dirac
equation in Sect.~\ref{ch09:sec9.23}, we obtain the rules for computation of probabilities and mean values of the
vector model observables using the Dirac equation. The time evolution implied by the rules (\ref{cq27}) turn out to be
different from the ingenuous prescription (\ref{4.2}). Hence the vector model of spin supports the point of view that
the operators of the Dirac equation do not represent directly measurable quantities.

\section{Spin-tensor of Frenkel}
\label{ch09:sec9.4}

To construct the relativistic spinning particle, we need a Lorentz-covariant description of the spin fiber bundle
(\ref{uf4.15}). We remind that our construction involves basic and target spaces as well as the map $f: ~
\mathbb{R}^6(\boldsymbol{\omega}, \boldsymbol{\pi})\rightarrow\mathbb{R}^3({\bf S})$, see Eqs.
(\ref{ss.1})-(\ref{ss.3}).  We embed this $SO(3)$\,-covariant construction into its suitably chosen $SO(1,
3)$\,-covariant extension. Let us start from the three-vector $\boldsymbol{\omega}$.  We assume that relativistic spin
can be described by a vector $\omega^\mu$ of Minkowski space such that $\omega^\mu=(0, \boldsymbol{\omega})$ for the
particle at rest in the laboratory frame. This condition expresses the Correspondence Principle: relativistic physics
should approximate to the Newton physics in the limit of small velocities. To represent this condition in a covariant
form in an arbitrary frame, we assume that in our model there exists a four-vector $p_\mu$ which for the particle at
rest has the components $(p_0,{\bf 0})$. For the case of a free particle, the natural candidate is a vector
proportional to the particle's four-velocity. For the particle in external field,  the form of this vector is dictated
by the structure of interaction, see below. With this $p^\mu$, the Lorentz-invariant statement $p\omega=0$ is
equivalent to the condition that $\omega^\mu=(0, \boldsymbol{\omega})$ for the particle at rest. Following the same
lines, we also assume the condition $p\pi=0$ for the conjugated momentum $\pi^\mu$ for $\omega^\mu$. Hence we replace
the basic space $\mathbb{R}^6(\boldsymbol{\omega}, \boldsymbol{\pi})$ by direct product of two Minkowski spaces with
the following natural action of the Lorentz group on it:
\begin{equation}\label{f1}
SO(1, 3): \left(
\begin{array}{c}
\omega\\
\pi
\end{array}
\right) \rightarrow\left(
\begin{array}{c}
\omega'\\
\pi'
\end{array}
\right)=\left(
\begin{array}{cc}
\Lambda  &0\\
0 &\Lambda
\end{array}
\right) \left(
\begin{array}{c}
\omega\\
\pi
\end{array}
\right) .
\end{equation}
The relativistic generalization of the surface (\ref{ss.3}) is given by the following $SO(1, 3)$\,-invariant surface of
the phase space $\mathbb{M}\times\mathbb{M}$
\begin{eqnarray}\label{f1.0}
\mathbb{T}^4=\left\{ ~ \omega\pi=0, \quad \pi^2-\frac{\alpha}{\omega^2}=0, \quad p\omega=0, \quad p\pi=0 ~\right\}.
\end{eqnarray}
Below we denote these constraints $T_2, T_5, T_3$ and $T_4$. As in non relativistic case, we have two first-class
constraints $\omega\pi=0$ and $\pi^2-\frac{\alpha}{\omega^2}=0$. The constraints $p\omega=0$ and $p\pi=0$ are of second
class, so we expect $8-2\times2-2=2$ physical degrees of freedom in the spin-sector.

It should be noted that $\omega^\mu$ and $\pi^\mu$ turn out to be space-like vectors. Indeed, in the frame where
$p^\mu=(p^ 0,{\bf 0})$ the constraints $p\omega=p\pi=0$ imply $\omega^0=\pi^0=0$. This implies $\omega^2\ge 0$ and
$\pi^2\ge 0$. Then from the constraint $\pi^2-\frac{\alpha}{\omega^2}=0$ we conclude $\omega^2>0$ and $\pi^2>0$.

Let us consider the target space. To generalize the map $S^i=\epsilon^{ijk}\omega^j\pi^k$ to the case of
four-dimensional quantities, we rewrite it in an equivalent form, using the known isomorphism among three-vectors and
antisymmetric $3\times 3$\,-matrices
\begin{eqnarray}\label{f1.1}
S^i=\frac14\epsilon^{ijk}S^{jk}, \quad \mbox{then} \quad S^{ij}=2\epsilon^{ijk}S^k.
\end{eqnarray}
Then
\begin{eqnarray}\label{f1.2}
S^i=\epsilon^{ijk}\omega^j\pi^k, \quad \mbox{is equivalent to} \quad S^{ij}=2(\omega^i\pi^j-\omega^j\pi^i).
\end{eqnarray}
The last equality has an evident generalization to the four-dimensional case:
$S^{\mu\nu}=2(\omega^\mu\pi^\nu-\omega^\nu\pi^\mu)$. Hence the target space $\mathbb{R}^3({\bf S})$ should be extended
to the six-dimensional space $\mathbb{R}^6({\bf D}, {\bf S})$ of antisymmetric $4\times 4$ matrices. We present them as
follows:
\begin{equation}\label{f1.4}
S^{\mu\nu}({\bf D},{\bf S})=\left(
\begin{array}{cccc}
 0    & -D_1    & -D_2    & -D_3 \\
D_1 & 0       &2 S_3  & -2 S_2 \\
D_2 & -2 S_3  & 0      & 2 S_1 \\
D_3 & 2 S_2   & -2 S_1 & 0
\end{array}
\right) \,,
\end{equation}
or, equivalently
\begin{eqnarray}\label{f1.5}
S^{\mu\nu}=(S^{i0}=D^i, S^{ij}=2\epsilon^{ijk}S^k).
\end{eqnarray}
Lorentz group naturally acts on this space
\begin{equation}\label{f1.6}
SO(1,3): ~  S^{\mu\nu}({\bf D},{\bf S}) \quad\rightarrow \quad S^{\mu\nu}({\bf D}',{\bf S}')=\Lambda^{\mu}{}_\alpha
\Lambda^{\nu}{}_\beta S^{\alpha\beta}({\bf D},{\bf S}).
\end{equation}
This equation determines transformation rules of the columns ${\bf D}$ and ${\bf S}$. They transform as three-vectors
under the subgroup of rotations of the Lorentz group. The embedding (\ref{f1.5}) of three-dimensional spin-vector ${\bf
S}$ into the four-dimensional spin-tensor has been suggested by Frenkel \cite{Frenkel}. So we call $S^{\mu\nu}$ the
Frenkel spin-tensor. The vector ${\bf D}$ is called dipole electric moment of the particle \cite{ba1}.

Now we are ready to define the covariant version of the map (\ref{ss.2})
\begin{eqnarray}\label{f1.7}
f: ~ \mathbb{M}(\omega^\mu)\times\mathbb{M}(\pi^\nu)\rightarrow \mathbb{R}^6(S^{\mu\nu}); \quad  (\omega^\mu,\pi^\nu)
\rightarrow S^{\mu\nu}=2(\omega^\mu\pi^\nu-\omega^\nu\pi^\mu).
\end{eqnarray}
It has rank equals $5$, and maps a point of $\mathbb{M}\times\mathbb{M}$ to a pair of orthogonal three-dimensional
vectors, ${\bf D}{\bf S}=0$. By construction, $f$ is compatible with the transformations (\ref{f1}) and (\ref{f1.6}) of
$SO(1,3)$: if $S'^{\mu\nu}({\bf D},{\bf S})=2(\omega'^\mu \pi'^\nu-\omega'^\nu\pi'^\mu)$, then $S^{\mu\nu}({\bf D},{\bf
S})=2(\omega^\mu\pi^\nu-\omega^\nu\pi^\mu)$.

\begin{figure}[t] \centering
\includegraphics[width=350pt, height=150pt]{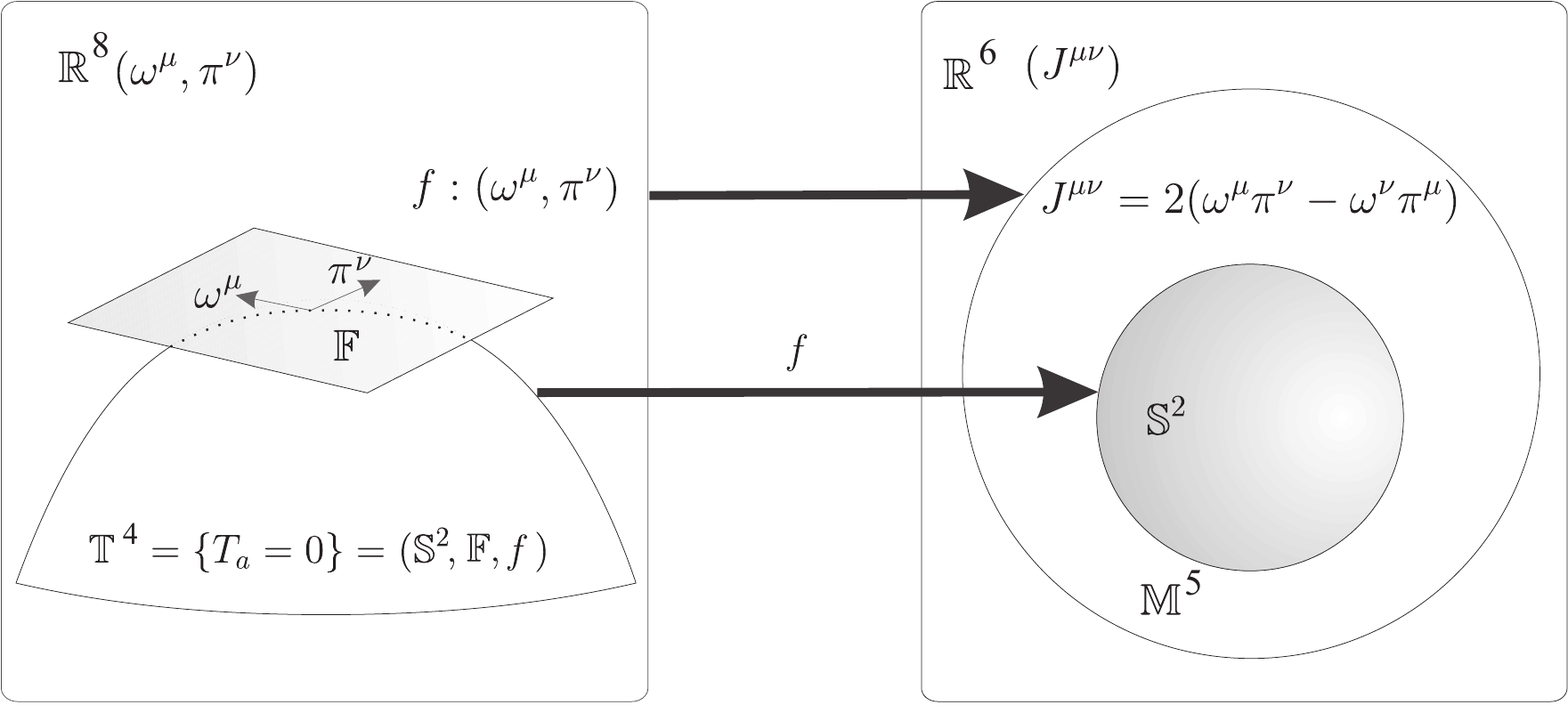}
\caption{Fiber bundle $\mathbb{T}^4$ associated with relativistic
spin}\label{ch09:fig9.3}
\end{figure}

If $\mathbb{M}\times\mathbb{M}$ is considered as a symplectic space with canonical Poisson bracket,
$\{\omega^\mu,\pi^\nu\}=\eta^{\mu\nu}$, the map $f$ induces $SO(1, 3)$\,-Lie-Poisson bracket on $\mathbb{R}^6$
\begin{equation}\label{f1.8}
\{S^{\mu\nu }(\omega, \pi),  S^{\alpha\beta}(\omega,
\pi)\}=2(\eta^{\mu\alpha}S^{\nu\beta}-\eta^{\mu\beta}S^{\nu\alpha}-\eta^{\nu\alpha}S^{\mu\beta}+\eta^{\nu\beta}S^{\mu\alpha}).
\end{equation}
Consider the image $S^{\mu\nu}(\omega, \pi)$ of a point of the surface (\ref{f1.0}). Using the identity
$S^{\mu\nu}S_{\mu\nu}=8(\omega^2\pi^2-(\omega\pi)^2)$ together with the equations (\ref{f1.0}), we obtain five
covariant equations which determine the spin-surface $\mathbb{S}^2$ in an arbitrary Lorentz frame
\begin{eqnarray}
S^{\mu\nu}S_{\mu\nu}=8\alpha=6\hbar^2,   \label{f1.9}  \\
S^{\mu\nu}p_\nu=0. \qquad \label{f1.10}
\end{eqnarray}
As $(S^{\mu\nu}p_\nu)p_\mu\equiv 0$, we have only four independent equations imposed on six variables, therefore the
spin-surface has dimension 2, as it should be. Denote $\mathbb{F}_S\in\mathbb{T}^4$ preimage of a point $S^{\mu\nu}$ of
the base, $\mathbb{F}_S=f^{-1}(S^{\mu\nu})$, that is the standard fiber, see Fig.~\ref{ch09:fig9.3} on page
\pageref{ch09:fig9.3}. Its points are related by the structure-group transformations (\ref{uf4.18}) and
(\ref{uf4.18.1}).

Consider the rest frame of the vector $p^\mu$, that is $p^\mu=(p^0, {\bf 0})$ in this frame. The surface (\ref{f1.0})
acquires the form
\begin{eqnarray}\label{f1.11}
\boldsymbol{\omega}\boldsymbol{\pi}= 0, \qquad \boldsymbol{\pi}^2-\frac{\alpha}{\boldsymbol{\omega}^2}=0, \qquad
 \pi^0    &=& 0\,,\qquad  \omega^0   = 0,
\end{eqnarray}
and can be identified with the non relativistic spin-surface (\ref{uf4.15}). Being restricted to this surface, the map
(\ref{f1.7})  reads
\begin{equation}\label{f1.12}
S^{\mu\nu}|_{\mathbb{T}^4}=\left(
\begin{array}{cccc}
0 & 0    & 0    & 0 \\
0 & 0    & 2 S_3  & -2 S_2 \\
0 & -2 S_3 & 0      & 2 S_1 \\
0 & 2 S_2  & -2 S_1 & 0
\end{array}
\right) \,,\qquad {\bf S}=\boldsymbol\omega\times\boldsymbol\pi.
\end{equation}
Hence in the rest frame the dipole electric moment vanishes, while the spatial part of spin-tensor coincides with the
non-relativistic spin. We conclude that $SO(3)$\,-construction (\ref{ss.1})-(\ref{uf4.15}) is embedded into $SO(1,
3)$\,-covariant scheme. As in non relativistic case, the basic variables $\omega^\mu$ and $\pi_\nu$ do not represent
observable quantities, only $S^{\mu\nu}$ does. This may be contrasted with \cite{hanson1974, Kunst14, Kunst16}, where
the equation $S^{\mu\nu}=0$ assumed to be the first-class constraint of the Dirac formalism. In the result,
$S^{\mu\nu}$ turns out to be unobservable quantity.

\section{Four-dimensional spin-vector, Pauli-Lubanski vector and Bargmann-Michel-Telegdi vector}
\label{ch09:sec9.4.1} On the pure algebraic grounds, spin-tensor of Frenkel turns out to be equivalent to a
four-dimensional vector. So the latter could also be used for the description of relativistic spin. Here we discuss the
relevant formalism.

Levi-Civita symbol\index{Symbol! of Levi-Civita} with $\epsilon^{0123}=1$ obeys the identities
\begin{eqnarray}
\epsilon^{abcd}\epsilon_{ab\mu\nu}=-2(\delta^c{}_\mu\delta^d{}_\nu-\delta^c{}_\nu\delta^d{}_\mu), \label{aa1} \qquad \qquad \\
\epsilon^{\mu abc}\epsilon_{\mu ijk}=-[\delta^a{}_i(\delta^b{}_j\delta^c{}_k-\delta^b{}_k\delta^c{}_j)-
\delta^a{}_j(\delta^b{}_i\delta^c{}_k-\delta^b{}_k\delta^c{}_i)+ \cr
\delta^a{}_k(\delta^b{}_i\delta^c{}_j-\delta^b{}_j\delta^c{}_i)]. \qquad \qquad \qquad \label{aa2}
\end{eqnarray}
Given an antisymmetric matrix $J^{\mu\nu}=-J^{\nu\mu}$ and a vector $p^\mu$, we define the vectors
\begin{eqnarray}\label{aa3}
s^\mu=\frac{1}{4\sqrt{-p^2}}\epsilon^{\mu\nu\alpha\beta}p_\nu J_{\alpha\beta}, \quad \mbox{then} \quad s^\mu p_\mu=0;
\end{eqnarray}
\begin{eqnarray}\label{aa4}
\Phi^\mu=J^{\mu\nu}p_\nu, \qquad \mbox{then} \qquad \Phi^\mu p_\mu=0.
\end{eqnarray}
When $p$ and $J$ represent generators of the Poincar\'{e} group, the vector (\ref{aa3}) is called Pauli-Lubanski
vector. It turns out to be useful for the classification of irreducible representations of the Poincar\'{e} group
\cite{wigner1939unitary, Bargmann1948, bib50}.

The tensor $J^{\mu\nu}$ and its dual, ${}^*J^{\mu\nu}\equiv\frac{1}{2} \epsilon^{\mu\nu ab}J_{ab}$, can be decomposed on
these vectors as follows:
\begin{eqnarray}\label{aa5}
J^{\mu\nu}=\frac{\Phi^\mu p^\nu-\Phi^\nu p^\mu}{p^2}-\frac{2}{\sqrt{-p^2}}\epsilon^{\mu\nu ab}p_as_b,
\end{eqnarray}
\begin{eqnarray}\label{aa6}
\epsilon^{\mu\nu ab}J_{ab}=4\frac{p^\mu s^\nu-p^\nu s^\mu}{\sqrt{-p^2}}-\frac{2}{p^2}\epsilon^{\mu\nu ab}p_a\Phi_b.
\end{eqnarray}
To prove (\ref{aa5}), we contract (\ref{aa2}) with $p_ap^iJ^{jk}$. Eq. (\ref{aa6}) follows from (\ref{aa5}) contracted
with $\epsilon_{ab\mu\nu}$. The definitions imply the identity relating the square of $s^\mu$ with a "square" of
$J^{\mu\nu}$
\begin{eqnarray}\label{aa7}
s^\mu s_\mu=\frac{1}{8}J^{\mu\nu}J_{\mu\nu}-\frac{1}{4p^2}(J^{\mu\nu}p_\nu)^2.
\end{eqnarray}
Frenkel spin-tensor obeys $S^{\mu\nu}p_\nu=0$, that is $\Phi^\mu=0$, and can be used to construct four-vector
of spin (below we also call it Pauli-Lubanski vector)
\begin{eqnarray}\label{1.6}
s^\mu(\tau)\equiv\frac{1}{4\sqrt{-p^2}}\epsilon^{\mu\nu\alpha\beta}p_\nu S_{\alpha\beta}.
\end{eqnarray}
In the free theory $p^\mu$ is independent on $S^{\mu\nu}$, so this equation is linear on $S^{\mu\nu}$ and can be
inverted. According to Eq. (\ref{aa5}) we have
\begin{eqnarray}\label{aa9}
S^{\mu\nu}=-\frac{2}{\sqrt{-p^2}}\epsilon^{\mu\nu\alpha\beta}p_\alpha s_\beta,
\end{eqnarray}
that is the two quantities are mathematically equivalent, and we could work with $s^\mu$ instead of $S^{\mu\nu}$. The
equation  (\ref{aa7}) implies proportionality of their magnitudes. In the interacting theory $p^\mu$ contains
$S^{\mu\nu}$, so (\ref{1.6}) becomes a non linear equation.

Let us compare spatial components of $s^\mu$ with the non-relativistic spin-vector ${\bf S}$. In the rest system of
$p^\mu$, $p^\mu=(p^0, {\bf 0})$, $\sqrt{-p^2}=|p^0|$, we have $s^0=0$ and
\begin{eqnarray}\label{aa12}
s^i=\frac{p^0}{4|p^0|}\epsilon^{ijk}S_{jk}=\frac{p^0}{|p^0|}S^i,
\end{eqnarray}
that is the two vectors coincide. This explains our normalization for $s^\mu$, Eq. (\ref{aa3}). Under the Lorentz
boost, ${\bf S}$ transforms as the spatial part of a tensor whereas $s^\mu$  transforms as a four-vector. So the two
spins are different in all Lorentz frames except the rest frame. The relation between them in an arbitrary frame
follows from Eq. (\ref{aa9})
\begin{eqnarray}\label{spin.1}
S^i=\frac{p^0}{\sqrt{-p^2}}\left(\delta_{ij}-\frac{p_ip_j}{(p^0)^2}\right)s^j\,.
\end{eqnarray}
A four-dimensional vector $s^\mu_{bmt}$ with the property $u_\mu s^\mu_{bmt}=0$, where $u_\mu$ represents a
four-velocity of the particle, has been successively used by Bargmann Michel and Telegdi to analyze the spin precession
in uniform magnetic field, see \cite{Jackson} for details. In our vector model, even in the case of interaction, the
condition $ps=0$ implies $us=0$. So we expect that our equations of motion for $s^\mu$ should represent a
generalization of the Bargmann-Michel-Telegdi equations to the case of an arbitrary electromagnetic field.

In summary, the relativistic spin can be described by the Frenkel spin-tensor (\ref{f1.4}) composed by the dipole
electric moment ${\bf D}$ and the spin ${\bf S}$. In our vector model the Frenkel tensor is a composite quantity, see
(\ref{f1.7}). In the rest frame of the vector $p^\mu$ we have ${\bf D}=0$, while ${\bf S}$ coincides with the vector of
non-relativistic spin. Intuitively, the Frenkel tensor  shows how the non relativistic spin looks like in an arbitrary
Lorentz frame.

\section{Lagrangian of relativistic spinning particle}
\label{ch09:sec9.5}
\subsection{Variational problem for the prescribed Dirac's constraints}
\label{ch09:sec9.5.1}

In the previous section we have discussed only the spin-sector of a spinning particle. To construct a complete theory,
we add the position $x^\mu(\tau)$ and its conjugated momentum $p^\mu(\tau)$ taken in an arbitrary parametrization
$\tau$. This implies that we deal with the reparametrization-invariant theory. So besides the spin-sector constraints
(\ref{f1.0}) we expect also the mass-shell condition $T_1=p^2+(mc)^2=0$. Let us look for the Hamiltonian action which
could produce these constraints. According to general theory \cite{bib08, bib10, AAD05}, it has the form $\int d\tau ~
p\dot x+\pi\dot\omega-(H_0+\lambda_iT_i)$ where $T_i$ are the primary constraints of the theory. We expect $H_0=0$ due
to the reparametrization invariance. As the suitable primary constraints, let us take
$p^2+(mc)^2+\pi^2-\frac{a}{\omega^2}$, $T_2, T_3$ and $T_4$. Thus we consider the Hamiltonian variational problem
\begin{eqnarray}\label{aa14}
S_H=\int d\tau ~ p\dot x+\pi\dot\omega-[\frac{\lambda_1}{2}(p^2+(mc)^2+\pi^2-\frac{\alpha}{\omega^2})+\cr
\lambda_2(\omega\pi)+\lambda_3(p\omega)+\lambda_4(p\pi)]. \qquad \qquad \quad
\end{eqnarray}
Due to the Poisson bracket $\{T_2, T_5\}=2T_5$, in this formulation $T_5=0$ appears as the secondary constraint. To
arrive at the Lagrangian action, we could follow the standard prescription. Excluding the conjugate
momenta from $S_H$ according to their equations of motion, we obtain an action with the auxiliary variables
$\lambda_i$. Excluding them, one after another, we obtain various equivalent forms of the Lagrangian action. To
simplify these  computations, we proceed as follows.  First, we note that the constraints $\omega\pi=0$ and $p\omega=0$
always appear from the Lagrangian which involves the projector $N$, that is we use $N\dot x$ and $N\dot\omega$ instead
of $\dot x$ and $\dot\omega$. So we set $\lambda_2=\lambda_3=0$ in Eq. (\ref{aa14}). Second, we present the
remaining terms in (\ref{aa14}) in the matrix form
\begin{eqnarray}\label{aa15}
S=\int d\tau ~
(p, \pi)\left(
\begin{array}{c}
\dot x\\
\dot\omega
\end{array}
\right)-\frac{\lambda_1}{2}(p, \pi)\left(
\begin{array}{cc}
\eta& \lambda\eta\\
\lambda\eta& \eta
\end{array}
\right)\left(
\begin{array}{c}
p\\
\pi
\end{array}
\right)-\frac{\lambda_1}{2}\left[(mc)^2-\frac{\alpha}{\omega^2}\right],
\end{eqnarray}
where $\lambda=\frac{\lambda_4}{\lambda_1}$.
The matrix appeared in (\ref{aa15}) is invertible, the inverse matrix is
\begin{eqnarray}\label{aa16}
\frac{1}{1-\lambda^2}\left(
\begin{array}{cc}
\eta& -\lambda\eta\\
-\lambda\eta& \eta
\end{array}\right).
\end{eqnarray}
Eq. (\ref{aa15}) is the Hamiltonian variational problem of the form $p\dot q-\frac{\lambda_1}{2}(pAp+M^2)$, the latter
follows from the Lagrangian $-M\sqrt{-\dot qA^{-1}\dot q}$. This allows us to exclude the variable $\lambda_1$. As it
was combined above, we then replace $\dot x$, $\dot\omega$ by $N\dot x$, $N\dot\omega$ and obtain
\begin{eqnarray}
S=-\int d\tau\sqrt{(mc)^2 -\frac{\alpha}{\omega^2}} ~ \sqrt{-(N\dot x, N\dot\omega)\left(
\begin{array}{cc}
\frac{\eta}{1-\lambda^2}& \frac{-\lambda\eta}{1-\lambda^2} \\
\frac{-\lambda\eta}{1-\lambda^2}& \frac{\eta}{1-\lambda^2}
\end{array}
\right)\left(
\begin{array}{c}
N\dot x\\
N\dot\omega
\end{array}
\right)}= \qquad \label{FF.1.2} \\
-\int d\tau\sqrt{(mc)^2 -\frac{\alpha}{\omega^2}} ~ \sqrt{(1-\lambda^2)^{-1}\left[-\dot x N \dot x - \dot\omega N
\dot\omega + 2\lambda\dot x N \dot\omega \right]}. \qquad \quad \label{FF.1.1}
\end{eqnarray}
To exclude the remaining auxiliary variable $\lambda$, we compute variation of (\ref{FF.1.1}) with respect to
$\lambda$, this gives the equation
\begin{eqnarray}\label{FF.1.3}
(\dot x N\dot\omega)\lambda^2-(\dot x N\dot x+\dot\omega N\dot\omega)\lambda+(\dot x N\dot\omega)=0,
\end{eqnarray}
which determines $\lambda$
\begin{eqnarray}\label{FF.1.4}
\lambda_{\pm}=\frac{(\dot x N\dot x+\dot\omega N\dot\omega)\pm \sqrt{(\dot x N\dot x+\dot\omega N\dot\omega)^2-4(\dot x
N\dot\omega)^2}} {2(\dot x N\dot\omega)}.
\end{eqnarray}
We substitute $\lambda_{+}$ into (\ref{FF.1.1}) and use $\lambda_{+}\lambda_{-}=1$.  Then (\ref{FF.1.1}) turns into the
following action
\begin{eqnarray}\label{Lfree}
S=-\frac{1}{\sqrt{2}} \int d\tau \sqrt{m^2c^2 -\frac{\alpha}{\omega^2}} \times \qquad \qquad \qquad\cr \sqrt{-\dot x N
\dot x - \dot\omega N \dot\omega +\sqrt{[\dot x N\dot x + \dot\omega N \dot\omega]^2- 4 (\dot x N \dot\omega )^2}}.
\end{eqnarray}
The matrix $N_{\mu\nu}$ is the projector on the plane orthogonal to $\omega^\nu$
\begin{equation}\label{FF0}
N_{\mu\nu}= \eta_{\mu\nu}-\frac{\omega_\mu \omega_\nu}{\omega^2}, \quad \mbox{then} \quad
N_{\mu\alpha}N^{\alpha\nu}=N_\mu{}^\nu, \quad N_{\mu\nu} \omega^\nu=0.
\end{equation}
In the spinless limit, $\alpha=0$ and $\omega^\mu=0$, the functional (\ref{Lfree}) reduces to the expected Lagrangian
of spinless particle, $-mc\sqrt{-\dot x^\mu\dot x_\mu}$. It is well known that the latter can be written in equivalent
form using the auxiliary variable $\lambda(\tau)$ as follows: $\frac{1}{2\lambda}\dot x^2-\frac{\lambda}{2}m^2c^2$.
Similarly to this, (\ref{Lfree}) can be presented in the equivalent form
\begin{eqnarray}\label{FF.1}
S=\int d\tau\frac{1}{4\lambda_1}\left[\dot xN\dot x+\dot\omega N\dot\omega-\sqrt{[\dot x N\dot x + \dot\omega N \dot\omega]^2- 4
(\dot x N \dot\omega )^2}\right]- \cr \frac{\lambda_1}{2}[(mc)^2-\frac{\alpha}{\omega^2}]. \qquad \qquad \qquad \qquad
\end{eqnarray}
In summary, besides the ``minimal" Lagrangian (\ref{Lfree}) we have obtained two its equivalent formulations given by
Eqs. (\ref{FF.1.1}) and (\ref{FF.1}). The Lagrangians provide the appearance of equation $p\pi=0$ as the primary
constraint. In turn, this seems crucial to introduce an interaction consistent with the constraints.

\subsection{Interaction and the problem of covariant formalism}
\label{ch09:sec9.5.2}

In the formulation (\ref{Lfree}) without auxiliary variables, our model admits the minimal interaction with
electromagnetic field and with gravity. As we detaily shown below, this does not spoil the number and algebraic
structure of constraints presented in the free theory. Interaction with an electromagnetic potential is achieved by
adding the standard term
\begin{equation}\label{FF0.1}
S_{int}=\frac{e}{c}\int d\tau A_\mu\dot x^\mu.
\end{equation}
The minimal interaction with gravity is achieved \cite{DPW2, DW2015.1} by covariantization of (\ref{Lfree}). We replace
$\eta_{\mu\nu}\rightarrow g_{\mu\nu}$, and usual derivative by the covariant one,
\begin{equation}\label{FF0.2}
\dot\omega^\mu\rightarrow \nabla\omega^\mu=\dot\omega^\mu +\Gamma^\mu_{\alpha\beta}\dot
x^\alpha\omega^\beta.
\end{equation}
Velocities $\dot x^\mu$, $\nabla\omega^\mu$ and projector $N_{\mu\nu}$ transform like contravariant vectors and
covariant tensor, so the action is manifestly invariant under the general-coordinate transformations.

To introduce an interaction of spin with electromagnetic field, we use  \cite{deriglazov2014Monster} the formulation
(\ref{FF.1}) with the auxiliary variable $\lambda_1$. We add to the action (\ref{FF.1}) the term (\ref{FF0.1}) and
replace
\begin{equation}\label{FF0.2}
\dot\omega^\mu \quad \rightarrow \quad D\omega^\mu\equiv\dot\omega^\mu-\lambda_1\frac{e\mu}{c}F^{\mu\nu}\omega_\nu.
\end{equation}
We have denoted $\mu=\frac{g}{2}$, where $g$ is gyromagnetic ratio, this agreement simplifies many of equations below.
So we restore $g$ only in the final answers. $\lambda_1$ in this expression provides the homogeneous transformation law
of $D\omega$ under the reparametrizations, $D_{\tau'}\omega'=\frac{\partial\tau}{\partial\tau'}D_\tau\omega$.

The interaction of spin with gravity through the gravimagnetic moment will be achieved in the formulation
(\ref{FF.1.1}), see below.

Concerning the interaction of spin with electromagnetic field, let us briefly discuss an issue with nearly a century of
history, that is not completely clarified so far. While the complete relativistic Hamiltonian of the covariant
formulation will be obtained below, its linear on spin part can be predicted from a symmetry considerations. Indeed,
the only Lorentz and $U(1)$\,-invariant term which involves $F$ and $S$ is $F_{\mu\nu}S^{\mu\nu}$. Using \textit{the covariant
condition} (\ref{f1.10}) we obtain
\begin{align}
H_{relspin}\sim-\frac{e}{4mc}F_{\mu\nu}S^{\mu\nu}=\frac{e}{mc}\left[\frac{1}{mc}{\rm {\bf S}}[{\bf p}\times{\bf E}]-
{\rm {\bf B}}{\rm {\bf S}}\right]. \label{FF0.2.1}
\end{align}
This can be compared with spin part of the Hamiltonian (\ref{ch08:eqn8.164}) with $g=2$
\begin{align}
H_{spin} =\frac{e}{mc}\left[\frac{1}{2mc}{\rm {\bf S}}[{\bf p}\times{\bf E}]- {\rm {\bf B}}{\rm {\bf S}}\right].
\label{FF0.2.2}
\end{align}
They differ by the famous and troublesome factor\footnote{In discussing this factor often refer to Thomas precession
\cite{Thomas1927}. We will not touch this delicate and controversial issue \cite{stepanov2012, Jackson, Nistor2016}
because of the covariant formalism automatically accounts the Thomas precession \cite {bib15}.} of $\frac12$. The same
conclusion follows from comparison of equations of motion of the two formulations \cite{Pomeranskii1998}.  As we saw in
Sect.~\ref{ch09:sec9.1}, the expression (\ref{FF0.2.2}) has very strong experimental support. The question, why a
covariant formalism does not lead directly to the correct result, has been raised in 1926 \cite{Frenkel}, and remain
under discussion to date \cite{khriplovich:1996, Khriplovich1989, Pomeranskii1998}.

Following the work \cite{DPM2016}, in Sect.~\ref{ch09:sec9.70} we show that the vector model provides an answer to this
question on a pure classical ground, without appeal to the Thomas precession, Dirac equation or Foldy-Wouthuysen
transformation. In a few words it can be described as follows. The relativistic vector model involves a second-class
constraints ($T_3$ and $T_4$ of Eq. (\ref{f1.0})), which we take into account by passing from the Poisson to Dirac
bracket. So in the covariant formulation we arrive at the relativistic Hamiltonian (\ref{FF0.2.1}) accompanied by non
canonical classical brackets. To construct the quantum mechanics, we could work with the relativistic Hamiltonian, but
in this case we need to find quantum realization of the non canonical brackets. Equivalently, we can find the variables
with canonical brackets and quantize them in the standard way. The relativistic Hamiltonian (\ref{FF0.2.1}), when
written in the canonical variables, just gives (\ref{FF0.2.2}).

\subsection{Particle with the fundamental length scale}
\label{ch09:sec9.5.3}

Our basic model yields the fixed value of spin, as it should be for
an elementary particle. Let us present the modification which leads to the theory with unfixed
spin, and, similarly to Hanson-Regge approach \cite{hanson1974}, with a mass-spin trajectory constraint. Consider the
following Lagrangian
\begin{eqnarray}\label{aaa1}
L =-\frac{mc}{\sqrt{2}}\sqrt{-\dot x N \dot x - l^2\frac{\dot\omega N \dot\omega}{\omega^2} +\sqrt{\left[\dot x N\dot x
+ l^2\frac{\dot\omega N \dot\omega}{\omega^2}\right]^2- 4l^2\frac{(\dot x N \dot\omega )^2}{\omega^2}}}, \qquad
\end{eqnarray}
where $l$ is a parameter with the dimension of length. The Dirac procedure yields the Hamiltonian
\begin{equation}\label{aaa2}
H=\frac{\lambda_1}{2}\left( p^2 + m^2 c^2 + \frac{\pi^2\omega^2}{l^2} \right) + \lambda_2 (\omega\pi) +\lambda_3
(p\omega) +\lambda_4 (p\pi )\,,
\end{equation}
which turns out to be combination of the first-class constraints $p^2 + m^2 c^2 + \frac{\pi^2\omega^2}{l^2}=0$,
$\omega\pi=0$ and the second-class constraints $p\omega=0$, $p\pi=0$. The Dirac procedure stops on the first stage,
that is there are no of secondary constraints. As compared with (\ref{Lfree}), the first-class constraint
$\pi^2-\frac{\alpha}{\omega^2}=0$ does not appear in the present model. Due to this, square of spin is not fixed,
$S^2=8(\omega^2\pi^2-\omega\pi)\approx8\omega^2\pi^2$. Using this equality, the mass-shell constraint acquires the
form similar to the string theory
\begin{equation}\label{aaa3}
p^2 + m^2 c^2 + \frac{1}{8l^2}S^2=0.
\end{equation}
It has a clear meaning: the energy of the particle grows with its spin.
The model has four physical degrees of freedom in the spin-sector. As the independent gauge-invariant degrees of
freedom, we can take three components $S^{ij}$ of the spin-tensor together with any one product of conjugate
coordinates, for instance, $\omega^0\pi^0$.

\subsection{Classification of vector models}
\label{ch09:sec9.6}

While we concentrate on the model specified by Eqs. (\ref{f1.0}), it is instructive to discuss other sets of
constraints that could be used for construction of a spinning particle. The equation (\ref{aa7}) relating the
Poincar\'{e} and Lorentz spins
\begin{eqnarray}\label{aa12.1}
s^\mu s_\mu=\frac{1}{8}S^{\mu\nu}S_{\mu\nu}-\frac{1}{4p^2}(S^{\mu\nu}p_\nu)^2= \qquad \qquad \quad \cr
\omega^2\pi^2-(\omega\pi)^2-\frac{1}{p^2}[\omega^2(p\pi)^2 +\pi^2(p\omega)^2-2(p\omega)(p\pi)(\omega\pi)],
\end{eqnarray}
turns out to be useful in what follows. \par

\noindent 1. Our basic model (\ref{f1.0}) with two degrees of freedom implies $S^{\mu\nu}p_\nu=0$. Then Eq.
(\ref{aa12.1}) implies proportionality of the two spins, $8s^2=S^2$, whereas their magnitudes are fixed due to the
constraints $T_2$ and $T_5$. The variables $x^\mu, p^\mu$ and $S^{\mu\nu}$ have vanishing brackets with first-class
constraints, so they are candidates for observables. \par

\noindent 2. The model with the constraints $\omega^2=\alpha^2$ and $\pi^2=\beta^2$ instead of
$\pi^2=\frac{\alpha}{\omega^2}$ is essentially equivalent to the basic model. The relationship between two models can
be probably established using the conversion scheme \cite{Dehghani15, Farmany14}. \par

\noindent 3. Let us replace $T_3\equiv p\omega=0$ by $T'_3\equiv p\omega-\sqrt{\omega^2}=0$ in the set (\ref{f1.0}).
These constraints appear in the model of rigid particle. $T_4$ and $T_5$ can be taken as the second-class constraints,
while $T_2$ and $T'_3$ form the first-class subset. As a consequence, the model has two degrees of freedom. The
Poincar\'{e} and Lorentz spins are proportional and have fixed magnitudes. The variables $x^\mu$ and $S^{\mu\nu}$ have
non vanishing brackets with first-class constraints. After canonical quantization, the constraint $T'_3=0$ turns into
the Dirac equation. Hence this semiclassical model can be used to study the relation among classical observables and
operators of the Dirac theory.  \par

\noindent 4. Hanson and Regge developed their model of a relativistic top \cite{hanson1974} on the base of
antisymmetric tensor $S^{\mu\nu}$ without making of any special assumptions on its inner structure. The tensor is
subject to first-class constraints $S^{\mu\nu}p_\nu=0$. This implies phase space with $2\times 6-2\times 3=6$ degrees
of freedom as well as proportional spins with unfixed magnitude. A similar vector model could be constructed starting
from the Hamiltonian action
\begin{eqnarray}\label{cv1}
S_H=\int d\tau ~ p\dot x+\pi\dot\omega-\frac{\lambda_1}{2}[ p^2 + m^2 c^2+f(S^2)]-
\lambda_\mu S^{\mu\nu}p_\nu,
\end{eqnarray}
where $S^{\mu\nu}=2(\omega^\mu\pi^\nu-\omega^\nu\pi^\mu)$.
The variables $x^\mu$ and $S^{\mu\nu}$ are not observables in this model. \par

\noindent 5. To avoid the unobservable character of original variables in the model (\ref{cv1}), we could replace
$S^{\mu\nu}p_\nu=0$ by the pair of second-class constraints $p\omega=p\pi=0$. They provide $S^{\mu\nu}p_\nu=0$ and
$8-2=6$ degrees of freedom. \par

\noindent 6. Adding the first-class constraint $\omega\pi=0$ to the model of Item 5 we arrive at the Lagrangian
(\ref{aaa1}) with four degrees of freedom. \par

\noindent 7. There are models based on the light-like vector $\omega^\mu$ \cite{Star, Shirzad16}. Consider the
first-class constraints
\begin{eqnarray}\label{cv2}
\omega^2=0, ~  \omega\pi=0, ~  \pi^2(p\omega)^2=\mbox{const}, \quad \mbox{then} \quad s^2=\mbox{const}, ~   S^2=0.
\end{eqnarray}
This implies two degrees of freedom. The Poincar\'{e} and Lorentz spins, while are fixed, do not correlate one with
another. The variables $x^\mu$, $S^{\mu\nu}$ and $s^\mu$ are not observable quantities. We note also that
$S^{\mu\nu}p_\nu\ne 0$, this complicates the analysis of non relativistic limit. \par

\noindent 8. Let us replace $\pi^2(p\omega)^2=\mbox{const}$ by $\sqrt{\pi^2}(p\omega)=\mbox{const}$ in the set (\ref{cv2}).
Similarly to Item 3, this constraint may be classical analog of the Dirac equation. This model still has not been studied.

A common for the models 5-8 is the problem whether they admit an interaction with external fields. Concerning the
Hanson-Regge model, in their work \cite{hanson1974} they analyzed whether the spin-tensor interacts directly with an
electromagnetic field, and concluded on impossibility to construct the interaction in a  closed form.  In our vector
model an electromagnetic field interacts with the part $\omega^\mu$ of the spin-tensor.

\section{Interaction with electromagnetic field}
\label{ch09:sec9.6}

In this rather technical section we demonstrate that our variational problem yields a model of spinning particle with
expected properties. In particular, our equations of motion generalize an approximate equations of Frenkel and
Bargmann-Michel-Telegdi to the case of an arbitrary electromagnetic field.

\subsection{Manifestly covariant Hamiltonian formulation}
\label{ch09:sec9.6.1}

As we saw in previous section, interaction with an arbitrary electromagnetic field can be described within the action
\begin{eqnarray}\label{m.1}
S=\int d\tau\frac{1}{4\lambda}\left[\dot xN\dot x+D\omega ND\omega-\sqrt{\left[\dot xN\dot x+D\omega
ND\omega\right]^2-4(\dot xND\omega)^2}\right]- \cr \frac{\lambda}{2}(m^2c^2-\frac{\alpha}{\omega^2})+\frac{e}{c}A\dot x,
\qquad \qquad \qquad \qquad \qquad
\end{eqnarray}
where the term
\begin{eqnarray}\label{m.2}
D\omega^\mu\equiv\dot\omega^\mu-\lambda\frac{e\mu}{c}F^{\mu\nu}\omega_\nu,
\end{eqnarray}
accounts the spin-field interaction.

Let us construct Hamiltonian formulation of the model. Conjugate momenta for $x^\mu$, $\omega^\mu$ and $\lambda$ are
denoted as $p^\mu$, $\pi^\mu$ and $p_\lambda$.  We use also the canonical momentum ${\cal P}^\mu\equiv
p^\mu-\frac{e}{c}A^\mu$. Contrary to $p^\mu$, the canonical momentum is $U(1)$ gauge-invariant quantity. Since
$p_\lambda=\frac{\partial L}{\partial\dot\lambda}=0$, the momentum $p_\lambda$ represents the primary constraint,
$p_\lambda=0$. Expressions for the remaining momenta, $p^\mu=\frac{\partial L}{\partial\dot x_\mu}$ and
$\pi^\mu=\frac{\partial L}{\partial\dot\omega_\mu}$, can be written in the form
\begin{eqnarray}\label{m.3}
{\cal P}^\mu=\frac{1}{2\lambda}(N\dot x^\mu-K^\mu), \qquad \qquad \qquad \cr K^\mu\equiv T^{-\frac12}\left[\left(\dot xN\dot x+D\omega
ND\omega\right)(N\dot x)^\mu-2(\dot xND\omega)(ND\omega)^\mu\right],
\end{eqnarray}
\begin{eqnarray}\label{m.4}
\pi^\mu=\frac{1}{2\lambda}(ND\omega^\mu-R^\mu), \qquad \qquad \qquad \cr R^\mu\equiv T^{-\frac12}\left[\left(\dot xN\dot x+D\omega
ND\omega\right)(ND\omega)^\mu-2(\dot xND\omega)(N\dot x)^\mu\right],
\end{eqnarray}
where $T=\left[\dot xN\dot x+D\omega ND\omega\right]^2-4(\dot xND\omega)^2$. The functions $K^\mu$ and $R^\mu$ obey the
following remarkable identities
\begin{eqnarray}\label{m.5}
K^2=\dot xN\dot x, \quad R^2=D\omega ND\omega, \quad KR=-\dot x ND\omega, \cr \dot xR+D\omega K=0,  \qquad \dot
xK+D\omega R=T^{\frac12}. \qquad
\end{eqnarray}
Due to Eq. (\ref{FF0}), contractions of the momenta with $\omega^\mu$ vanish, that is we have the primary constraints
$\omega\pi=0$ and ${\cal P}\omega=0$. One more primary constraint, ${\cal P}\pi=0$, is implied by (\ref{m.5}).

Hence we deal with a theory with four primary constraints. Hamiltonian is obtained excluding velocities from the expression
\begin{eqnarray}\label{m.6}
H=p\dot x+\pi\dot\omega-L+\lambda_iT_i,
\end{eqnarray}
where $\lambda_i$ are the Lagrangian multipliers for the primary constraints $T_i$. To obtain its manifest form, we
note the equalities ${\cal P}^2=\frac{1}{2\lambda^2}[\dot xN\dot x-\dot xK]$, $\pi^2=\frac{1}{2\lambda^2}[D\omega
ND\omega-D\omega R]$, and ${\cal P}\dot x+\pi D\omega=2L_1$, where $L_1$ is the first line in Eq. (\ref{m.1}). Then,
using (\ref{m.5}) we obtain
\begin{eqnarray}\label{m.8}
({\cal P}^2+\pi^2)=\frac{2}{\lambda}L_1.
\end{eqnarray}
Further, using Eqs. (\ref{m.5}) we have
\begin{eqnarray}\label{m.9}
p\dot x+\pi\dot\omega\equiv{\cal P}\dot x+\frac{e}{c}A\dot x+\pi D\omega+\lambda\frac{e\mu}{c}(\pi
F\omega)= \cr 2L_1+\frac{e}{c}A\dot x-\lambda\frac{e\mu}{4c}(FS), \qquad \qquad
\end{eqnarray}
where appeared the Frenkel spin-tensor $S^{\mu\nu}$. Using (\ref{m.9}) and (\ref{m.8}) in (\ref{m.6}), the Hamiltonian reads
\begin{eqnarray}\label{m.11}
H=\frac{\lambda}{2}\left({\cal P}^2-\frac{e\mu}{2c}(FS)+m^2c^2+\pi^2-\frac{\alpha}{\omega^2}\right)+ \cr
\lambda_2(\omega\pi)+ \lambda_3({\cal P}\omega)+\lambda_4({\cal P}\pi)+\lambda_0p_\lambda. \qquad
\end{eqnarray}
The fundamental Poisson brackets $\{x^\mu, p^\nu\}=\eta^{\mu\nu}$ and $\{\omega^\mu, \pi^\nu\}=\eta^{\mu\nu}$ imply
\begin{eqnarray}\label{m.12}
\{x^\mu, {\cal P}^\nu\}=\eta^{\mu\nu}, \quad \{{\cal P}^\mu, {\cal P}^\nu\}=\frac{e}{c}F^{\mu\nu},
\end{eqnarray}
\begin{eqnarray}\label{m.13}
\{S^{\mu\nu},S^{\alpha\beta}\}= 2(\eta^{\mu\alpha} S^{\nu\beta}-\eta^{\mu\beta} S^{\nu\alpha}-\eta^{\nu\alpha}
S^{\mu\beta} +\eta^{\nu\beta} S^{\mu\alpha}).
\end{eqnarray}
\begin{eqnarray}\label{m.13.1}
\{S^{\alpha\beta}, \omega^\mu\}=2\eta^{\mu[\alpha}\omega^{\beta]}, \qquad \{S^{\alpha\beta}, \pi^\mu\}=2\eta^{\mu[\alpha}\pi^{\beta]}.
\end{eqnarray}
According to Eq. (\ref{m.13}), the spin-tensor is generator of Lorentz algebra $SO(1,3)$. As $\omega\pi$, $\omega^2$
and $\pi^2$ are Lorentz-invariants, they have vanishing Poisson brackets with $S^{\mu\nu}$. To reveal the higher-stage
constraints we write the equations $\dot T_i=\{T_i, H\}=0$. The Dirac procedure stops on  third stage with the
following equations
\begin{eqnarray}
p_\lambda=0 ~ ~ \quad &\Rightarrow& \quad  T_1\equiv{\cal P}^2-\frac{e\mu}{2c}(FS)+m^2c^2+\pi^2-\frac{\alpha}{\omega^2}=0  \nonumber \\
&\Rightarrow& \quad \lambda_3C+\lambda_4D=0\,,\label{m.14.1} \\
T_2\equiv (\omega\pi)=0 ~ ~ \quad &\Rightarrow& \quad ~  T_5\equiv\pi^2-\frac{\alpha}{\omega^2}=0\,,\label{m.14.2} \\
T_3\equiv({\cal P}\omega)=0 ~ \quad  &\Rightarrow & \quad ~  \lambda_4=-\frac{2\lambda c}{e}aC\,,\label{m.14.3} \\
T_4\equiv({\cal P}\pi)=0 ~ \quad &\Rightarrow&\quad ~  \lambda_3=\frac{2\lambda c}{e}aD\,.\label{m.14.4}
\end{eqnarray}
We have denoted
\begin{eqnarray}\label{m.15}
C=-\frac{e(\mu-1)}{c}(\omega F{\cal P})+\frac{e\mu}{4c}(\omega\partial)(FS), \cr  D=-\frac{e(\mu-1)}{c}(\pi F{\cal
P})+\frac{e\mu}{4c}(\pi\partial)(FS),
\end{eqnarray}
and the function $a$ is written in (\ref{uf4.71}).
The last equation from (\ref{m.14.1}) turns out to be a consequence of (\ref{m.14.3}) and (\ref{m.14.4}) and can be
omitted. Due to the secondary constraint $T_5$ appeared in (\ref{m.14.2}) we can replace the constraint $T_1$ on the equivalent one
\begin{eqnarray}\label{gmm120}
T_1\equiv{\cal P}^2-\frac{e\mu}{2c}(FS)+m^2c^2=0.
\end{eqnarray}
This can be compared with Eq. (\ref{1.1_1}).
The Dirac procedure revealed two secondary constraints written in Eqs. (\ref{gmm120}) and
(\ref{m.14.2}), and fixed the Lagrangian multipliers $\lambda_3$ and $\lambda_4$, the latter can be substituted into the Hamiltonian.
The multipliers $\lambda_0$,
$\lambda_2$ and the auxiliary variable $\lambda$ have not been determined. $H$ vanishes on the complete constraint surface,
as it should be in a reparametrization-invariant theory.

We summarized the algebra of Poisson brackets between constraints in the Table \ref{tabular:monster-algebra1}.
\begin{table}
\begin{center}
\begin{tabular}{c|c|c|c|c|c}
                              & $\qquad T_1 \qquad$  & $T_5$                   & $T_2$                 & $T_3$    & $T_4$     \\  \hline  \hline
$T_1=\mathcal{P}^2- $         & 0             & 0             & 0                                   & -2C   & -2D     \\
$\frac{\mu e}{2c}(FS)+m^2c^2$
                              &               &               &           &                           &             \\
\hline
$T_5=\pi^2-\frac{\alpha}{\omega^2}$               & 0             & 0                & $-2T_5\approx 0$          & $-2T_4\approx 0$   & $\frac{2\alpha}{(\omega^2)^2}T_3\approx 0$\\
& ${}$ &      &           &
&        \\
\hline
$T_2=\omega\pi$               & 0     & $2T_5\approx 0$       &0     & $-T_3\approx 0$     & $T_4\approx 0$\\
& ${}$ &      &           &
&        \\
\hline
$T_3={\cal P}\omega$       &$2C$  &$2T_4\approx 0$       &$T_3\approx 0$            & 0       & $T_1+\frac{e}{2ca}$  \\
&               &               &           &                                                                               &$\approx \frac{e}{2ca}$\\
\hline
$T_4={\cal P}\pi$          & $2D$      &$-\frac{2\alpha}{(\omega^2)^2}T_3\approx 0$           &$-T_4\approx 0$          &$-T_1-\frac{e}{2ca}$  &0\\
&    &       &                                                                                                                                      &$\approx -\frac{e}{2ca}$  &\\
\hline
\end{tabular}
\end{center}
\caption{Algebra of constraints.} \label{tabular:monster-algebra1}
\end{table}
The constraints $p_\lambda$, $T_1$, $T_2$ and $T_5$ form the first-class subset, while $T_3$ and $T_4$ represent a pair
of second class. The presence of two primary first-class constraints $p_\lambda$ and $T_2$ is in correspondence with
the fact that two lagrangian multipliers remain undetermined within the Dirac procedure.

Below we will use the following notation. In the equation which relates velocity and canonical
momentum will appear the matrix $T$
\begin{eqnarray}\label{uf4.71}
T^{\mu\nu}=\eta^{\mu\nu}-(\mu-1)a(SF)^{\mu\nu}, \qquad
a=\frac{-2e}{4m^2c^3-e(2\mu+1)(SF)}.
\end{eqnarray}
Using the identity $S^{\mu\alpha}F_{\alpha\beta}S^{\beta\nu}=-\frac{1}{2}(S^{\alpha\beta}F_{\alpha\beta})S^{\mu\nu}$ we
find the inverse matrix
\begin{eqnarray}\label{pp7}
\tilde T^{\mu\nu}=\eta^{\mu\nu}+(\mu-1)b(SF)^{\mu\nu}, \qquad
b=\frac{-2e}{4m^2c^3-3e\mu(SF)},
\end{eqnarray}
The two functions are related as follows: $b=2a[2+(\mu-1)a(SF)]^{-1}$.  The vector $Z^\mu$ is defined by
\begin{eqnarray}\label{gmm12}
Z^\mu=\frac{b}{4c}S^{\mu\sigma}(\partial_\sigma F_{\alpha\beta})S^{\alpha\beta}\equiv
\frac{b}{4c}S^{\mu\sigma}\partial_\sigma(FS).
\end{eqnarray}
This vanishes for homogeneous field, $\partial F=0$. The evolution of the basic variables obtained according the
standard rule $\dot z=\{z, H\}$. The equations read
\begin{eqnarray}\label{uf4.5}
\dot x^\mu=\lambda(T^{\mu}{}_\nu{\cal P}^\nu+\frac{\mu ca}{b}Z^\mu), \qquad \dot{\cal P}^\mu=\frac{e}{c}(F\dot x)^\mu+\lambda\frac{\mu
e}{4c}\partial^\mu(FS),
\end{eqnarray}
\begin{eqnarray}\label{uf4.6}
\dot\omega^\mu=\lambda\frac{e\mu}{c}(F\omega)^\mu-\lambda\frac{2caC}{e}{\cal P}^\mu+\pi^\mu+\lambda_2\omega^\mu, \cr
\dot\pi^\mu=\lambda\frac{e\mu}{c}(F\pi)^\mu-\lambda\frac{2caD}{e}{\cal
P}^\mu-\frac{\alpha}{(\omega^2)^2}\omega^\mu-\lambda_52\pi^\mu,
\end{eqnarray}
Neither constraints nor equations of motion do not determine the variables $\lambda$ and $\lambda_2$, that is the
interacting theory preserves both reparametrization and spin-plane symmetries of the free theory. As a consequence, all
the basic variables have ambiguous evolution. $x^\mu$ and ${\cal P}^\mu$ have one-parametric ambiguity due to $\lambda$
while $\omega$ and $\pi$ have two-parametric ambiguity due to $\lambda$ and $\lambda_2$. The variables with ambiguous
dynamics do not represent observable quantities, so we need to search  for the variables that can be candidates for
observables. We note that (\ref{uf4.6}) imply an equation for $S^{\mu\nu}$ which does not contain $\lambda_2$
\begin{eqnarray}\label{uf4.9}
\dot S^{\mu\nu}&=& \lambda\frac{e\mu}{c}(FS)^{[\mu\nu]}+2{\cal P}^{[\mu}\dot x^{\nu]}\,.
\end{eqnarray}
This proves that the spin-tensor is invariant under local spin-plane symmetry. The remaining ambiguity due to $\lambda$
contained in Eqs. (\ref{uf4.5}) and (\ref{uf4.9}) is related with reparametrization invariance and disappears when we
work with physical dynamical variables $x^i(t)$. So we will work with $x^\mu$, ${\cal P}^\mu$ and $S^{\mu\nu}$. We
remind that our constraints imply the algebraic restrictions on spin-tensor
\begin{eqnarray}\label{uf4.12.1}
S^{\mu\nu}{\cal P_\nu}=0, \qquad
S^{\mu\nu}S_{\mu\nu}=8\alpha.
\end{eqnarray}
Equations (\ref{uf4.5}) and (\ref{uf4.9}), together with (\ref{uf4.12.1}), form a closed system which determines
evolution of a spinning particle.

The quantities $x^\mu$, $P^\mu$ and $S^{\mu\nu}$, being invariant under spin-plane symmetry, have vanishing brackets
with the corresponding first-class constraints $T_2$ and $T_5$. So, obtaining equations for these quantities, we can
omit the corresponding terms in the Hamiltonian (\ref{m.11}). Further, we can construct the Dirac bracket for the
second-class pair $T_3$ and $T_4$. Since the Dirac bracket of a second-class constraint with any quantity vanishes, we
can now omit $T_3$ and $T_4$ from (\ref{m.11}). Then the relativistic Hamiltonian acquires an expected form (compare it
with the square of Dirac equation (\ref{1.1_1}))
\begin{eqnarray}\label{uf4.12}
H=\frac{\lambda}{2}\left({\cal P}^2-\frac{e\mu}{2c}(FS)+m^2c^2\right).
\end{eqnarray}
The equations (\ref{uf4.5}) and (\ref{uf4.9}) follow from this $H$ with use of Dirac bracket, $\dot z=\{z, H\}_{DB}$.
The Dirac brackets in physical-time parametrization will be computed in Sect.~\ref{ch09:sec9.7}. The brackets in
arbitrary parametrization can be found in \cite{DPM3}.

We could also use the constraint $S^{\mu\nu}{\cal P}_\nu=0$ to represent $S^{0i}$ through $S^{ij}$, then
\begin{eqnarray}\label{uf4.12.2}
H=\frac{\lambda}{2}\left({\cal P}^2+\frac{eg}{c}\left[\frac{1}{{\cal P}^0}{\rm {\bf S}}[{\bf p}\times{\bf E}]- {\rm
{\bf B}}{\rm {\bf S}}\right]+m^2c^2\right).
\end{eqnarray}

\subsection{Comparison with approximate equations of Frenkel and Bargmann-Michel-Telegdi}
\label{ch09:sec9.6.2}

{\bf Lagrangian equations.} We can exclude the momenta ${\cal P}$ and the auxiliary variable $\lambda$ from the
equations of motion. This yields second-order equation for the particle's position. To achieve this, we solve the first
equation from (\ref{uf4.5}) with respect to ${\cal P}$ and use the identities $(SFZ)^\mu=-\frac12 (SF)Z^\mu$, $\tilde
T^\mu{}_\nu Z^\nu=\frac{b}{a}Z^\mu$, this gives ${\cal P}^\mu=\frac{1}{\lambda}\tilde T^\mu{}_\nu\dot x^\nu-\mu
cZ^\mu$. Then the condition $S^{\mu\nu}{\cal P}_\nu=0$ reads $\frac{1}{\lambda}(S\tilde T\dot x)^\mu=\mu c(SZ)^\mu$.
Using this equality, ${\cal P}^2$ can be presented as ${\cal P}^2=\frac{1}{\lambda^2}(\dot x G\dot x)+\mu^2c^2Z^2$,
where appeared the symmetric matrix
\begin{eqnarray}\label{L.13}
G_{\mu\nu}=(\tilde T^T\tilde T)_{\mu\nu}=[\eta+b(\mu-1)(SF+FS)+b^2(\mu-1)^2FSSF]_{\mu\nu}. \qquad
\end{eqnarray}
The matrix $G$ is composed from the Minkowski metric $\eta_{\mu\nu}$ plus spin and field-dependent contribution,
$G_{\mu\nu}=\eta_{\mu\nu}+h_{\mu\nu}(S)$. So we call $G$ the effective metric induced \textit{along the world-line} by
interaction of spin with electromagnetic field. We substitute ${\cal P}^2$ into the constraint (\ref{gmm120}), this
gives $\lambda$
\begin{eqnarray}\label{L.13.1}
\lambda=\frac{\sqrt{-\dot x G\dot x}}{m_rc}, \qquad  m_r^2=m^2-\frac{\mu e}{2c^3}(FS)-\mu^2 Z^2\,.
\end{eqnarray}
This shows that the presence of $\lambda$ in Eq. (\ref{m.2}) implies highly non-linear interaction of spinning particle
with electromagnetic field. The final expression of canonical momentum through velocity is
\begin{eqnarray}\label{L.13.2}
{\cal P}^\mu=\frac{m_rc}{\sqrt{-\dot x G\dot x}}\left[\delta^\mu{}_\nu+(\mu-1)b(SF)^\mu{}_\nu\right]\dot x^\nu-\mu cZ^\mu.
\end{eqnarray}
Using (\ref{L.13.1}) and (\ref{L.13.2}), we exclude ${\cal P}^\mu$ and $\lambda$ from the Hamiltonian equations
(\ref{uf4.5}), (\ref{uf4.9}) and (\ref{uf4.12.1}). This gives closed system of Lagrangian equations for the set $x, S$.
It is convenient to work with reparametrization-invariant derivative
\begin{eqnarray}\label{L.13.2.1}
D=\frac{1}{\sqrt{-\dot x G\dot x}}\frac{d}{d\tau}.
\end{eqnarray}
Then we have the dynamical equations
\begin{eqnarray}\label{FF.6}
D\left[m_r(\tilde T Dx)^\mu\right]=\frac{e}{c^ 2}(FDx)^\mu+ \frac{\mu
e}{4m_rc^3}\partial^\mu(SF)+\mu DZ^\mu,
\end{eqnarray}
\begin{eqnarray}\label{FF.7}
D S^{\mu\nu}=\frac{e\mu}{m_rc^2}F^{[\mu\alpha}S_{\alpha}{}^{\nu]}-2bm_rc(\mu-1)Dx^{[\mu}(SFDx)^{\nu]}+2\mu cD
x^{[\mu}Z^{\nu]}\,,
\end{eqnarray}
the Lagrangian counterpart of the condition $S^{\mu\nu}{\cal P}_\nu=0$,
\begin{eqnarray}\label{FF.8}
S^{\mu\nu}\left[\dot x^\nu+(\mu-1)b(SF\dot x)_\nu-\frac{\mu\sqrt{-\dot xG\dot x}}{m_r}Z_\nu\right]=0,
\end{eqnarray}
as well as the value-of-spin condition, $S^{\mu\nu}S_{\mu\nu}=8\alpha$. The equations contains the effective (spin and
position-dependent) mass $m_r$, this can lead to certain geometric effects \cite{Ball2017}.

In the absence of interaction we obtain an expected dynamics
\begin{eqnarray}\label{FF.8.1}
\frac{d}{d\tau}\frac{\dot x^\mu}{\sqrt{-\dot x^2}}=0, \qquad \dot S^{\mu\nu}=0, \qquad S^{\mu\nu}\dot x_\nu=0.
\end{eqnarray}
The trajectory is a straight line, while $S^{\mu\nu}$ is a constant tensor.

{\bf Discussion.} Eq. (\ref{FF.6}) and (\ref{FF.8}) show how spin modifies the classical equation of a point particle
subject to Lorentz force
\begin{align}
m\frac{d}{d\tau}\left( {\frac{\dot x^\mu }{\sqrt {-\dot x^2 } }} \right)= \frac{e}{c^2}(F\dot x)^\mu.
\label{ch01:eqn1.278}
\end{align}
Let us discuss qualitatively the corresponding contributions. Canonical momentum ${\cal P}^\mu=p^\mu-\frac{e}{c}A^\mu$
of a spinless particle is proportional to its velocity, ${\cal P}^\mu=\frac{mc}{\sqrt{-\dot x^2}}\dot x^\mu$.
Interaction of spin with electromagnetic field modifies the relation between the two quantities, see Eq.
(\ref{L.13.2}). Contribution of anomalous magnetic moment $\mu\ne 1$ to the difference between $\dot x^\mu$ and ${\cal
P}^\mu$ is proportional to $\frac{J}{c^3}\sim\frac{\hbar}{c^3}$, while the term with a gradient of field is
proportional to $\frac{J^2}{c^3}\sim\frac{\hbar^2}{c^3}$. The interaction also modifies the constraints. In particular,
the condition $S^{\mu\nu}\dot x_\nu=0$ of a free theory turns into $S^{\mu\nu}{\cal P}_\nu=0$ with ${\cal P}_\nu\ne\dot
x_\nu$. This has an important consequence. If we adopt the standard special relativity notions of time and distance,
the components $S^{0i}$ vanish in the frame ${\cal P}^\mu=({\cal P}^0, \vec 0)$ instead of the rest frame. Hence our
model predicts small \textit{dipole electric moment} of the particle immersed in an external field (for an experimental
estimations, see \cite{dm}).

Other important point is the emergence of an effective metric (\ref{L.13}) for the particle in flat space. As we saw
above,  the incorporation of the constraints (\ref{uf4.12.1}) into a variational problem, as well as the search for an
interaction consistent with them turn out to be rather non trivial tasks, and the action (\ref{m.1}) is probably the
only solution of the problem. So, the appearance of effective metric (\ref{L.13}) in equations of motion seems to be
unavoidable in a systematically constructed model of spinning particle. An important consequences will be discussed in
Sect.~\ref{ch09:sec9.8}.

Summing up, in general case the Lorentz force is modified due to the presence of (time-dependent) radiation mass $m_r$
(\ref{L.13.1}), the tetrad field $\tilde T$, the effective metric $G$ and due to two extra-terms on right hand side of
(\ref{FF.6}).

Consider the ``classical"  value of magnetic moment $\mu=1$. Then $\tilde T=\eta$ and $G=\eta$. The Lorentz force is
modified due to the presence of time-dependent radiation mass $m_r$, and two extra-terms on right hand side of
(\ref{FF.6}).

{\bf Homogeneous field.} The structure of our equations simplifies significantly for the homogeneous field
$\partial_\alpha F^{\mu\nu}=0$, then $Z^\mu=0$. Contraction of (\ref{FF.8}) with $F_{\mu\nu}$ yields $(SF)\dot{}=0$,
that is $S^{\mu\nu}F_{\mu\nu}$ turns out to be the conserved quantity. This implies $\dot m_r=\dot a=\dot b=0$. Hence
the Lorentz force is modified due to the presence of time-independent radiation mass $m_r$, the tetrad field $\tilde T$
and the effective metric $G$.  The equations (\ref{FF.6}) and (\ref{FF.8}) read
\begin{eqnarray}\label{pp20}
\frac{d}{d\tau}\frac{\dot x^\mu}{\sqrt{-\dot xG\dot x}}=\frac{e}{m_rc^2}(TF\dot x)^\mu-(T\dot{\tilde T}\dot x)^\mu,
\end{eqnarray}
\begin{equation}\label{pp19-J-F0}
\dot S^{\mu\nu} =\frac{e\mu\sqrt{-\dot xG\dot x}}{m_rc^2}F^{[\mu}{}_\alpha
S^{\alpha\nu]}-\frac{2bm_rc(\mu-1)}{\sqrt{-\dot xG\dot x}}\dot x^{[\mu}(SF\dot x)^{\nu]}.
\end{equation}
They simplify more in the parametrization which implies
\begin{eqnarray}\label{pp19}
G_{\mu\nu}\dot x^\mu\dot x^\nu=-c^2\,.
\end{eqnarray}
Since  $G\dot x\dot x=\dot x^2+O(S^2)$, in the linear approximation on $S$ this is just the proper-time
parametrization.

The equations become even more simple when $\mu=\frac{g}{2}=1$. Let us specify the equation of precession of spin to
this case, taking physical time as the parameter, $\tau=t$. Then (\ref{FF.8}) reduces to the Frenkel condition,
$S^{\mu\nu}\dot x_\nu=0$, while (\ref{FF.7}) reads $\dot S^{\mu\nu}=\frac{e\sqrt{-\dot x^2}}{m_rc^2}(FS)^{[\mu\nu]}$.
We decompose spin-tensor on electric dipole moment ${\bf D}$ and Frenkel spin-vector ${\bf S}$ according to
(\ref{f1.5}), then ${\bf D}=-\frac{2}{c}{\bf  S}\times{\bf v}$, while variation rate of ${\bf S}$ is given by
\begin{eqnarray}\label{L.14.5}
\frac{d{\bf S}}{dt}=\frac{e\sqrt{c^2-{\bf v}^2}}{m_rc^3}\left\{c{\bf S}\times{\bf B}-[{\bf E}\times[{\bf v}\times{\bf
S}]]\right\}.
\end{eqnarray}
Interaction with magnetic field yields precession of ${\bf S}$ around the vector ${\bf B}$, while interaction with
electric field leads to an extra variation rate of ${\bf S}$ in the plane of vectors ${\bf v}$ and ${\bf S}$.

{\bf Comparison with Frenkel equations.} Frenkel found equations of motion consistent with the condition
$S^{\mu\nu}\dot x_\nu=0$ up to order $O^3(S, F, \partial F)$. Besides, he considered the case $\mu=1$. Taking these
approximations in our equations in the proper-time parametrization $\sqrt{-\dot x^2}=c$, we arrive at those of Frenkel
(our $S$ is $\frac{2mc}{e}$ of Frenkel $S$)
\begin{eqnarray}\label{pp11f}
\frac{d}{d\tau}\left[(m-\frac{e}{4mc^3}(SF))\dot
x^\mu+\frac{e}{8m^2c^3}S^{\mu\alpha}\partial_\alpha(SF)\right]=\frac{e}{c}(F\dot x)^\mu+\frac{e}{4mc}\partial^\mu(SF),
\quad
\end{eqnarray}
\begin{eqnarray}\label{pp11.1f}
\dot S^{\mu\nu} =\frac{e}{mc}\left[F^{[\mu}{}_\alpha S^{\alpha\nu]}-\frac{1}{4mc^2}\dot
x^{[\mu}S^{\nu]\alpha}\partial_\alpha(SF)\right]\,, \qquad S^{\mu\nu}\dot x_\nu=0.
\end{eqnarray}

{\bf Comparison with Bargmann-Michel-Telegdi equations.}  BMT-equations are
\begin{eqnarray}\label{pp22}
\ddot x^\mu=\frac{e}{mc}(F\dot x)^\mu,
\end{eqnarray}
\begin{eqnarray}\label{pp23}
\dot s^\mu=\frac{e\mu}{mc}(Fs)^\mu-\frac{e}{mc^3}(\mu-1)(sF\dot x)\dot x^\mu, \qquad s^\mu\dot x_{\mu}=0.
\end{eqnarray}
Obtaining their equations in homogeneous field, Bargmann, Michel and Telegdi supposed that the motion of a particle is
independent from the motion of spin. Besides they looked for the equation linear on  $s^\mu$ and $F^{\mu\nu}$. It is
convenient to introduce BMT-tensor dual to $s^\mu$
\begin{eqnarray}\label{pp23.1}
S_{BMT}^{\mu\nu}=\frac{2}{c}\epsilon^{\mu\nu\alpha\beta}s_\alpha\dot x_\beta.
\end{eqnarray}
Due to (\ref{pp23}) this obeys the equation
\begin{eqnarray}\label{pp22-J}
\dot S^{\mu\nu}_{BMT}=\frac{e}{mc}F^{[\mu}{}_\alpha S^{\alpha\nu]}_{BMT}+\frac{\mu-1}{c^2}\dot x^{[\mu}(S_{BMT}F\dot x)^{\nu]}.
\end{eqnarray}
Taking the proper-time parametrization and neglecting non linear on $F$ and $S$ terms in our equations (\ref{pp20}) and
(\ref{pp19-J-F0}), we obtain (\ref{pp22}) and (\ref{pp22-J}).

{\bf Exact solution to equations of motion in a constant magnetic field.} Comparing Eqs. (\ref{pp20}) and
(\ref{pp22}) we conclude that spin-field interaction modifies the Lorentz-force equation even for the homogeneous
magnetic field. To estimate the influence, it is convenient to work with four-dimensional spin-vector (\ref{1.6})
instead of spin-tensor. The constraint $S^{\mu\nu}{\cal P}_\nu=0$ implies $s^\mu\dot x_\mu=0$, so $s^\mu$ can be
identified with BMT-vector of spin. As a consequence of Eqs. (\ref{uf4.5}) and (\ref{uf4.9}), it obeys the equation
\begin{eqnarray}\label{uf4.14}
\label{uf4.15} \dot s^\mu &=&\lambda\frac{e\mu}{c}\left[(Fs)^\mu+\frac{1}{{\cal P}^2}(sF{\cal P}){\cal
P}^\mu\right]-\frac{1}{{\cal P}^2}(\dot{\cal P}s){\cal P}^\mu.
\end{eqnarray}
For the homogeneous magnetic field the equations (\ref{uf4.5}) and (\ref{uf4.14}) has been solved exactly \cite{DPM3},
a qualitative picture of motion for $\mu\ne 1$ can be described as follows.
Besides oscillations of spin first calculated by Bargmann, Michel and Telegdi, the particle with anomalous magnetic
moment experiences an effect of magnetic {\it Zitterbewegung} of the trajectory. Usual circular motion in the plane
orthogonal to ${\bf B}$ is perturbed by slow oscillations along ${\bf B}$ with the amplitude of order of Compton
wavelength, $\frac{\vec{\cal P}}{{\cal P}^0}\lambda_C$. The Larmor frequency and the frequency of spin oscillations are
also shifted by small corrections.

\subsection{Parametrization of physical time and physical Hamiltonian}
\label{ch09:sec9.7}

Equations for physical variables $x^i(t)$, ${\cal P}^i(t)$ and $S^{\mu\nu}(t)$ follow from the formula of derivative of
parametric function, $\frac{d z}{dt}=c\frac{\dot z}{\dot x^0}$, after the substitution of (\ref{uf4.5}) and (\ref{uf4.9})
on the right hand side. Our task here is to find a conventional Hamiltonian for these equations. Consider the
Hamiltonian action associated with the Hamiltonian (\ref{m.11}), $\int d\tau ~ p\dot x+\pi\dot\omega-\lambda_iT_i$. The
variational problem provides both equations of motion and constraints of the vector model in arbitrary parametrization.
Using the reparametrization invariance of the functional, we take physical time as the evolution parameter,
$\tau=\frac{x^0}{c}=t$, then the functional reads
\begin{eqnarray}\label{ch09:eqn9.7.1}
S_H=\int dt ~  c\tilde{\cal P}_0-eA^0+p_i\dot x^i+\pi_\mu\dot\omega^\mu- \qquad \cr
\frac{\lambda}{2}\left(-\tilde{\cal P}_0^2+{\cal P}_i^2-\frac{e\mu}{2c}(FS)+m^2c^2+\pi^2-\frac{\alpha}{\omega^2}\right)-
\lambda_iT_i,
\end{eqnarray}
where it is convenient to denote $\tilde{\cal P}_0=p_0-\frac{e}{c}A_0$. We can treat the term associated with $\lambda$
as a kinematic (that is velocity-independent) constraint of the problem. According to the standard classical-mechanics
prescription \cite{deriglazov2010classical}, we solve the constraint
\begin{eqnarray}\label{ch09:eqn9.7.2}
\tilde{\cal P}_0=-\tilde{\cal P}^0=-\sqrt{{\cal P}_i^2-\frac{e\mu}{2c}(FS)+m^2c^2+\pi^2-\frac{\alpha}{\omega^2}},
\end{eqnarray}
and substitute the result back into Eq. (\ref{ch09:eqn9.7.1}), this gives an equivalent form of the functional
\begin{eqnarray}\label{ch09:eqn9.7.3}
S_H=\int dt ~  p_i\dot x^i+\pi_\mu\dot\omega^\mu- \left[c\sqrt{{\cal P}_i^2-\frac{e\mu}{2c}(FS)+
m^2c^2+\pi^2-\frac{\alpha}{\omega^2}}+eA^0+ \right. \cr \left. \lambda_2\omega_\mu\pi^\mu+\lambda_3{\cal
P}_\mu\omega^\mu+\lambda_4{\cal P}_\mu\pi^\mu\right], \qquad \qquad \qquad \qquad
\end{eqnarray}
where the substitution (\ref{ch09:eqn9.7.2}) is implied in the last two terms as well. The
expression in square brackets is the Hamiltonian. The sign in front of the square
root in (\ref{ch09:eqn9.7.2}) was chosen according to the right spinless limit, $H=c\sqrt{{\cal P}_i^2+
m^2c^2}+eA^0$. The variational problem implies the first-class constraints
$T_2=\omega\pi=0$, $T_5=\pi^2-\frac{\alpha}{\omega^2}=0$ and the second-class constraints
\begin{eqnarray}\label{ch09:eqn9.7.4}
T_3=-{\cal P}^0\omega^0+{\cal P}^i\omega^i=0, \qquad
T_4=-{\cal P}^0\pi^0+{\cal P}^i\pi^i=0,
\end{eqnarray}
where
\begin{eqnarray}\label{ch09:eqn9.7.5}
{\cal P}^0\equiv\sqrt{{\cal P}_i^2-\frac{e\mu}{2c}(FS)+m^2c^2}.
\end{eqnarray}
In all expressions below the symbol ${\cal P}^0$ represents the function (\ref{ch09:eqn9.7.5}).

To represent the Hamiltonian from (\ref{ch09:eqn9.7.3}) in a more familiar form, we take into account the second-class
constraints by passing from Poisson to Dirac bracket
\begin{eqnarray}\label{pht.13}
\{A, B\}_{D}=\{A, B\}-\{A, T_3\}\{T_4, T_3\}^{-1}\{T_4, B\}-\cr
\{A, T_4\}\{T_3, T_4\}^{-1}\{T_3, B\}. \qquad \qquad
\end{eqnarray}
To compute the Dirac brackets of our variables, we use an auxiliary Poisson brackets shown in table
\ref{tabular:poissonbr2}.
\begin{table}
\caption{Auxiliary Poisson brackets} \label{tabular:poissonbr2}
\begin{center}
\begin{tabular}{c|c|c|c}
${}$                                & $\{{\cal P}^0, *\}$                     & $\{T_3, *\}$             & $\{T_4, *\}$   \\

\hline \hline

$x^i$       & $-\frac{{\cal P}^i}{{\cal P}^0}$    & $-\omega^i+\frac{\omega^0{\cal P}^i}{{\cal P}^0}$    & $-\pi^i+\frac{\pi^0{\cal P}^i}{{\cal P}^0}$ \\

\hline

$\mathcal{P}^i$  & $-\frac{e}{{\cal P}^0c}[(F\vec {\cal P})^i+\frac{\mu}{4}\partial^i(SF)]$      & $\frac{e\omega^0}{{\cal
P}^0c}[(F\vec {\cal P})^i+\frac{\mu}{4}\partial^i(SF)]-$       & $\frac{e\pi^0}{{\cal P}^0c}[(F\vec
{\cal P})^i+\frac{\mu}{4}\partial^i(SF)]-$ \\

  &  & $\frac{e}{c}(F\vec\omega)^i$   &  $\frac{e}{c}(F\vec\pi)^i$  \\

\hline

${\cal P}^0$  & $0$ & $\frac{e}{{\cal P}^0c}[(\mu-1)(\vec{\cal P}F\vec\omega)+$
& $\frac{e}{{\cal P}^0c}[(\mu-1)(\vec{\cal P} F\vec\pi)+$ \\

     &      & $\frac{\mu}{4}\omega^i\partial^i(SF)-\mu F^{0i}{\cal P}^{[0}\omega^{i]}]$  & $\frac{\mu}{4}\pi^i\partial^i(SF)-\mu F^{0i}{\cal P}^{[0}\pi^{i]}]$ \\

\hline

$\omega^\mu$   & $-\frac{e\mu}{{\cal P}^0c}(F\omega)^\mu$  & $\frac{\omega^0e\mu}{{\cal P}^0c}(F\omega)^\mu$  &  $-{\cal
P}^\mu+\frac{\pi^0e\mu}{{\cal P}^0c}(F\omega)^\mu$ \\

\hline

$\pi^\mu$  & $-\frac{e\mu}{{\cal P}^0c}(F\pi)^\mu$  &  ${\cal P}^\mu+\frac{\omega^0e\mu}{{\cal P}^0c}(F\pi)^\mu$  &    $\frac{\pi^0e\mu}{{\cal P}^0c}(F\pi)^\mu$   \\

\hline

$J^{\mu\nu}$  & $-\frac{e\mu}{{\cal P}^0c}(FS)^{[\mu\nu]}$  & $\frac{\omega^0e\mu}{{\cal P}^0c}(FS)^{[\mu\nu]}-2{\cal P}^{[\mu}\omega^{\nu]}$   & $\frac{\pi^0e\mu}{{\cal P}^0c}(FS)^{[\mu\nu]}-2{\cal P}^{[\mu}\pi^{\nu]}$ \\

\hline
\end{tabular}
\end{center}
\end{table}
We will use the notation (\ref{uf4.71}) and
\begin{eqnarray}\label{pht15.1}
u^0=T^0{}_\mu{\cal P}^\mu+\frac{\mu ca}{b}Z^0, \qquad \qquad \qquad \qquad \cr
\triangle^{\mu\nu}=-\frac{2ca}{eu^0}{\cal P}^{(0}S^{\mu\nu)}, \quad {\cal P}^{(0}S^{\mu\nu)}={\cal
P}^0J^{\mu\nu}+{\cal P}^\mu S^{\nu 0}+{\cal P}^\nu S^{0\mu}, \cr
K^{\mu\nu}=-\frac{\mu ca}{2eu^0}S^{0\mu}\partial^\nu(SF), \qquad
L^{\mu\nu\alpha}=-\frac{2\mu a}{u^0}(FS)^{[\mu\nu]}S^{0\alpha}, \cr
g^{\mu\nu}=\eta^{\mu\nu}-\frac{2ca{\cal P}^0}{eu^0}{\cal P}^\mu{\cal P}^\nu. \qquad \qquad \qquad
\end{eqnarray}
Using the table, we obtain $\{T_3, T_4\}=\frac{eu^0}{2ca{\cal P}^0}$. Then Dirac brackets among the physical
variables $x^i(t)$, ${\cal P}^i(t)$ and $S^{\mu\nu}(t)$ are
\begin{eqnarray}\label{pht14}
\{x^i, x^j\}_{D}=\frac12\triangle^{ij}, \qquad  \{x^i, {\cal
P}^j\}_{D}=\delta^{ij}-\frac{e}{2c}\left[\triangle^{ik}F^{kj}-K^{ij}\right],
\end{eqnarray}
\begin{eqnarray}\label{pht14.1}
\{{\cal P}^i, {\cal
P}^j\}_{D}=\frac{e}{c}F^{ij}-\frac{e^2}{2c^2}\left[F^{ik}\triangle^{kn}F^{nj}-F^{[ik}K^{kj]}\right],
\end{eqnarray}
\begin{eqnarray}\label{pht14.2}
\{S^{\mu\nu}, S^{\alpha\beta}\}_{D}= 2(g^{\mu\alpha} S^{\nu\beta}-g^{\mu\beta} S^{\nu\alpha}-g^{\nu\alpha} S^{\mu\beta}
+g^{\nu\beta} S^{\mu\alpha})+L^{\mu\nu[\alpha}{\cal P}^{\beta]}, \qquad \quad
\end{eqnarray}
\begin{eqnarray}\label{pht14.3}
\{S^{\mu\nu}, x^j\}_{D}={\cal P}^{[\mu}\triangle^{\nu]j}+\frac12 L^{\mu\nu j},
\end{eqnarray}
\begin{eqnarray}\label{pht14.4}
\{S^{\mu\nu}, {\cal P}^j\}_{D}=\frac{e}{c}\left[-{\cal P}^\mu(\triangle^{\nu k}F^{kj}-K^{\nu
j})-(\mu\leftrightarrow\nu)+\frac12 L^{\mu\nu k}F^{kj}\right].
\end{eqnarray}
To continue, let us restrict to the case of a stationary electromagnetic field, then constraints do not depend
explicitly on time. Dirac bracket of any quantity with second-class constraints vanish, so they can be omitted from the
Hamiltonian. So we omit the last two terms in (\ref{ch09:eqn9.7.3}). The first-class constraints $T_2$ and $T_5$ can be
omitted as well, as they do not contribute into equations of motion for physical variables. In the result we obtain the
physical Hamiltonian
\begin{eqnarray}\label{pht.16}
H_{ph}=c\sqrt{\vec{\cal P}^2-\frac{e\mu}{2c}F_{\mu\nu}S^{\mu\nu}+m^2c^2}+eA^0.
\end{eqnarray}
The equations of motion that we discussed at the beginning of this section follow from this Hamiltonian according the
rule $\frac{d z}{dt}=\{z, H_{ph}\}_D$.

Note that the Dirac brackets encode the most part of spin-field interaction, on this reason we have arrived at a rather
simple form of physical Hamiltonian. The inclusion of an interaction into the geometry of phase-space and the resulting
non commutative geometry is under intensive investigation in various models \cite{Wilson10, Bruno15, Gorji16,
Daszkiewicz15, Daszkiewicz16, Mendez16, Ghosh14, Tkachuk16, Ghosh16, Quangh16, 1, 2, Dasz12017, Rossi, Kai1, Kai2,
Majid, Liu1, Liu2, Abreu12017, Abreu22017}.

\section{Spin-induced non commutativity of position and fine structure of hydrogen spectrum}
\label{ch09:sec9.70} Here we discuss how the vector model resolves \cite{DPM2016} the problem of covariant formalism
described in Sect.~\ref{ch09:sec9.5.2}.

To quantize our relativistic theory we need to find quantum realization of highly non linear classical brackets
(\ref{pht14})-(\ref{pht14.4}). They remain non canonical even in absence of interaction. For instance, Eq.
(\ref{pht14}) in a free theory reads $\{x^i, x^j\}=\frac{1}{2mcp^0}S^{ij}$. We emphasize that non relativistic model
has canonical brackets (\ref{qq7}), so the deformation arises as a relativistic correction induced by spin of a
particle. Technically, the deformation is due to the fact that the constraints $p\omega=p\pi=0$ of relativistic theory,
used to construct the Dirac bracket, mixes up space-time and inner-space coordinates.

Quantum realization of the brackets in a free theory will be obtained in Sect.~\ref{ch09:sec9.20}, while in an
interacting theory its explicit form is unknown. Therefore we quantize the interacting theory perturbatively,
considering $c^{-1}$ as a small parameter and expanding all quantities in a power series. Let us consider the
approximation $O(c^{-2})$, that is we neglect $c^{-3}$ and higher order terms. For the Hamiltonian (\ref{pht.16}) we
have $H_{ph}\approx mc^2+\frac{\boldsymbol{\cal P}^2}{2m}-\frac{\boldsymbol{\cal P}^4}{8m^3c^2}-\frac{e\mu}{4mc}(FS)$.
Since the last term is of order $c^{-1}$, resolving the constraint $S^{\mu\nu}{\cal P}_\nu=0$ with respect to $S^{i0}$
we can approximate ${\cal P}^0=mc$, then $S^{i0}=\frac{1}{mc}S^{ij}{\cal P}^j$. Using this expression together with Eq.
(\ref{f1.1}) we obtain, up to order $c^{-2}$
\begin{eqnarray}\label{pht.16.0}
H_{ph}\approx mc^2+\frac{\boldsymbol{\cal P}^2}{2m}-\frac{\boldsymbol{\cal P}^4}{8m^3c^2}+eA^0+\frac{e\mu}{mc}
\left[\frac{1}{mc}{\rm {\bf S}}[{\boldsymbol{\cal P}}\times{\bf E}]- {\rm {\bf B}}{\rm {\bf S}}\right].
\end{eqnarray}
Due to the second and fourth terms, we need to know the operators $\hat{\cal P}^i$ and $\hat x^i$ up to order $c^{-2}$,
while $\hat S^{ij}$ should be found up to order $c^{-1}$. With this approximation, the commutators $[\hat x, \hat x]$,
$[\hat x, \hat{\cal P}]$, and $[\hat{\cal P}, \hat{\cal P}]$ can be computed up to order $c^{-2}$, while the remaining
commutators can be written only up to $c^{-1}$. Therefore,  we expand the right hand sides of Dirac brackets
(\ref{pht14})-(\ref{pht14.4}) in this approximation
\begin{eqnarray}\label{16.1}
\{x^i, x^j\} & = & \frac{1}{2m^2c^2}S^{ij}+O\left(\frac{1}{c^3}\right), \nonumber \\
\label{16.2}
\{x^i, {\cal P}^j\} & = & \delta^{ij}+O\left(\frac{1}{c^3}\right), \nonumber \\
\label{16.3}
\{x^i, S^{jk}\} &= & 0+O\left(\frac{1}{c^2}\right),\\
\label{16.4} \{{\cal P}^i, {\cal P}^j\} &= &\frac{e}{c} F^{ij}+O\left(\frac{1}{c^4}\right), \nonumber \\
\label{16.5} \{{\cal P}^i, S^{jk}\} &= & O\left(\frac{1}{c^3}\right),  \nonumber \\
\label{16.6} \{S^{ij}, S^{kl}\} &= &
2(\delta^{ik}S^{jl}-\delta^{il}S^{jk}-\delta^{jk}S^{il}+\delta^{jl}S^{ik})+O\left(\frac{1}{c^2}\right). \nonumber
\end{eqnarray}
An operator realization of these brackets reads
\begin{eqnarray}
\hat{P}_i &=& -i\hbar\frac{\partial}{\partial x^i}-\frac{e}{c}{A}_i({\bf x}),  \label{16.7} \\
\hat{x}_i & =& x_i-\frac{\hbar}{4m^2c^2}\epsilon_{ijk}\hat{P}^j\sigma^k,  \label{16.8} \\
\hat{S}^{ij}& = &\hbar\epsilon_{ijk}\sigma_k,  \label{16.9}  \qquad
\end{eqnarray}
then
\begin{eqnarray}
\hat S^i=\frac14\epsilon_{ijk}S^{jk}=\frac{\hbar}{2}\sigma^i, \label{pht.16.9} \\
\hat{S}^{i0} =\frac{\hbar}{mc}\epsilon_{ijk}\hat{P}^j\sigma^k. \label{pht.16.10}
\end{eqnarray}
By construction of a Dirac bracket, the operator $\hat S^{i0}$ automatically obeys the desired commutators up to order
$c^{-1}$. So we do not worried on this operator in the computations above.

We substitute these operators into the classical Hamiltonian (\ref{pht.16.0}). Expanding $A^0(\hat{\bf x})$ in a power
series, we obtain an additional contribution of order $c^{-2}$ to the potential due to non commutativity of the
position operator
\begin{eqnarray}\label{17}
eA^0\left(x_i-(2mc)^{-2}\epsilon_{ijk}\hat{P}^j\hat S^k\right)\approx eA^0({\bf x})-\frac{e}{2m^2c^2}\hat{\bf
S}[\hat{\bf P}\times\hat{\bf E}].
\end{eqnarray}
The contribution has the same structure as fifth term in the Hamiltonian (\ref{pht.16.0}). In the result, the quantum
Hamiltonian up to order $c^{-2}$ reads (we remind that $\mu=\frac{g}{2}$)
\begin{equation}\label{18}
\hat H_{ph}= mc^2+\frac{\hat{\bf P}^2}{2m}-\frac{\hat{\bf P}^4}{8m^3c^2}+eA^0+\frac{e(g-1)}{2m^2c^2}
\hat{\bf S}[\hat{\bf P}\times{\bf E}]-\frac{eg}{2mc}{\bf B}\hat{\bf S}.
\end{equation}
The first three terms corresponds to an increase of relativistic mass. The last two terms coincides with those in Eq.
(\ref{ch08:eqn8.164}). In the result, we have shown that non commutativity of electron's position in the vector model
of spin is responsible for the fine structure of hydrogen atom.

We could carry out the same reasoning in classical theory, by asking on the new variables $z'$ that obey the canonical
brackets (\ref{qq7}) as a consequence of equations (\ref{16.3}). In the desired approximation they are ${\cal
P}^i={\cal P}'^i$, $x^i=x'^i-\frac{1}{4m^2c^2}S'^{ij}{\cal P}'^j$ and $S^{ij}=S'^{ij}$, that is the first relativistic
corrections modify only the position variable.

\section{Ultra-relativistic spinning particle in electromagnetic background}
\label{ch09:sec9.8}

Let us compare the Lagrangian equations of spinning (\ref{FF.6}) and spinless (\ref{ch01:eqn1.278}) particle. For the
spinning particle with $\mu\ne 1$, the relativistic-contraction factor (see (\ref{L.13.2.1})) contains the effective
metric (\ref{L.13}) instead of the Minkowski metric $\eta_{\mu\nu}$. In the result,  equations for trajectory
(\ref{FF.6}) and for precession of spin (\ref{FF.7}) became singular at the critical velocity which obeys the equation
$\dot xG\dot x=0$.
As we saw above, the singularity determines behavior of the particle in
ultra-relativistic limit. The effective metric is composed from the Minkowski one plus (spin and field-dependent)
contribution, $G=\eta+h(S)$. So we need to decide, which of them should be used to construct the three-dimensional
geometry discussed in Sect.~\ref{ch06:sec6.9}. We first test the usual special-relativity notions,
$v^i=\frac{dx^i}{dt}$, $a^i=\frac{dv^i}{dt}$ and ${\bf v}{\bf a}=v^ia^i$, that is we suppose that the particle sees
$\eta$ as the space-time metric. We show that in this case acceleration vanishes at the critical speed which is
different from the speed of light. Then we estimate the ultra-relativistic limit using $G$ to define the
three-dimensional geometry (\ref{La.3.0})-(\ref{La.5}). Then $v_{cr}=c$, but since $G$ depends on spin, particles with
different spins will probe slightly different three-dimensional geometries.

{\bf Ultra relativistic limit within the usual special-relativity notions.} It will be sufficient to estimate the
acceleration in the uniform and stationary field. We take $\tau=t$ in equations (\ref{FF.6})-(\ref{FF.8}) and compute
the time derivative derivative on l. h. s. of equations (\ref{FF.6}) with $\mu=1, 2, 3$. Then the equations read
\begin{eqnarray}\label{FF.15.0}
a^i-\frac{v^i}{2(-vGv)}\frac{d}{dt}(-vGv)=T^i{}_\nu\left[\frac{e\sqrt{-vGv}}{m_rc^ 2}(Fv)^i-\frac{d}{dt}\tilde
T^\nu{}_\alpha v^\alpha\right]\,,
\end{eqnarray}
\begin{eqnarray}\label{FF.16}
\frac{d}{dt}S^{\mu\nu}=\frac{e\mu\sqrt{-vGv}}{m_rc^2}(FS)^{[\mu\nu]}- \frac{2bm_rc(\mu-1)}{\sqrt{-vGv}}v^{[\mu}(SFv)^{\nu]}\,,
\end{eqnarray}
\begin{eqnarray}\label{FF.17}
(Sv)^\mu+b(\mu-1)(SSFv)^\mu=0,
\end{eqnarray}
where $v^\mu=(c, {\bf v})$.
Eqs. (\ref{FF.17}) and (\ref{L.13}) imply
\begin{eqnarray}\label{FF.17.1}
-vGv=-v\tilde Tv=c^2-{\bf v}^2-(\mu-1)b(vSFv).
\end{eqnarray}
We compute the time-derivatives in Eq. (\ref{FF.15.0})
\begin{eqnarray}\label{FF.15.1}
\frac{d}{dt}(-vGv)=-2({\bf v}{\bf a})-(\mu-1)b\left\{ [v(FS+SF)]_ia^i+ \right.\cr\left.
\frac{e\mu\sqrt{-vGv}}{m_rc^2}[(vFFSv)+(vFSFv)]-\qquad \right. \cr \left.
\frac{2bm_rc(\mu-1)}{\sqrt{-vgv}}[v^2(vFSFv)-(vSFv)(vFv)]\right\},
\end{eqnarray}
\begin{eqnarray}\label{FF.15.2}
-T^i{}_\nu\frac{d}{dt}\tilde T^\nu{}_\alpha v^\alpha=-\frac{e\sqrt{-vGv}}{m_rc^2}\left\{\mu(\mu-1)b(FSFv)^i- \right. \cr\left.
\mu(\mu-1)a(SFFv)^i-\mu(\mu-1)^2ab(SFFSFv)^i\right\}+ \cr
\frac{2bm_rc(\mu-1)}{\sqrt{-vGv}}T^i{}_\nu[v^\nu(vFSFv)-(SFv)^\nu(vFv)].
\end{eqnarray}
We note that all the potentially divergent terms (two last terms in (\ref{FF.15.1}) and in (\ref{FF.15.2})), arising due
to the contribution from $\dot S\sim\frac{1}{\sqrt{-vGv}}$, disappear on the symmetry grounds. We substitute non
vanishing terms into (\ref{FF.15.0}) obtaining the expression
\begin{eqnarray}\label{FF.15}
M^i{}_ja^j= \frac{e\sqrt{-vGv}}{m_rc^2}\left\{(Fv)^i-\mu(\mu-1)b(FSFv)^i+ \qquad \right. \cr\left. (\mu-1)^2a(SFF[\eta+\mu bSF]v)^i-
v^i\frac{\mu(\mu-1)b}{2(-vGv)}(vFFSv)\right\},
\end{eqnarray}
where the matrix
\begin{eqnarray}\label{FF.15.3}
M^i{}_j=\delta^i{}_j+\frac{v^iv^\mu\Omega_{\mu j}}{2(-vGv)},
\quad \mbox{with} \quad  \Omega_{\mu j}=2\delta_{\mu j}+(\mu-1)b(FS+SF)_{\mu j}, \qquad
\end{eqnarray}
has the inverse
\begin{eqnarray}\label{FF.15.4}
\tilde M^i{}_j=\delta^i{}_j-\frac{v^iv^\mu\Omega_{\mu j}}{2c^2-(\mu-1)bv^\mu(FS+SF)_{\mu 0}v^0},
\end{eqnarray}
with the property
\begin{eqnarray}\label{FF.15.5}
\tilde M^i{}_jv^j=v^i\frac{2(-vGv)}{2c^2-(\mu-1)bv^\mu(FS+SF)_{\mu 0}v^0}.
\end{eqnarray}
Applying the inverse matrix we obtain the acceleration
\begin{eqnarray}\label{FF.15.6}
a^i= \frac{e\sqrt{-vGv}}{m_rc^2}\left\{\tilde M^i{}_j[(Fv)^j-\mu(\mu-1)b(FSFv)^j+ \right. \cr\left.
(\mu-1)^2a(SFF[\eta+\mu bSF]v)^j]-\right. \cr \left. v^i\frac{\mu(\mu-1)b(vFFSv)}{2c^2-(\mu-1)bv^\mu(FS+SF)_{\mu
0}v^0}\right\}.
\end{eqnarray}
For the particle with non anomalous magnetic moment ($\mu=1$), the right hand side reduces to the Lorentz force, so the
expression in braces is certainly non vanishing in the ultra-relativistic limit. Thus the acceleration vanishes only
when $v\rightarrow v_{cr}$, where the critical velocity is determined by the equation $vGv=0$.

Let us estimate the critical velocity.  Using the consequence $(\dot xSF\dot x)=-b(\mu-1)(\dot xFSSF\dot x)$ of the
supplementary spin condition, and the expression $S^{\mu}{}_\alpha
S^{\alpha\nu}=-4\left[\pi^2\omega^\mu\omega^\nu+\omega^2\pi^\mu\pi^\nu\right]$, we write
\begin{eqnarray}\label{FFF.6.1}
-(\dot xG\dot x)=c^2-{\bf v}^2+4b^2(\mu-1)^2\left[\pi^2(\omega F\dot
x)^2+\omega^2(\pi F\dot x)^2\right].
\end{eqnarray}
As $\pi$ and $\omega$ are space-like vectors, the last term is non-negative, so $v_{cr}\ge c$. We show that generally
this term is nonvanishing function of velocity, then $v_{cr}> c$. Assume the contrary, that this term vanishes at some
velocity, then
\begin{eqnarray}\label{FFF.7.1}
\omega F\dot x=-\omega^0({\bf E}{\bf v})+({\boldsymbol{\omega}}, c{\bf E}+{\bf v}\times{\bf B})=0\,, \cr \pi F\dot
x=-\pi^0({\bf E}{\bf v})+({\boldsymbol{\pi}}, c{\bf E}+{\bf v}\times{\bf B})=0\,.
\end{eqnarray}
This implies $c({\bf D}{\bf E})+({\bf D}, {\bf v}\times{\bf B})=0$. Consider the case ${\bf B}=0$, then it should be
$({\bf D}{\bf E})=0$. On other hand, for the homogeneous field the quantity $S^{\mu\nu}F_{\mu\nu}=2\left[({\bf D}{\bf
E})+2({\bf S}{\bf B})\right]=2({\bf D}{\bf E})$ is a constant of motion. Hence we can take the initial
conditions for spin such, that $({\bf D}{\bf E})\ne 0$ at any instant, this implies $v_{cr}>c$.

{\bf Ultra-relativistic limit within the geometry determined by effective metric.}\label{sec7.3}
As we saw above, if we insist to preserve the usual special-relativity definitions of time and
distance, the speed of light does not represent special point of the equation for trajectory.
Acceleration of the particle with anomalous magnetic moment generally vanishes at the speed slightly higher than the speed of
light. Hence we arrive at a rather surprising result that speed of light does not represent maximum velocity of the
manifestly relativistic equation (\ref{FF.15}). This state of affairs is unsatisfactory because the Lorentz
transformations have no sense above $c$, so two observers with relative velocity $c<v<v_{cr}$ will not be able to
compare results of their measurements.

To keep the condition $v_{cr}=c$, we use formal similarity of the matrix $G$ appeared in (\ref{L.13}) with space-time
metric. Then we can follow the general-relativity prescription of Sect.~\ref{ch06:sec6.9} to define time and distance in the
presence of electromagnetic field. That is we use $G$ of Eq. (\ref{L.13}) to define the three-dimensional geometry
(\ref{La.3.0})-(\ref{La.5}). The effective metric depends on $x^i$ via the field strength $F(x^0, x^i)$, and on $x^0$
via the field strength as well as via the spin-tensor $S(x^0)$. So the effective metric is time-dependent even in
stationary electromagnetic field. With these definitions we have, by construction, $-\dot x G \dot x=
(\frac{dt}{dx^0})^2(c^2 - ({\bf v}\gamma{\bf v}))$, so the critical speed coincides with the speed of light. The
intervals of time and distance are given now by Eq. (\ref{La.3.0}) and (\ref{La.3}), they slightly differ from those in
empty space.

In the present case, the expression for three-acceleration can be obtained in closed form in an arbitrary
electromagnetic field. We present Eq. (\ref{FF.6}) in the form (\ref{La.20.7})
\begin{eqnarray}\label{FF.16}
DDx^\mu={\mathcal F}^\mu=-Dx^\mu\frac{Dm_r(S)}{m_r}- T^\mu{}_\nu D\tilde
T^\nu{}_\alpha(S)Dx^\alpha+ \cr T^\mu{}_\nu\left\{\frac{e}{m_rc^ 2}(FDx)^\nu+ \frac{\mu
e}{4m^2_rc^3}\partial^\nu(SF)+\frac{\mu}{m_r} DZ^\nu\right\}\,.
\end{eqnarray}
Then the acceleration is given by (\ref{La.20.9}). The first two terms on right hand side of (\ref{FF.16}) give
potentially divergent contributions arising from the piece $\dot S\sim\frac{1}{\sqrt{c^2-{\bf v}\gamma{\bf v}}}$ of Eq.
(\ref{FF.7}). In the previous section we have seen that the dangerous contribution contained in the second term
disappears. To analyze the first term, we substitute ${\mathcal F}^i$ from (\ref{FF.16}) into (\ref{La.20.9}). With use
the property $\tilde M^i{}_jv^j=v^i\frac{c^2-{\bf v}\gamma{\bf v}}{c^2}$, we obtain the acceleration
\begin{eqnarray}\label{FF.170}
a^i=(c^2-{\bf v}\gamma{\bf v})\left[-v^i\frac{\dot m_r}{m_rc^2}-\frac{\tilde M^i{}_jT^j{}_\nu\dot{\tilde
T}^\nu{}_\alpha v^\alpha}{c^2-{\bf v}\gamma{\bf v}}+\right. \qquad \qquad \qquad \quad \cr \left. \tilde
M^i{}_jT^j{}_\nu\left\{\frac{e}{m_rc^2\sqrt{c^2-{\bf v}\gamma{\bf v}}}(Fv)^\nu+ \frac{\mu
e}{4m^2_rc^3}\partial^\nu(SF)+\frac{\mu}{m_r\sqrt{c^2-{\bf v}\gamma{\bf v}}} \dot Z^\nu\right\}\right]+ \cr \tilde
M^i{}_j\tilde\Gamma^j{}_{kl}(\gamma)v^kv^l+\frac12\left(\frac{dt}{dx^0}\right)^{-1}\left[({\bf
v}\partial_0\gamma\gamma^{-1})^i-\frac{v^i}{c^2}({\bf v}\partial_0\gamma{\bf v})\right], \qquad \quad
\end{eqnarray}
so the divergency due to $\dot m_r\sim\frac{1}{\sqrt{c^2-{\bf v}\gamma{\bf v}}}$ is cancelled by the factor in front of
this term. In the result, the acceleration is finite as $v\rightarrow c$. Besides, taking into account the property
$({\bf v}\gamma)_i\tilde M^i{}_j=({\bf v}\gamma)_j\frac{c^2-{\bf v}\gamma{\bf v}}{c^2}$, we conclude that the
longitudinal acceleration (\ref{La.20.10})
\begin{eqnarray}\label{FF.18}
{\bf v}\gamma{\bf a}=\frac{(c^2-{\bf v}\gamma{\bf v})^2}{c^2}({\bf v}\gamma{\boldsymbol{\mathcal F}})+\qquad \qquad
\qquad \cr \frac{c^2-{\bf v}\gamma{\bf v}}{c^2} \left[({\bf
v}\gamma)_i\tilde\Gamma^i{}_{kl}(\gamma)v^kv^l+\frac12\left(\frac{dt}{dx^0}\right)^{-1}({\bf v}\partial_0\gamma{\bf
v})\right],
\end{eqnarray}
vanishes in this limit.

In summary, to preserve the equality $v_{cr}=c$, we are forced to assume that particle in electromagnetic field probes
the three-dimensional geometry determined with respect to the effective metric instead of the Minkowski metric.
Similarly to Sect.~\ref{ch09:sec9.11},  this implies rather unusual picture of the Universe filled with spinning
matter. Since $G$ depends on spin, each particle will probe his own three-dimensional geometry. In principle this could
be an observable effect. With the effective metric (\ref{L.13}), the equation (\ref{La.3.0}) implies that the time of
life of muon in electromagnetic field and in empty space should be different.

\section{Interaction with Gravitational Field}
\label{ch09:sec9.9}

\subsection{Lagrangian of spinning particle with gravimagnetic Moment}
\label{ch09:sec9.9}

As we saw in Sect.~\ref{ch09:sec9.5.2}, minimal interaction with gravitational field can be achieved by direct
covariantization of the action (\ref{Lfree}). Remarkably, this leads to MPTD equations, see Sect.~\ref{ch09:sec9.9.5}
below. Since they become problematic in ultra-relativistic regime, we are forced to look for a non minimal interaction
that could suitably modify the equations in this regime. To understand, how they might look, we use the remarkable
analogy existing between the gravitational and electromagnetic  fields. Hamiltonian formulations of the two minimally
interacting theories become very similar if we identify electromagnetic field strength with the Riemann tensor
contracted with spin,  $F_{\mu\nu}\sim R_{\mu\nu\alpha\beta}S^{\alpha\beta}$. In particular, Hamiltonian action for
both theories is
\begin{eqnarray}\label{ag1}
S_H=\int d\tau ~ p_\mu\dot x^\mu+\pi_\mu\dot \omega^\mu-\left[\frac{\lambda}{2}\left( P^2+(mc)^2
+ \pi^2 - \frac{\alpha}{\omega^2}\right)+\lambda_aT_a\right], \quad
\end{eqnarray}
where $P_\mu=p_\mu-\frac{e}{c}A_\mu$ for electromagnetic field and
$P_\mu=p_\mu-\Gamma^\beta_{\alpha\mu}\omega^\alpha\pi_\beta$ for gravitational field. According to Eq. (\ref{m.11}),
non minimal interaction through gyromagnetic ratio $2\mu$ implies the contribution
$-\frac{e\mu}{2c}F_{\mu\nu}S^{\mu\nu}$ into the third term. So we expect that non minimal interaction with gravity
could be achieved replacing this term by $\frac{\lambda_1}{32}\kappa R_{\alpha\beta\mu\nu}S^{\alpha\beta}S^{\mu\nu}$.
By analogy with the magnetic moment, the coupling constant $\kappa$ is called gravimagnetic moment
\cite{Khriplovich1989, Pomeranskii1998}. Thus we consider the variational problem \cite{DWGR2016S}
\begin{eqnarray}\label{gmm1}
S_{\kappa}=\int d\tau ~ p_\mu\dot x^\mu+\pi_\mu\dot \omega^\mu-\left[\frac{\lambda_1}{2}\left( P^2+\kappa
R_{\alpha\beta\mu\nu}\omega^\alpha \pi^\beta \omega^\mu \pi^\nu + (mc)^2 + \pi^2 -
\frac{\alpha}{\omega^2}\right)+\right.\cr\left. \lambda_2 (\omega\pi) +\lambda_3 (P\omega)+\lambda_4  (P\pi)\right]. \qquad \qquad \qquad
\end{eqnarray}
on the space of independent variables $x^\mu, p_\nu$, $\omega^\mu, \pi_\nu$ and $\lambda_a$.

Let us look for the Lagrangian which in the phase space implies the variational problem (\ref{gmm1}). First, we note
that the constraints $\pi\omega=P\omega=0$ always appear from the Lagrangian which depends on $N\dot x$ and
$N\dot\omega$ instead of $\dot x$ and $\dot\omega$. So we set $\lambda_2=\lambda_3=0$ in (\ref{gmm1}). Second, we
present the remaining terms in (\ref{gmm1}) in the form
\begin{eqnarray}\label{acc-7}
S_{\kappa}=\int d\tau ~ p_\mu\dot x^\mu+\pi_\mu\dot \omega^\mu-\frac{\lambda_1}{2}(P, \pi)\left(
\begin{array}{cc}
g& \lambda g\\
\lambda g& \sigma
\end{array}
\right)\left(
\begin{array}{c}
P\\
\pi
\end{array}
\right)-\frac{\lambda_1}{2}\left[(mc)^2-\frac{\alpha}{\omega^2}\right],
\end{eqnarray}
where we have introduced the symmetric matrix
\begin{eqnarray}\label{acc-8}
\sigma^{\mu\nu}=g^{\mu\nu}+\kappa R_\alpha{}^\mu{}_\beta{}^\nu\omega^\alpha\omega^\beta, \quad \mbox{then} \quad
\sigma^{\mu\nu}\omega_\nu=\omega^\mu.
\end{eqnarray}
The matrix appeared in (\ref{acc-7}) is invertible, the inverse matrix is
\begin{eqnarray}\label{acc-9}
\left(
\begin{array}{cc}
K\sigma& -\lambda K\\
-\lambda K& K
\end{array}\right), \quad \mbox{where} \quad K=(\sigma-\lambda^2g)^{-1}.
\end{eqnarray}
When $\kappa=0$ we have $K^{\mu\nu}=(1-\lambda^2)^{-1}g^{\mu\nu}$, and (\ref{acc-9}) coincides with the matrix appeared
in the free Lagrangian (\ref{FF.1.2}). Third, we remind that the Hamiltonian variational problem of the form $p\dot
q-\frac{\lambda_1}{2}pAp$ follows from the reparametrization-invariant Lagrangian $\sqrt{\dot qA^{-1}\dot q}$. So, we
tentatively replace the matrix appeared in (\ref{FF.1.2}) by (\ref{acc-9}) and switch on the minimal interaction of
spin with gravity, $\dot\omega\rightarrow\nabla\omega$. This gives the following Lagrangian formulation of spinning
particle with gravimagnetic moment \cite{DWGR2017}:
\begin{eqnarray}
L=-\sqrt{(mc)^2 -\frac{\alpha}{\omega^2}} ~ \sqrt{-(N\dot x, N\nabla\omega)\left(
\begin{array}{cc}
K\sigma& -\lambda K\\
-\lambda K& K
\end{array}
\right)\left(
\begin{array}{c}
N\dot x\\
N\nabla\omega
\end{array}
\right)}=  \quad \label{acc-10}  \\
-\sqrt{(mc)^2 -\frac{\alpha}{\omega^2}} ~ \sqrt{-\dot xNK\sigma N\dot x-\nabla\omega NKN\nabla\omega+2\lambda\dot
xNKN\nabla\omega}. \quad  \label{acc-11}
\end{eqnarray}
Let us show that it does give the desired Hamiltonian formulation (\ref{gmm1}). The matrixes $\sigma$, $K$ and $N$ are
symmetric and mutually commuting. Canonical momentum for $\lambda$ vanishes and hence represents the primary
constraint, $p_\lambda=0$. Conjugate momenta for $x^\mu$ and $\omega^\mu$
are $p_\mu=\frac{\partial L}{\partial\dot x^\mu}$ and  $\pi_\mu=\frac{\partial L}{\partial\dot\omega^\mu}$
respectively. Due to the presence of Christoffel symbols in $\nabla\omega^\mu$, the conjugated momentum $p_\mu$ does
not transform as a vector, so it is convenient to introduce the canonical momentum
\begin{equation}\label{ag2}
P_\mu\equiv p_\mu-\Gamma^\beta_{\alpha\mu}\omega^\alpha\pi_\beta,
\end{equation}
the latter  transforms as a vector under general transformations of coordinates. Manifest form of the momenta is as
follows:
\begin{eqnarray}\label{acc-12}
P_\mu& =& \frac{1}{L_0}\left[m^2c^2 -\frac{\alpha}{\omega^2}\right]^{\frac12}\left[(\dot xNK\sigma
N)_\mu-\lambda(\nabla\omega NKN)_\mu\right],
\end{eqnarray}
\begin{eqnarray}\label{acc-13}
\pi_\mu &=& \frac{1}{\sqrt{2}L_0}\left[m^2c^2 -\frac{\alpha}{\omega^2}\right]^{\frac12} \left[(\nabla\omega
NKN)_\mu-\lambda(\dot xNKN)_\mu\right],
\end{eqnarray}
where $L_0$ is the second square root in (\ref{acc-11}). They immediately imply the primary constraints $\omega\pi=0$
and $P\omega=0$. From the expressions
\begin{eqnarray}\label{acc-14}
P^2=\frac{1}{L_0^2}\left[(mc)^2 -\frac{\alpha}{\omega^2}\right]\left[(\dot xNK\sigma K\sigma N\dot
x)+\lambda^2(\nabla\omega NKKN\nabla\omega)- \right.\cr\left. 2\lambda(\dot xNK\sigma KN\nabla\omega)\right], \qquad
\qquad \qquad \qquad \cr \pi\sigma\pi=\frac{1}{L_0^2}\left[(mc)^2 -\frac{\alpha}{\omega^2}\right]\left[\lambda^2(\dot
xNK\sigma K N\dot x)+(\nabla\omega NK\sigma KN\nabla\omega)-  \right.\cr\left.  2\lambda(\dot xNK\sigma
KN\nabla\omega)\right], \qquad \qquad \qquad\qquad \cr 2\lambda P\pi=\frac{1}{L_0^2}\left[(mc)^2
-\frac{\alpha}{\omega^2}\right]\left[-2\lambda^2(\dot xNK\sigma K N\dot x)-2\lambda^2(\nabla\omega NK KN\nabla\omega)+
\right.\cr\left. 2\lambda(\dot xNK\sigma KN\nabla\omega)+ 2\lambda^3(\dot xNKKN\nabla\omega)\right], \qquad \qquad
\qquad
\end{eqnarray}
we verify that their sum does not depend on velocities and hence gives one more constraint
\begin{eqnarray}\label{acc-15}
P^2+\pi\sigma\pi+2\lambda P\pi=-\left[(mc)^2 -\frac{\alpha}{\omega^2}\right].
\end{eqnarray}
Then Hamiltonian is $H=p \dot x +\pi \dot\omega - L + \lambda_i T_i\equiv P \dot x +\pi \nabla\omega - L + \lambda_i
T_i$, where the first and second terms have been identically rewritten in the general-covariant form. From
(\ref{acc-12}) and (\ref{acc-13}) we obtain $H_0=P \dot x +\pi \nabla\omega-L=0$, so the Hamiltonian is composed from primary
constraints
\begin{eqnarray}\label{acc-16}
H=\frac{\lambda_1}{2}\left[P^2+\kappa
R_{\alpha\mu\beta\nu}\omega^\alpha\pi^\mu\omega^\beta\pi^\nu+(mc)^2+\pi^2-\frac{\alpha}{\omega^2}+
2\lambda(P\pi)\right]+ \cr \lambda_2(\omega\pi)+\lambda_3(P\omega). \qquad \qquad \qquad \qquad \qquad
\end{eqnarray}
After the change of variables $\lambda\rightarrow\lambda_4=\frac12\lambda_1\lambda$, we arrive at the Hamiltonian
appeared in the variational problem (\ref{gmm1}).

{\bf Hamiltonian equations of motion.} Variation of the Hamiltonian action
(\ref{gmm1}) with respect to $\lambda_a$ gives the algebraic equations
\begin{eqnarray}\label{gmm2}
P^2+\kappa R_{\alpha\beta\mu\nu}\omega^\alpha \pi^\beta \omega^\mu \pi^\nu + (mc)^2 + \pi^2 -
\frac{\alpha}{\omega^2}=0,
\end{eqnarray}
\begin{eqnarray}\label{gmm3}
\omega\pi=0 , \qquad P\omega=0, \qquad  P\pi=0,
\end{eqnarray}
while variations with respect to the remaining variables yield dynamical equations which can be written in the
covariant form as follows
\begin{eqnarray}\label{gmm4}
\frac{\delta S_{\kappa}}{\delta p_\mu}=0 \quad \Leftrightarrow \quad \dot
x^\mu=\lambda_1P^\mu+\lambda_3\omega^\mu+\lambda_4\pi^\mu,
\end{eqnarray}
\begin{eqnarray}\label{gmm5}
\frac{\delta S_{\kappa}}{\delta x^\mu}=0 \quad \Leftrightarrow \quad \nabla P_\mu= -R_{\mu\nu\alpha\beta}\dot
x^\nu\omega^\alpha\pi^\beta-\frac12\lambda_1\kappa\nabla_\mu
R_{\sigma\nu\alpha\beta}\omega^\sigma\pi^\nu\omega^\alpha\pi^\beta, \qquad
\end{eqnarray}
\begin{eqnarray}\label{gmm6}
\frac{\delta S_{\kappa}}{\delta\pi_\mu}=0 \quad \Leftrightarrow \quad \nabla\omega^\mu=
\lambda_1\pi^\mu-\lambda_1\kappa R^\mu{}_{\alpha\beta\nu}\omega^\alpha\omega^\beta\pi^\nu+\lambda_2\omega^\mu+\lambda_4
P^\mu, \qquad
\end{eqnarray}
\begin{eqnarray}\label{gmm7}
\frac{\delta S_{\kappa}}{\delta\omega^\mu}=0 \quad \Leftrightarrow \quad \nabla\pi_\mu=
-\frac{\lambda_1\alpha}{\omega^4}\omega_\mu-\lambda_1\kappa
R_{\mu\nu\alpha\beta}\pi^\nu\omega^\alpha\pi^\beta-\lambda_2\pi_\mu-\lambda_3 P_\mu. \qquad
\end{eqnarray}
Eq. (\ref{gmm4}) has been repeatedly used to obtain the final form of equations (\ref{gmm5})-(\ref{gmm7}).  Computing
time-derivative of the algebraic equations (\ref{gmm3}) and using (\ref{gmm4})-(\ref{gmm7}) we obtain the consequences
\begin{eqnarray}\label{gmm8}
\pi^2 - \frac{\alpha}{\omega^2}=0,
\end{eqnarray}
\begin{eqnarray}\label{gmm9}
\lambda_3 = 4a\lambda_1\left[2(1-\kappa)R_{\alpha\beta\mu\nu}\omega^\alpha \pi^\beta \pi^\mu P^\nu
+\kappa\pi^\sigma(\nabla_\sigma R_{\mu\nu\alpha\beta}) \omega^\mu\pi^\nu\omega^\alpha\pi^\beta \right]   \, , \cr
\lambda_4 =-4a\lambda_1\left[2(1-\kappa)R_{\alpha\beta\mu\nu} \omega^\alpha \pi^\beta \omega^\mu P^\nu
+\kappa\omega^\sigma(\nabla_\sigma R_{\mu\nu\alpha\beta}) \omega^\mu\pi^\nu\omega^\alpha\pi^\beta \right] \, ,
\end{eqnarray}
Here and below we use the following notation. The gravitational analogy of electromagnetic field strength is denoted
\begin{equation}\label{ga3}
\theta_{\mu\nu}=R_{\mu\nu\alpha\beta}S^{\alpha\beta}.
\end{equation}
In the equation which relates velocity and momentum will appear the
matrix
\begin{equation}\label{gmm11}
T^{\alpha}{}_{\nu} \equiv \delta^\alpha{}_\nu -(\kappa-1)aS^{\alpha \sigma}\theta_{\sigma\nu} \, , \qquad
a=\frac{2}{16m^2c^2+(\kappa+1)(S\theta)}\,.
\end{equation}
Using the identity
\begin{eqnarray}\label{Id0.0}
(S\theta S)^{\mu\nu}=-\frac{1}{2} (S\theta)S^{\mu\nu}, \quad \mbox{where} \quad
S\theta=S^{\alpha\beta}\theta_{\alpha\beta},
\end{eqnarray}
we find inverse of the matrix $T$
\begin{eqnarray}\label{gmm14}
\tilde T^{\alpha}{}_{\nu} \equiv \delta^\alpha{}_\nu+(\kappa-1)bS^{\alpha \sigma}\theta_{\sigma\nu} \, ,
\qquad b=\frac{1}{8m^2c^2+\kappa(S\theta)} \,.
\end{eqnarray}
The vector $Z^\mu$ is defined by
\begin{eqnarray}\label{gmm12}
Z^\mu=\frac{b}{8c}S^{\mu\sigma}(\nabla_\sigma R_{\alpha\beta\rho\delta})S^{\alpha\beta}S^{\rho\delta}\equiv
\frac{b}{8c}S^{\mu\sigma}\nabla_\sigma(S\theta).
\end{eqnarray}
This vanishes in a space with homogeneous curvature, $\nabla R=0$.

The time-derivatives of (\ref{gmm2}), (\ref{gmm8}) and (\ref{gmm9}) do not yield new algebraic equations. Due to
(\ref{gmm8}) we can replace the constraint  (\ref{gmm2}) on $P^2+\kappa R_{\alpha\beta\mu\nu}\omega^\alpha\pi^\beta
\omega^\mu \pi^\nu + (mc)^2=0$. The obtained expressions for $\lambda_3$ and $\lambda_4$ can be used to exclude these
variables from the equations (\ref{gmm4})-(\ref{gmm7}). The constraints $T_1$, $T_2$ and $T_5$ form the first-class
subset, while $T_3$ and $T_4$ represent a pair of second class.

Neither constraints nor equations of motion do not determine the functions $\lambda_1$ and $\lambda_2$, that is the
non-minimal interaction preserves both reparametrization and spin-plane symmetries of the theory.  The presence of
$\lambda_1$ and $\lambda_2$ in the equations (\ref{gmm6}) and (\ref{gmm7}) implies that evolution of the basic
variables is ambiguous, so they are not observable.  To find the candidates for observables, we note once again that
(\ref{gmm6}) and (\ref{gmm7}) imply an equation for $S^{\mu\nu}$ which does not contain $\lambda_2$. So we rewrite
(\ref{gmm4}) and (\ref{gmm5}) in terms of spin-tensor and add to them the equation for $S^{\mu\nu}$, this gives the
system
\begin{eqnarray}
\dot x^\mu &=&\lambda_1\left[T^{\mu}{}_{\nu} P^\nu +\kappa \frac{ac}{b}Z^\mu\right] \, , \label{gmx-2}\\
\nabla P_\mu &=&- \frac{1}{4}\theta_{\mu\nu}\dot x^\nu-\frac{\lambda_1\kappa}{32}\nabla_\mu(S\theta)\, , \label{gmP-2} \\
\nabla S^{\mu\nu} &=&-\frac{\kappa\lambda_1}{4}(\theta S)^{[\mu\nu]}+2P^{[\mu}\dot x^{\nu]}   \, . \label{gmJ.3-2}
\end{eqnarray}
Besides, the constraints (\ref{gmm2}), (\ref{gmm3}) and (\ref{gmm8}) imply
\begin{eqnarray}
P^2+\frac{\kappa}{16}\theta S + (mc)^2=0, \label{gmm13} \\
S^{\mu\nu}P_\nu=0, \qquad S^{\mu\nu}S_{\mu\nu}=8\alpha \,.  \label{gmm13.1}
\end{eqnarray}
The equations (\ref{gmm13.1}) imply that only two components of spin-tensor are independent, as it should be for spin
one-half particle. Eq. (\ref{gmJ.3-2}), contrary to the equations for $\omega$ and $\pi$, does not depend on
$\lambda_2$. This proves that the spin-tensor is invariant under local spin-plane symmetry. The remaining ambiguity due
to  $\lambda_1$ is related with reparametrization invariance and disappears when we work with physical dynamical
variables $x^i(t)$. Equations (\ref{gmx-2})-(\ref{gmJ.3-2}), together with (\ref{gmm13}) and (\ref{gmm13.1}), form  a
closed system which determines evolution of a spinning particle with gravimagnetic moment.

The Hamiltonian equations can be equally obtained computing $\dot z=\{z , H\}$, where $z=(x, p, \omega, \pi)$, with the
Hamiltonian given in square brackets of Eq. (\ref{gmm1}).  Our original variables fulfill the usual Poisson brackets
$\{x^\mu ,  p_\nu\}=\delta^\mu_{\nu}$ and $\{\omega^\mu , \pi_\nu\}=\delta^\mu_\nu$, then $\quad \{P_\mu , P_\nu \} =
R^\sigma_{\ \lambda \mu\nu} \pi_\sigma \omega^\lambda$, $\{P_\mu , \omega^\nu \}=\Gamma^\nu_{\mu\alpha}\omega^\alpha$,
$\{ P_\mu , \pi_\nu \}=- \Gamma^\alpha_{\mu\nu} \pi_\alpha$. For the quantities $x^\mu$, $P^\mu$ and $S^{\mu\nu}$ these
brackets imply
\begin{equation}\label{br}
\{ x^\mu , P_\nu \} =\delta^\mu_\nu,  \quad \{ P_\mu , P_\nu\}=-\frac14 R_{\mu\nu\alpha\beta}S^{\alpha\beta}, \quad \{
P_\mu, S^{\alpha\beta} \}=\Gamma^{\alpha}_{\mu\sigma}S^{\sigma\beta}-\Gamma^\beta_{\mu\sigma}S^{\sigma\alpha}.
\end{equation}
\begin{eqnarray}\label{br1}
\{S^{\mu\nu},S^{\alpha\beta}\}= 2(g^{\mu\alpha} S^{\nu\beta}-g^{\mu\beta} S^{\nu\alpha}-g^{\nu\alpha} S^{\mu\beta}
+g^{\nu\beta} S^{\mu\alpha}).
\end{eqnarray}
We can simplify the Hamiltonian introducing the Dirac bracket constructed with help of second-class constraints
\begin{equation}\label{DB}
\{A , B \}_D = \{A , B\} -\frac{1}{8a} \left[ \{ A , T_3 \} \{T_4 , B\} - \{ A , T_4\}\{ T_3 , B\} \right].
\end{equation}
Since the Dirac bracket of a second-class constraint with any quantity vanishes, we can now omit $T_3$ and $T_4$ from
the Hamiltonian. The quantities $x^\mu$, $P^\mu$ and $S^{\mu\nu}$, being invariant under spin-plane symmetry, have
vanishing brackets with the first-class constraints $T_2$ and $T_5$. So, obtaining equations for these quantities, we
can omit the last two terms in (\ref{gmm1}), arriving at the relativistic Hamiltonian
\begin{eqnarray}\label{Hamiltonian.0.02}
H_1= \frac{\lambda_1}{2} \left( P^2 +\frac{\kappa}{16}(\theta S) + m^2c^2\right).
\end{eqnarray}
The equations (\ref{gmx-2})-(\ref{gmJ.3-2}) can be obtained according the rule $\dot z=\{z, H_1\}_D$.

We have obtained a rather simple expression for the Hamiltonian because of the most part of spin-gravity interaction is
encoded now in the Dirac brackets. The expression (\ref{Hamiltonian.0.02}) together with the Dirac brackets could be an
alternative starting point for computation of post-Newton corrections due to spin \cite{Kunst14, Kunst16}.

{\bf Lagrangian equations.} Let us exclude $P^\mu$ and $\lambda_1$ from the equations (\ref{gmP-2}) and
(\ref{gmJ.3-2}). Using (\ref{gmm14}) we solve (\ref{gmx-2}) with respect to $P^\mu$. Using the resulting expression in
the constraint (\ref{gmm13}) we obtain $\lambda_1$
\begin{eqnarray} \label{gmm15}
\lambda_1=\frac{\sqrt{-\dot x G\dot x}}{m_r c}, \quad \mbox{with} \quad  m_r^2  \equiv m^2 +
\frac{\kappa}{16 c^2} (S\theta) -\kappa^2 Z^2,
\end{eqnarray}
where the effective metric now is  given by
\begin{eqnarray}\label{gmm16}
G_{\mu\nu} = \tilde T^\alpha{}_{\mu} g_{\alpha\beta} \tilde T^\beta{}_{\nu}.
\end{eqnarray}
Then the expression for momentum in terms of velocity implied by (\ref{gmx-2})  is
\begin{eqnarray}\label{gmm17}
P^\mu = \frac{m_r c}{\sqrt{-\dot x G \dot x}} \tilde T^{\mu}{}_{\nu} \dot x^\nu-\kappa c Z^\mu.
\end{eqnarray}
We substitute this $P^\mu$ into (\ref{gmP-2}),  (\ref{gmJ.3-2})
\begin{eqnarray}
\nabla\left[ \frac{m_r }{\sqrt{-\dot x G \dot x}} \tilde T^\mu{}_\nu \dot x^\nu \right] =
-\frac{1}{4c} \theta^\mu{}_\nu\dot x^\nu  -  \kappa\frac{\sqrt{-\dot x G\dot x}}{32m_r c^2}
\nabla^\mu(S\theta)+\kappa\nabla Z^\mu, \qquad \label{gml1} \\
\nabla S^{\mu\nu} =- \frac{\kappa\sqrt{-\dot x G\dot x}}{4m_r c }(\theta S)^{[\mu\nu]} -\frac{2m_r c
(\kappa-1)b}{\sqrt{-\dot x
G\dot x}}\dot x^{[\mu} (S\theta\dot x)^{\nu]}  +2\kappa c\dot x^{[\mu}Z^{\nu]}. \qquad \label{gml2}
\end{eqnarray}
Together with (\ref{gmm13.1}), this gives us the Lagrangian equations for the spinning particle with gravimagnetic
moment.

Comparing our equations to those of spinning particle on electromagnetic background (\ref{FF.6})-(\ref{FF.8}), we see
that the two systems have the same structure after the identification $\kappa\sim\mu$ and  $\theta_{\mu\nu}\equiv
R_{\mu\nu\alpha\beta}S^{\alpha\beta}\sim F_{\mu\nu}$, where $\mu$ is the magnetic moment. That is a curvature
influences trajectory of a spinning particle in the same way as an electromagnetic field with the strength
$\theta_{\mu\nu}$.

\subsection{MPTD particle as the spinning particle without gravimagnetic moment}
\label{ch09:sec9.9.5}
Let us compare MPTD equations (\ref{r003})-(\ref{r005}) with equations of our
spinning particle. Imposing $\kappa=0$ in Eqs. (\ref{gmx-2})-(\ref{gmm13.1}), we write them in the form
\begin{eqnarray}
P^\mu =\frac{mc}{\sqrt{-\dot x G \dot x}}(\tilde T\dot x)^\mu, \qquad \nabla P^\mu =-\frac 14 (\theta\dot x)^\mu,
\cr \nabla S^{\mu\nu} = 2P^{[\mu} \dot x^{\nu]}, \qquad
S^{\mu\nu}P_\nu  =0, \qquad \label{m003} \\
P^2+(mc)^2=0, \qquad  \qquad \qquad\label{m004} \\
S^2=8\alpha, \qquad \qquad \qquad \qquad  \label{m005}
\end{eqnarray}
with $\tilde T$ from Eq. (\ref{gmm14}) with $\kappa=0$. Comparing the systems, we see that our spinning particle has
fixed values of square of spin and canonical momentum, while for MPTD-particle these quantities represent constants of
motion. We conclude that all the trajectories of a body with given $m$ and $S^2=\beta$ are described by our spinning
particle with spin $\alpha=\frac{\beta}{8}$ and with the mass equal to $\sqrt{m^2+\frac{f^2(\beta)}{c^2}}$. In this
sense our spinning particle is equivalent to MPTD-particle. We point out that our final conclusion remains true even we
do not add the equation (\ref{r9.2}) to MPTD-equations: to study the class of trajectories of a body with
$\sqrt{-P^2}=k$ and $S^2=\beta$ we take our spinning particle with $m=\frac{k}{c}$ and $\alpha=\frac{\beta}{8}$.
MPTD-equations in the Lagrangian form (\ref{r9})-(\ref{r11}) can be compared with (\ref{gml1})-(\ref{gml2}).

Summing up, we demonstrated that MPTD-equations correspond to the minimal interaction of spinning particle with gravity.

\subsection{Consistency in ultra relativistic regime implies quantized gravimagnetic moment}
\label{ch09:sec9.12}

The paradoxical behavior of MPTD-particle originates from the fact that variation rate of spin (\ref{motionJ-5})
diverges in the ultra-relativistic limit, $\nabla S\sim\frac{1}{\sqrt{\dot xG\dot x}}$, and contributes into the
expression for acceleration (\ref{pe12}) through the tetrad field $\tilde T(S)$. Remarkably, for the non minimal
interaction with $\kappa=1$, the undesirable term in Eq. (\ref{gml2}) vanishes. Besides this implies $\tilde
T^\mu{}_\nu=\delta^\mu{}_\nu$, $ G_{\mu\nu}=g_{\mu\nu}$, and crucially simplifies the equations of
motion\footnote{Besides $S^{\mu\nu}P_\nu=0$, there are known others supplementary spin conditions \cite {Tulc,
Dixon1964, pirani:1956}. In this respect we point out that the MPTD theory implies this condition with certain $P_\nu$
written in Eq. (\ref{r003}). Introducing $\kappa$, we effectively changed $P_\nu$ and hence changed the supplementary
spin condition. For instance, when $\kappa=1$ and in the space with $\nabla R=0$, we have $P^\mu=\frac{\tilde
mc}{\sqrt{-\dot xg\dot x}}\dot x^\mu$ instead of (\ref{r003}).}. The Hamiltonian equations
(\ref{gmx-2})-(\ref{gmJ.3-2}) read
\begin{eqnarray}
\frac{m_r c}{\sqrt{-\dot x g \dot x}}\dot x^\mu =P^\mu +cZ^\mu, \label{gmm18} \\
\nabla P_\mu =- \frac{1}{4}\theta_{\mu\nu}\dot x^\nu-\frac{\sqrt{-\dot x g \dot x}}{32m_r c}\nabla_\mu(S\theta) \, ,  \label{gmm19} \\
\nabla S^{\mu\nu} = -\frac{\sqrt{-\dot x g \dot x}}{4m_r c}(\theta S)^{[\mu\nu]}+2P^{[\mu} \dot x^{\nu]},
\label{gmm20}
\end{eqnarray}
Particle with unit gravimagnetic moment and MPTD particle have a qualitatively different behavior at low velocities.
Indeed, keeping only the terms which may give a contribution in the leading post-Newton approximation,
$\sim\frac{1}{c^2}$, we obtain from (\ref{gmm19}), (\ref{gmm20}) the approximate equations
\begin{eqnarray}\label{xk.1}
\nabla P_\mu = -\frac 14 \theta_{\mu\nu}\dot x^\nu -\frac{\sqrt{-\dot x g \dot x}}{32m c} (\nabla_\mu
R_{\alpha\beta\sigma\lambda}) S^{\alpha\beta}S^{\sigma\lambda}  \, , \quad \nabla S^{\mu\nu} =-\frac{\sqrt{-\dot x g
\dot x}}{4m c}(\theta S)^{[\mu\nu]}  \,,
\end{eqnarray}
while MPTD equations in the same approximation read
\begin{eqnarray}\label{x.111}
\nabla P_\mu =-\frac{1}{4}\theta_{\mu\nu}\dot x^\nu \, , \qquad \nabla S^{\mu\nu} = 0 \, .
\end{eqnarray}
Lagrangian equations are composed now by the equation for trajectory
\begin{eqnarray}\label{acc-5}
\nabla \left[ \frac{m_r\dot x^\mu}{\sqrt{-\dot x g \dot x}} \right] = -\frac{1}{4c} \theta ^\mu{}_{\nu}\dot x^\nu -
\frac{\sqrt{-\dot x g \dot x}}{32m_r c^2}\nabla^\mu(S\theta) +\nabla Z^\mu,
\end{eqnarray}
and by the equation for precession of spin-tensor
\begin{equation}\label{motionS}
\nabla S^{\mu\nu} =-\frac{\sqrt{-\dot x g \dot x}}{4m_r c}(\theta S)^{[\mu\nu]}+2c\dot x^{[\mu}Z^{\nu]}.
\end{equation}
These equations can be compared with (\ref{byM13}) and (\ref{motionJ-5}).  In the modified theory:

\par \noindent 1. Time interval and distance should be unambiguously defined within the original space-time
metric $g_{\mu\nu}$. So the critical speed is equal to the speed of light.

\par \noindent 2. Covariant precession of spin (\ref{motionS}) has a smooth behavior, in particular, for
covariantly-constant curvature, $\nabla R=0$, we have $\nabla S\sim \sqrt{-\dot x g\dot x}$ contrary to $\nabla
S\sim\frac{1}{\sqrt{-\dot x g\dot x}}$ in the equation (\ref{motionJ-5}).

\par \noindent 3.  Spin ceases to affect the trajectory in
ultra-relativistic limit: the trajectory of spinning particle becomes more and more close to that of spinless particle
as $v\rightarrow c$. Besides, the spin precesses with finite angular velocity in this limit.

\par \noindent 4. The equation (\ref{acc-5}) in the space with covariantly-constant curvature has the structure similar to
(\ref{ch01:eqn1.278}), hence we expect that longitudinal acceleration vanishes as $v\rightarrow c$. Let us  confirm
this by direct computations.

To find the acceleration, we separate derivative of the radiation mass $m_r$ and write equation (\ref{acc-5}) in the form
\begin{eqnarray}\label{acc-6}
\frac{d}{d\tau}\left[ \frac{\dot x ^\mu}{\sqrt{-\dot x g \dot x}} \right] = \frac{f^\mu}{\sqrt{-\dot x g \dot x}},
\end{eqnarray}
where the force is
\begin{eqnarray}\label{force-6}
f^\mu \equiv -\Gamma^\mu_{\alpha\beta}\dot x^\alpha \dot x^\beta  -\frac{\sqrt{-\dot x g \dot x}}{4m_r c} \theta^\mu_{\
\nu} \dot x^\nu+ \frac{\dot xg\dot x}{32m_r^2c^2}\nabla^\mu(S\theta)+ \cr \frac{\sqrt{-\dot xg\dot x}}{m_r}\nabla Z^\mu-\dot
x^\mu\frac{\dot m_r}{m_r}. \qquad \qquad \quad
\end{eqnarray}
While this expression contains derivatives of spin due to $\dot m_r$-term, the resulting expression is non singular function of velocity
because  $\nabla S$ is a smooth function. Hence, contrary to Eq. (\ref{pe7}), the force now is non singular function of velocity.
We take $\tau=x^0$ in the spatial part of the system (\ref{acc-6}), this gives
\begin{equation}
\left(\frac{dt}{dx^0}\right)^{-1}\frac{d}{dx^0}\left[ \frac{v^i}{\sqrt{c^2-({\bf v}\gamma {\bf v})}}\right]
=\frac{f^i(v)}{\sqrt{c^2-({\bf v}\gamma {\bf v})}},
\end{equation}
where $f^i(v)$ is obtained from (\ref{force-6}) replacing $\dot x^\mu$ by $v^\mu$ of equation (\ref{La.5.1}).
This system is of the form (\ref{La.12}), so the acceleration is given by (\ref{La.11}) and (\ref{La.20})
\begin{eqnarray}\label{accc1}
a^i=\tilde M^i{}_j[f^j+\tilde\Gamma^j{}_{kl}(\gamma)v^kv^l]+\frac12\left(\frac{dt}{dx^0}\right)^{-1}\left[({\bf
v}\partial_0\gamma\gamma^{-1})^i-\frac{({\bf v}\partial_0\gamma{\bf v})}{c^2}v^i\right], \qquad
\end{eqnarray}
\begin{eqnarray}\label{accc2}
{\bf v}\gamma{\bf a}=\left(1-\frac{{\bf v}\gamma{\bf v}}{c^2}\right)\left[ ({\bf
v}\gamma)_i[f^i(v)+\tilde\Gamma^i{}_{kl}(\gamma)v^kv^l]+\frac12\left(\frac{dt}{dx^0}\right)^{-1}({\bf
v}\partial_0\gamma{\bf v})\right]. \qquad
\end{eqnarray}
With the smooth $f^i$ given in Eq. (\ref{force-6}),  the acceleration (\ref{accc1}) remains finite while the
longitudinal acceleration (\ref{accc2}) vanishes in the limit $v\rightarrow c$. Due to the identities (\ref{La.17.1}),
we have $({\bf v}\gamma)_if^i\stackrel{v\rightarrow c}{\longrightarrow}-({\bf v}\gamma)_i\Gamma^i{}_{\alpha\beta}\dot
x^\alpha \dot x^\beta$, that is  the trajectory tends to that of spinless particle in the limit.

In resume, contrary to MPTD-equations, the modified theory is consistent with respect to the original metric
$g_{\mu\nu}$. Hence the modified equations could be more promising for description of the rotating objects in
astrophysics.

\section{One-particle relativistic quantum mechanics with positive energies, canonical formalism}
\label{ch09:sec9.20}

As we have seen above, on the classical level our vector model adequately describes spinning particle in an arbitrary
gravitational and electromagnetic fields. Moreover, taking into account the leading relativistic corrections in
quantized theory with interaction, we have explained the famous one-half factor in non-relativistic Hamiltonian (\ref
{FF0.2.2}), see Sect.~\ref{ch09:sec9.7}. Now we turn to a systematic discussion of our model on the quantum level. In
this section we construct quantum mechanics of the free theory (\ref{FF.1}) in the physical-time parametrization. This
yields the Schr\"{o}dinger equation (\ref{pha.21.1}), with the Hamiltonian $c\sqrt{{\bf p}^2+(mc)^2}$ acting on a space
of two-component wave functions. Note that all the solutions have positive energy. The novel point is that the naive
expressions, $x^i$ and $\sigma^i$, do not represent operators of position and spin of our theory. This is due to the
second-class constraints $p\omega=p\pi=0$ of the relativistic theory, which guarantee the supplementary spin condition
$S^{\mu\nu}p_\nu=0$.  The constraints should be taken into account with help of Dirac bracket, this implies a
deformation of classical brackets which are subject to quantization. In the result, the position and spin of a spinning
particle are represented by the operators (\ref{pha.20.2}) and (\ref{spin.1}). The remaining sections are devoted to
establishing of Lorentz covariance of the obtained quantum mechanics.

In the free Lagrangian (\ref{FF.1}) it is
convenient to rescale $\omega\rightarrow\sqrt{\lambda}\omega$, then
\begin{eqnarray}\label{mq.1}
S=\int d\tau\frac{1}{4\lambda}\left[\dot xN\dot x+\lambda\dot\omega N\dot\omega-\sqrt{\left[\dot xN\dot x+\lambda\dot\omega
N\dot\omega\right]^2-4\lambda(\dot xN\dot\omega)^2}\right]- \cr \frac{\lambda}{2}m^2c^2+\frac{\alpha}{2\omega^2}.
\qquad \qquad \qquad \qquad \qquad \qquad
\end{eqnarray}
Repeating the computations made in Sect. \ref{ch09:sec9.6.1}, we arrive at the Hamiltonian action $\int d\tau ~ p\dot
x+\pi\dot\omega-H$ with the Hamiltonian
\begin{eqnarray}\label{mq.11}
H=\frac{\lambda}{2}\left(p^2+m^2c^2\right)+\frac12\left(\pi^2-\frac{\alpha}{\omega^2}\right)+ \cr
\lambda_2(\omega\pi)+ \lambda_3(p\omega)+\lambda_4(p\pi)+\lambda_0p_\lambda.
\end{eqnarray}
This can be compared with (\ref{m.11}). Recall that the constraint $\pi^2-\frac{\alpha}{\omega^2}=0$ arises as a
secondary constraint, from the condition of preservation in time of the primary constraint $\omega\pi=0$. The
Hamiltonian action provides both equations of motion and constraints of the vector model in an arbitrary
parametrization. Using the reparametrization invariance, we take physical time as the evolution parameter,
$\tau=\frac{x^0}{c}=t$, then the Hamiltonian action reads
\begin{eqnarray}\label{ch09:eqn9.20.1}
S_H=\int dt ~  cp_0+p_i\dot x^i+\pi_\mu\dot\omega^\mu-
\frac{\lambda}{2}\left(-p_0^2+p_i^2+m^2c^2\right)-\cr \frac12\left(\pi^2-\frac{\alpha}{\omega^2}\right)-
\lambda_iT_i, \qquad \qquad \qquad
\end{eqnarray}
We can treat the term associated with $\lambda$ as a kinematic constraint of the variational problem. According to
known prescription of classical mechanics, we solve the constraint,
\begin{eqnarray}\label{ch09:eqn9.20.2}
-p_0=p^0=\sqrt{{\bf p}^2+m^2c^2},
\end{eqnarray}
and substitute the result back into Eq. (\ref{ch09:eqn9.7.1}), this gives an equivalent form of the functional
\begin{eqnarray}\label{ch09:eqn9.20.3}
S_H=\int dt ~  p_i\dot x^i+\pi_\mu\dot\omega^\mu- \left[c\sqrt{{\bf p}^2+
m^2c^2}+\frac12\left(\pi^2-\frac{\alpha}{\omega^2}\right)+ \right. \cr \left.
\lambda_2\omega_\mu\pi^\mu+\lambda_3p_\mu\omega^\mu+\lambda_4p_\mu\pi^\mu\right], \qquad \qquad \quad \qquad
\end{eqnarray}
where the substitution (\ref{ch09:eqn9.20.2}) is implied in the last two terms as well. We have excluded the variables
$x^0$ and $p_0$, the remaining variables are $x^i(t), p_i(t), \omega^\mu(t)$ and $\pi^\mu(t)$. The expression in square
brackets represents the Hamiltonian. The sign in front of the square root in (\ref{ch09:eqn9.20.2}) was chosen
according to the right spinless limit. We have excluded non physical variables of the position sector and work now with
$x^i(t), p^i(t), \omega^\mu(t)$ and $\pi^\mu(t)$.

The variational problem implies the first-class constraints $T_2=\omega\pi=0$, $T_5=\pi^2-\frac{\alpha}{\omega^2}=0$
and the second-class constraints
\begin{eqnarray}\label{ch09:eqn9.7.4}
T_3=-p^0\omega^0+p^i\omega^i=0, \qquad
T_4=-p^0\pi^0+p^i\pi^i=0.
\end{eqnarray}
In all expressions below the symbol $p^0$ represents the function (\ref{ch09:eqn9.20.2}).

The action (\ref{ch09:eqn9.20.3}) implies the Hamiltonian equations
\begin{eqnarray}\label{pha.8}
\frac{dx^i}{dt}=c\frac{p^i}{p^0}, \qquad \frac{dp^i}{dt}=0,
\end{eqnarray}
\begin{eqnarray}\label{pha.10}
\dot\omega^\mu=\pi^\mu+\lambda_2\omega^\mu+\lambda_4p^\mu, \qquad
\dot\pi^\mu=-\frac{\omega^\mu}{\omega^4}-\lambda_2\pi^\mu-\lambda_3p^\mu.
\end{eqnarray}
Equations (\ref{pha.8}) describe free-moving particle with the speed less then speed of light
\begin{eqnarray}\label{pha.11}
x^i=x^i_0+v^it, \quad v^i=c\frac{p^i}{\sqrt{{\bf p}^2+(mc)^2}}, \quad p^i=\mbox{const}.
\end{eqnarray}
The spin-sector variables have ambiguous evolution, because a general solution to (\ref{pha.10}) depends on an arbitrary
function $\lambda_2$. So they do not represent the observable quantities. As candidates for the physical variables of
spin-sector, we can take either the Frenkel spin-tensor,
\begin{eqnarray}\label{pha.12}
\frac{dS^{\mu\nu}}{dt}=0, \quad S^{\mu\nu}p_\nu=0, \quad S^2=6\hbar^2,
\end{eqnarray}
or, equivalently, the Pauli-Lubanski vector (\ref{1.6})
\begin{eqnarray}\label{pha.14}
\frac{ds^\mu}{dt}=0, \quad s^\mu p_\nu=0, \quad s^2=\frac{3\hbar^2}{4}.
\end{eqnarray}
For the attempt to impose the first-class constraints on state-vectors, see \cite{Rempel2016}.

To take into account the second-class constraints $T_3$ and $T_4$, we construct the Dirac bracket (\ref{pht.13}). The
non vanishing Dirac brackets are
\begin{equation}\label{pha.15}
\{x^i,x^j\}_{D}=\frac{\epsilon^{ijk}s_k}{mcp^0}=\frac{S^{ij}}{2mcp^0}, \qquad \{x^i,p^j\}_{D}=\delta^{ij},
\end{equation}
\begin{equation}\label{cq3}
\{S^{\mu\nu}, S^{\alpha\beta}\}_D= 2\left(g^{\alpha \left[\mu\right.}
S^{\left.\nu\right]\beta}-g^{\beta\left[\mu\right.} S^{\left.\nu\right]\alpha}\right)\,,
\end{equation}
\begin{eqnarray}\label{xmJab-bmt-diracD}
\{x^\mu, S^{\alpha\beta}\}_{D}= \frac{1}{(mc)^2}(S^{\mu\left[\beta\right.}p^{\left.\alpha\right]}
-\frac{p^\mu}{p^0} S^{0\left[\beta\right.}p^{\left.\alpha\right]}) \,,
\end{eqnarray}
\begin{eqnarray}\label{pha.16}
\{s^i,s^j\}_D=\frac{p^0}{mc}\epsilon^{ijk}\left(s_k- \frac{({\bf{s}\,\bf{p}})p_k}{p_0^2}\right)\,,
\end{eqnarray}
\begin{eqnarray}\label{pha.17}
\{x^i,s^j\}_D=\left(s^i-\frac{({\bf{s}\, \bf{p}})p^i}{p_0^2}\right)\frac{p^j}{(mc)^2}\,,
\end{eqnarray}
where
\begin{eqnarray}\label{pha.7.1}
g^{\mu\nu}\equiv\eta^{\mu\nu}-\frac{p^{\mu}p^\nu}{p^2}.
\end{eqnarray}
After transition to the Dirac brackets, the
second-class constraints can be used as strong equalities. In particular, we can present $s^0$ in terms of independent
variables
\begin{eqnarray}\label{pha.18}
s^0=\frac{(\bf{s}\, \bf{p})}{\sqrt{{\bf p}^2+(mc)^2}},
\end{eqnarray}
and in the expression for Hamiltonian (\ref{ch09:eqn9.20.3}) we can omit the last two terms. Besides, we omit the
second and third terms, as they do not contribute into equations for observables. In the result, we obtain the physical
Hamiltonian
\begin{eqnarray}\label{pha.19}
H_{ph}=cp^0=c\sqrt{{\bf p}^2+(mc)^2}.
\end{eqnarray}
As it should be, the equations (\ref{pha.8}), (\ref{pha.12}) and (\ref{pha.14}) follow from physical Hamiltonian with
the use of Dirac bracket, $\dot z=\{ z, H_{ph}\}_D$, where $z=({\bf p}, {\bf x}, S^{\mu\nu}, s^\mu)$.

Both operators (except $\hat p_i$) and abstract state-vectors of the physical-time formalism we denote by capital
letters, $\hat Z$, $\Psi(t, {\bf x})$. In order to quantize the model, classical Dirac-bracket algebra should be
realized by operators, $[\hat Z_i , \hat Z_j]=i\hbar \left.\{ z_i , z_j \}_D\right|_{z_i\rightarrow\hat Z_i}$. To find
the quantum realization, we first look for classical variables which have canonical Dirac brackets, thus simplifying
the quantization procedure. We introduce \cite{DPM2} the variables $\tilde x^j$, $\tilde p^j=p^j$ and $\tilde{s}_j$ as follows
\begin{eqnarray}\label{pha.18.1}
\tilde{x}^j=x^j-\frac{1}{mc(p^0+mc)}\epsilon^{jkm}s_kp_m, \quad
\tilde{s}_j=\left(\delta_{jk}-\frac{p_jp_k}{p^0(p^0+mc)}\right)s^k, \quad
\end{eqnarray}
then the inverse transformation is
\begin{eqnarray}\label{pha.20}
x^j=\tilde{x}^j+\frac{1}{mc(p^0+mc)}\epsilon^{jkm}\tilde{s}_kp_m, \quad
s_j=\left(\delta_{jk}+\frac{p_jp_k}{mc(p^0+mc)}\right)\tilde{s}^k. \quad
\end{eqnarray}
Note that in the expression for $x$ and $\tilde x$ we can replace $\tilde s\leftrightarrow s$. We point out that the
original and new variables obey the same equations of motion (\ref{pha.8}) and (\ref{pha.14}), so they are
indistinguishable in the free theory. In an interacting theory their dynamics will be different.

The new variables have a canonical algebra with respect to Dirac brackets
\begin{eqnarray}\label{pha.20.1}
\{\tilde{x}^j,\tilde{x}^i\}_D=0, \quad \{\tilde{x}^i,p^j\}_D=\delta^{ij},\quad \{\tilde{x}^j,\tilde{s}^i\}_D=0, \quad
\{\tilde{s}^i,\tilde{s}^j\}_D=\epsilon^{ijk}\tilde{s}_k. \qquad
\end{eqnarray}
Besides, the constraints (\ref{pha.14}) on $s^\mu$ imply ${\bf\tilde{s}}^2=\frac{3}{4}\hbar^2$. So the corresponding operators
$\hat{\tilde{S}}^j$ should realize an irreducible representation of $SO(3)$ with spin $s=1/2$. Quantization in terms of
these variables becomes straightforward. The Hilbert space consists from two-component functions $\Psi_a(t, {\bf x})$,
$a=1, 2$. A realization of the Dirac brackets by operators has the standard form
\begin{eqnarray}\label{pha.20.1}
p_j\to \hat{p}_j=-i\hbar \partial_j\,,\qquad \tilde x^j\to\hat{\tilde{X}}^j=x^j\,, \qquad \tilde{s}^j\to
\hat{\tilde{S}}^j=\frac{\hbar}{2}\sigma^j.
\end{eqnarray}
The conversion formulas (\ref{pha.20}) between canonical and original variables have no ordering ambiguities, so we
immediately obtain the operators $\hat{X}^j$ and $\hat{S}_{PL}^j$ corresponding to position and Pauli-Lubanski vector
of classical theory
\begin{equation}\label{pha.20.2}
x^i ~  \rightarrow ~ \hat{X}^{i}=x^i+\frac{\hbar}{2mc(\hat p^0+mc)}\epsilon^{ijk}\sigma_j\hat{p}_k,
\end{equation}
\begin{equation}\label{pha.20.3}
s^j ~ \rightarrow ~ \hat{S}_{PL}^j=\frac{\hbar}{2}\left(\sigma^j+\frac{1}{mc(\hat p^0+mc)}(\hat{\bf{p}}{\boldsymbol{\sigma}})\hat
p^j\right), \quad
\hat{S}_{PL}^0=\frac{\hbar}{2mc}({\hat{\bf{p}}\boldsymbol{\sigma}}),
\end{equation}
where the expression for $\hat{S}_{PL}^0$ follows from (\ref{pha.18}) and (\ref{pha.20}). Using equations (\ref{aa9})
and (\ref{f1.1}) relating the Pauli-Lubanski vector with Frenkel spin-tensor and three-vector of spin, we obtain their
quantum realization as follows:
\begin{equation}\label{pha.21}
\hat{S}^{0i}=-\frac{\hbar}{mc}\epsilon^{ijk}\hat p_j \sigma_k\,, \qquad
\hat S^{ij}=\frac{\hbar}{mc}\epsilon^{ijk}\left(\hat p^0\sigma_k-\frac{1}{(\hat
p^0+mc)}(\hat{\bf{p}}{\boldsymbol{\sigma}})\hat p_k\right) \,,
\end{equation}
\begin{equation}\label{spin.1}
\hat S^i=\frac{1}{4}\epsilon^{ijk}\hat S_{jk}=\frac{\hbar}{2mc}\left(\hat p^0\sigma^i-\frac{1}{(\hat
p^0+mc)}(\hat{\bf{p}}{\boldsymbol{\sigma}})\hat p^i\right).
\end{equation}
The energy operator (\ref{pha.19})
determines the evolution of a state-vector according to the Schr\"{o}dinger equation
\begin{equation}\label{pha.21.1}
i\hbar\frac{d\Psi}{dt}=c\sqrt{\hat{{\bf p}}^{\,2}+(mc)^2}\Psi,
\end{equation}
as well the evolution of operators by Heisenberg equations. The scalar product we define as follows
\begin{equation}\label{pha21.2}
\langle \Psi,\Phi \rangle=\int d^3x\Psi^\dagger \Phi,
\end{equation}
then
\begin{equation}\label{pha21.3}
P=\Psi^\dagger \Psi,
\end{equation}
is a probability density for $\tilde x^i$. We emphasize that an abstract vector $\Psi(t, {\bf x})$ of Hilbert space
represents an amplitude of probability density of canonical coordinate $\tilde x^i$. The wave function for the original
coordinate $x^i$ should be constructed according to known rules of quantum mechanics.

Let us introduce the operator $\hat p^0=-i\hbar\frac{d}{dx^0}$. Then the Schr\"{o}dinger equation equation reads $(\hat
p^0+\sqrt{\hat{{\bf p}}^{\,2}+(mc)^2})\Psi=0$, and applying the operator $\hat p^0-\sqrt{\hat{{\bf p}}^{\,2}+(mc)^2}$
we obtain Klein-Gordon equation
\begin{equation}\label{pha.21.4}
(\hat p^2+m^2c^2)\Psi=0, \qquad \hat p^2=\hat p^\mu\hat p_\mu.
\end{equation}
Hence all solutions to the Schr\"{o}dinger equation form the subspace of positive-energy solutions to the
manifestly-covariant Klein-Gordon equation for a two-component wave functions.

\begin{center}
\begin{table}
\caption{Position/spin operators for the relativistic electron \cite{pryce1948mass}.} \qquad \qquad \qquad \qquad $\beta=\left(
\begin{array}{cc}
1& 0\\
0& -1
\end{array}
\right)$, $\alpha^i=\left(
\begin{array}{cc}
0& \sigma^i\\
\sigma^i& 0
\end{array}
\right)$,  $\Sigma^i=\left(
\begin{array}{cc}
\sigma^i& 0\\
0& \sigma^i
\end{array}
\right)$. \label{tabular:Pryce-cm} \vspace{2mm}
\begin{center}
\begin{tabular}{c|c}
{}  & Dirac representation, $i\hbar\partial_t\Psi_D=c(\alpha^ip_i+mc\beta)\Psi_D$  \\
\hline \hline $\hat {X}_{P(d)}^j$           &
$x^j+\frac{i\hbar}{2mc}\beta\left(\alpha^j-\frac{\alpha^k\hat{p}_k\hat{p}^j}{(\hat p^0)^2}\right)$
\\
$\hat{S}_{P(d)}^{j}$       &  $\frac{1}{2m^2c^2}\left(m^2c^2 \Sigma^j -imc\beta \epsilon^{jkl}\alpha_k\hat{p}_{l}
 \right)$
\\
\hline $\hat{X}_{P(e)}^j=\hat{x}_{FW}^j$  & $x^j+\frac{\hbar}{2\hat p^0}\left(i \beta\alpha^j +\frac{1}{\hat
p^0+mc}\epsilon^{jkm}\hat{p}_k \Sigma_m -\frac{1}{\hat p^0(\hat p^0+mc)}i\beta \alpha^k\hat{p}_k\hat{p}^j\right) $
\\
$\hat{S}_{P(e)}^{j}=\hat{S}_{FW}^{j}$       & $\frac{\hbar}{2\hat p^0}\left(mc \Sigma^j -im\beta
\epsilon^{jkl}\alpha_k\hat{p}_{l}+\frac{\Sigma^k\hat{p}_k\hat{p}^j}{\hat p^0+mc}
 \right)$
\\
\hline $\hat X_{P(c)}^j$           & $x^j+\frac{\hbar}{2(\hat p^0)^2}\left(\epsilon^{jkm}\hat{p}_k \Sigma_m +i mc
\beta\alpha^j \right)$
\\
$\hat{S}_{P(c)}^{j}$       & $\frac{\hbar}{2(\hat p^0)^2}\left(m^2c^2 \Sigma^j -imc\beta
\epsilon^{jkl}\alpha_k\hat{p}_{l}+\Sigma^k\hat{p}_k\hat{p}^j
 \right)$
\end{tabular}
\end{center}
\end{table}
\end{center}
\begin{center}
\begin{table}
\caption{Position/spin operators for the relativistic electron \cite{pryce1948mass}.}
\qquad \qquad \qquad \qquad \qquad \qquad $\beta=\left(
\begin{array}{cc}
1& 0\\
0& -1
\end{array}
\right)$,  $\Sigma^i=\left(
\begin{array}{cc}
\sigma^i& 0\\
0& \sigma^i
\end{array}
\right)$. \label{tabular:Pryce-cm1} \vspace{2mm}
\begin{center}
\begin{tabular}{c|c|c}
{}     & F-W representation, $i\hbar\partial_t\Psi=c\beta\hat p^0\Psi$
&Vector model \\
\hline \hline $\hat {X}_{P(d)}^j$              &
$x^j-\frac{\hbar}{2mc(\hat p^0+mc)}\epsilon^{jkm}\hat{p}_k \Sigma_m$ &position $x^j\rightarrow\hat X^j$, (\ref{pha.20.2})
\\
$\hat{S}_{P(d)}^{j}$        & $\frac{\hbar}{2mc}\beta\left(\hat p^0 \Sigma^j-\frac{1}{(\hat p^0+mc)}\hat{p}^k\Sigma_k \hat p^j\right)$ &Frenkel spin $S^j\rightarrow\hat S^j$, (\ref{spin.1})
\\
\hline $\hat{X}_{P(e)}^j=\hat{x}_{FW}^j$   &
$x^j$ &$\tilde x^j\rightarrow\hat{\tilde X}^j$
\\
$\hat{S}_{P(e)}^{j}=\hat{S}_{FW}^{j}$       & $\frac{\hbar}{2}\Sigma^j $ &$\tilde s^j\rightarrow\hat{\tilde S}^j$
\\
\hline $\hat X_{P(c)}^j$              & $x^j+\frac{\hbar}{2\hat p^0(\hat p^0+mc)}\epsilon^{jkm}\hat{p}_k \Sigma_m$
\\
$\hat{S}_{P(c)}^{j}$       &
$\frac{\hbar}{2\hat p^0}\beta\left(mc\Sigma^j+\frac{1}{(\hat p^0+mc)}\hat{p}^k\Sigma_k \hat p^j\right)$ &$s^j\rightarrow\hat S_{PL}^j$, (\ref{pha.20.3})
\\
\end{tabular}
\end{center}
\end{table}
\end{center}

Let us compare our operators with known in the literature. Pryce\index{Operators! of Pryce} \cite{pryce1948mass}
studied possible candidates for observables of the Dirac equation, they are marked as $P(d)$, $P(e)$ and $P(c)$ in the
Tables \ref{tabular:Pryce-cm} and \ref{tabular:Pryce-cm1}. He wrote his operators acting on space of Dirac spinor
$\Psi_D$, see Table \ref{tabular:Pryce-cm}. Foldy and Wouthuysen \cite{foldy:1978} found unitary transformation which
maps the Dirac equation $i\hbar\partial_t\Psi_D=c(\alpha^ip_i+mc\beta)\Psi_D$ into the pair of square-root equations
$i\hbar\partial_t\Psi=c\beta\hat p^0\Psi$. Applying the FW transformation, the Pryce operators acquire block-diagonal
form on space $\Psi$, see Table \ref{tabular:Pryce-cm1}. Our operators act on space of solutions of square-root
equation (\ref{pha.21.1}), so we compared them with positive-energy parts (upper-left blocks) of Pryce operators in the
Table \ref{tabular:Pryce-cm1}.

Our operators of canonical variables $\hat{\tilde{X}}^j=x^j$ and $\hat{\tilde{S}}^j$ correspond to the Pryce (e)
($\sim$ Foldy-Wouthuysen $\sim$ Newton-Wigner \cite{newton1949localized}) position and spin operators.

However, operators of position $x^j$ and spin $S^j$ of our model are $\hat{X}^j$ and $\hat S^j$. They correspond to the
Pryce (d)-operators.

Operator of Pauli-Lubanski vector $\hat{S}_{PL}^j$ is the Pryce (c) spin (we remind the normalization of our $s^j$, see (\ref{1.6})).

\section{Relativistic covariance of canonical formalism}
\label{ch09:sec9.21} While we have started from the relativistic theory (\ref{mq.1}), working in the physical-time
parametrization we have lost, from the beginning, the manifest relativistic covariance. Is the quantum mechanics thus
obtained is a relativistic theory? Are the scalar product (\ref{pha21.2}) and probability (\ref{pha21.3}) the
Lorentz-invariant quantities? Are the mean values $\langle \Psi,\hat X^i\Phi \rangle$ and  $\langle \Psi,\hat S^i\Phi
\rangle$ the Lorentz-covariant quantities? To answer these questions, we follow the standard ideology of quantum
theory.

First, we associate with our theory the manifestly covariant Hilbert space of states $\psi$ which carries
 a representation of Poincaré group and admits conserved four-vector $\partial_\mu I^\mu(\psi)=0$ with positive
null-component. We define the invariant integral over space-like surface $\Omega$
\begin{eqnarray}\label{0003}
P_{\Omega}=-\frac 16\int_{\Omega} \epsilon_{\mu\alpha\beta\gamma}J^\mu dx^\alpha dx^\beta dx^\gamma, \quad \Rightarrow
\quad \int_{\Omega_1}=\int_{\Omega_2}, \quad \mbox{then} \quad P_{t=const}=\int J^0 d^3x
\end{eqnarray}
can be identified with probability.

Second, using the covariant formulation (\ref{mq.11}) of the classical theory, we find quantum realization of basic
variables by means of covariant operators acting in this space. The resulting construction is called a covariant
formalism.

Third, we establish a correspondence between the canonical and covariant pictures which respect the scalar products
(\ref{pha21.2}) and (\ref{0003}), and show that the scalar products, mean values and transition amplitudes of canonical
formalism can be computed using the covariant formalism. This proves the relativistic covariance of quantum mechanics
constructed in Sect.~\ref{ch09:sec9.20}.

As we saw above, state-vectors of spinning particle belong to space of solutions to the covariant two-component
Klein-Gordon equation. So it is natural to construct the covariant formalism on this base. We do this in the next
subsection. The covariant formalism based on the space of solutions to the Dirac equation will be discussed in
Sect.~\ref{ch09:sec9.23}.

We emphasize that quantum mechanics of Sect.~\ref{ch09:sec9.20} already has a clear physical interpretation: the state
vector $\Psi$ describes a spinning particle with positive energy in $\tilde x$\,-representation, the operator $\hat X$
represents a position, $\hat S$ represents a spin and so on. Therefore there is no need to search physical
interpretation of the covariant formalisms (negative energy states of the Dirac equation and so on), and we will not do
it. We consider the covariant formalisms as an auxiliary construction that has the only aim to prove the relativistic
covariance of the quantum mechanics formulated in Sect.~\ref{ch09:sec9.20}.

\subsection{Relativistic quantum mechanics of two-component Klein-Gordon equation}
\label{ch09:sec9.22}
We denote states and operators of covariant formalism by small letters, to distinguish them from the quantities of
canonical formalism.

According to Wigner \cite{wigner1939unitary, Bargmann1948, bib50}, with an elementary particle in quantum-field theory
we associate the Hilbert space of representation of Poincare group. The space can be described in a manifestly
covariant form as a space of solutions to Klein-Gordon equation for properly chosen multicomponent field
$\psi_a(x^\mu)$. The space of one-component fields corresponds to spin-zero particle. It is well-known, that this space
has no quantum-mechanical interpretation. In contrast, two-component Klein-Gordon equation does admit the probabilistic
interpretation. As we show below, the four-vector (\ref{cq17}) represents positively defined conserved current of this
equation. Using the current, we can define an invariant scalar product and hence the covariant rules for computing mean
values of covariant operators defined on  this space.

The two-component KG equation has been considered by Feynman and Gell-Mann \cite{feynman1958fermi-interaction} to
describe weak interaction of spin one-half particle.

Using the Pauli matrices (\ref{ch08:eqn8.157}) we form the two sets
\begin{eqnarray}\label{cq8}
\sigma^\mu=({\bf 1}, \sigma^i), \qquad \bar\sigma^\mu=(-{\bf 1}, \sigma^i).
\end{eqnarray}
All the matrices are hermitian and obey the following rules of permutation of indexes:
\begin{eqnarray}\label{cq9}
\sigma^\mu\bar\sigma^\nu=-\sigma^\nu\bar\sigma^\mu+2\eta^{\mu\nu}, \qquad
\bar\sigma^\mu\sigma^\nu=-\bar\sigma^\nu\sigma^\mu+2\eta^{\mu\nu}.
\end{eqnarray}
Further we define two more sets of $2\times 2$\,-matrices
\begin{eqnarray}\label{cq9.0}
\sigma^{\mu\nu}=-\frac{i}{2}(\sigma^{\mu}\bar\sigma^{\nu}-\sigma^{\nu}\bar\sigma^{\mu})=(0, -i\sigma^i, i\sigma^i, \epsilon^{ijk}\sigma^k), \cr
\bar\sigma^{\mu\nu}=-\frac{i}{2}(\bar\sigma^{\mu}\sigma^{\nu}-\bar\sigma^{\nu}\sigma^{\mu})=(0, i\sigma^i, -i\sigma^i, \epsilon^{ijk}\sigma^k),
\end{eqnarray}
We have shown the explicit form of their components $00$, $0i$, $i0$ and $ij$. They are related by hermitian
conjugation, $\sigma^{\mu\nu\dagger}=\bar\sigma^{\mu\nu}$, and obey the identities
\begin{eqnarray}\label{cq9.01}
\bar\sigma^\alpha\sigma^{\mu\nu}=\bar\sigma^{\mu\nu}\bar\sigma^\alpha-2i\eta^{\alpha[\mu}\bar\sigma^{\nu]}, \quad
\sigma^\alpha\bar\sigma^{\mu\nu}=\sigma^{\mu\nu}\sigma^\alpha-2i\eta^{\alpha[\mu}\sigma^{\nu]}.
\end{eqnarray}
Using them, we verify that both $\sigma^{\mu\nu}$ and $\bar\sigma^{\mu\nu}$ obey $SO(1,3)$\,-algebra, for instance
\begin{eqnarray}\label{cq9.03}
[\sigma^{\mu\nu}, \sigma^{\alpha\beta}]=2i(\eta^{\mu\alpha}\sigma^{\nu\beta}-\eta^{\mu\beta}\sigma^{\nu\alpha}-
\eta^{\nu\alpha}\sigma^{\mu\beta}+\eta^{\nu\beta}\sigma^{\mu\alpha}),
\end{eqnarray}
so they can be taken as generators of linear representation of the Lorentz group on space of two-component complex
columns (Weyl spinors) $\psi=(\psi_1, \psi_2)$. Under an infinitesimal Lorentz transformation with the parameters
$\omega^\mu{}_\nu$, $\delta x^\mu=\omega^{\mu}{}_\nu x^\nu$, the column $\psi$ transforms as follows:
\begin{eqnarray}\label{cq9.04}
\delta\psi=\frac{i}{4}\omega_{\mu\nu}\sigma^{\mu\nu}\psi, \quad \mbox{then} \quad
\delta\psi^\dagger=-\frac{i}{4}\psi^\dagger\bar\sigma^{\mu\nu}\omega_{\mu\nu}.
\end{eqnarray}
Note that the contraction $\psi^\dagger_1\psi_1+\psi^\dagger_2\psi_2$ is not an invariant of the transformation. Using
(\ref{cq9.01}), we verify that the quantity $\phi^\dagger\bar\sigma^\mu\psi$ is a four-vector\footnote{Under the finite
transformations $x'^\mu=\Lambda^\mu{}_\nu(\omega) x^\nu$, $\Psi'=D(\omega)\Psi$ and $\Psi'^\dagger=\Psi^\dagger
D^\dagger(\omega)$, where $D=e^{\frac{i}{4}\omega_{\mu\nu}\sigma^{\mu\nu}}$, the $\bar\sigma^\mu$ is an invariant
tensor, that is $(D^\dagger\bar\sigma_\mu D)\Lambda^\mu{}_\nu=\bar\sigma_\nu$. For the proof, see \cite{ba1, Kyzya}.}.
If $v_\alpha$ is a vector, the combination $\chi=v_\alpha\bar\sigma^\alpha\psi$ transforms with help of
$\bar\sigma^{\mu\nu}$, $\delta\chi=\frac{i}{4}\omega_{\mu\nu}\bar\sigma^{\mu\nu}\chi$, then
$\delta\chi^\dagger=-\frac{i}{4}\chi^\dagger\sigma^{\mu\nu}\omega_{\mu\nu}$. So the quantity
$(v_\beta\bar\sigma^\beta\phi)^\dagger\sigma^\mu(v_\alpha\bar\sigma^\alpha\psi)$ turns out to be a vector.

Introducing the space of two-component complex functions $\psi(x^\mu)=(\psi_1, \psi_2)$, the generators of Poincar\'{e}
transformations in this space read
\begin{eqnarray}\label{cq9.1}
{j}^{\mu\nu}=\frac12(x^\mu\partial^\nu-x^\nu\partial^\mu)+ \frac{i}{4}\sigma^{\mu\nu}\,, \qquad \partial^\mu=\frac{\partial}{\partial x_\mu}.
\end{eqnarray}
On the Poincar\'{e}-invariant subspace selected by two-component KG equation
\begin{eqnarray}\label{cq16}
(\hat p^2+m^2c^2)\psi=0\,,
\end{eqnarray}
we define an invariant and positive-defined scalar product as follows. The four-vector
\begin{eqnarray}\label{cq17}
I^\mu[\psi,\phi]=\frac{1}{m^2c^2}(\bar\sigma\hat p\psi)^\dagger\sigma^\mu\bar\sigma\hat p\phi- \psi^\dagger\bar\sigma^\mu\phi\,,
\end{eqnarray}
represents a conserved current of Eq. (\ref{cq16}), that is $\partial_\mu I^{\mu}=0$, when $\psi$ and $\phi$ satisfy to
Eq. (\ref{cq16}).  We define the scalar product by means of invariant integral
\begin{equation}\label{cq-inv-scalar-product}
(\psi,\phi)=\int\limits_{\Omega} d\Omega_\mu I^\mu\,,\qquad
d\Omega_\mu=-\frac16\epsilon_{\mu\nu\alpha\beta}dx^\nu dx^\alpha dx^\beta,
\end{equation}
computed over a space-like three-surface $\Omega$. Using the Gauss theorem for the four-volume contained between the
surfaces $\Omega_1$ and $\Omega_2$, we conclude that the scalar product does not depend on the choice of the surface,
$\int_{\Omega_1}=\int_{\Omega_2}$. In particular, it does not depend on time.  So we can restrict ourselves to the
hyperplane defined by the equation $x^0=\mbox{const}$, then
\begin{equation}\label{sp1}
(\psi,\phi)=\int d^3x I^0.
\end{equation}
Besides, the scalar product is positive-defined, since
\begin{equation}\label{sp2}
I^0[\psi,\psi]=\frac{1}{m^2c^2}(\bar\sigma\hat p\psi)^\dagger\bar\sigma\hat p\psi+\psi^\dagger \psi >0.
\end{equation}
So, this can be considered as a probability density of operator $\hat{\bf x}={\bf x}$. We point out that transformation
properties of the column $\psi$ are in agreement with this scalar product: only if $\psi$ transforms as a (right) Weyl
spinor, the quantity $I^\mu$ represents a four-vector.

Now we can confirm relativistic invariance of scalar product (\ref{pha21.2}) of canonical formalism. We write
\begin{eqnarray}\label{sp3}
(\psi, \phi)=\int d^3x \frac{1}{m^2c^2}(\bar\sigma\hat p\psi)^\dagger\bar\sigma\hat p\phi+\psi^\dagger\bar\phi = \qquad \cr \int
d^3x \left[\left(\frac{1}{mc}\bar\sigma\hat p+i\right)\psi\right]^\dagger\left(\frac{1}{mc}\bar\sigma\hat
p+i\right)\phi=\langle W\psi, W\phi \rangle,
\end{eqnarray}
where the operator $W=\frac{1}{mc}\bar\sigma\hat p+i$ has an inverse, $W^{-1}=\frac{1}{2\hat p^0}\left(i\sigma\hat
p+mc\right)$. The operator $\frac{1}{\hat p^0}$ is well-defined in momentum representation, see the next section. Note
also that $W$ and $W^{-1}$ commutes with the Schr\"{o}dinger operator (\ref{pha.21.1}). Equation (\ref{sp3}) suggests
the map of canonical space $\{\Psi\}$ onto subspace of positive-energy solutions of covariant space $\{\psi\}$,
$W^{-1}: \{\Psi\}  \rightarrow  \{\psi\}, \psi=W^{-1}\Psi$.

The map respects the scalar products (\ref{pha21.2}) and (\ref{sp1}), and thus proves relativistic invariance
of the scalar product $\langle \Psi,\Phi \rangle$, $\langle \Psi,\Phi \rangle=\langle W\psi, W\phi \rangle=(\psi, \phi)$.

We note that map $W$ is determined up to an isometry, we can multiply $W$ from the left by an arbitrary unitary
operator $U$, $W\to W'=U W$, $U^\dagger U=1$. Here $\dagger$ denotes Hermitian conjugation with respect to scalar
product $\langle ~ ,~ \rangle$. It is convenient to remove the ambiguity \cite{conway1990course} by requiring the
Hermiticity of the operator. Positively defined operator $W^\dagger W>0$ has a unique square root, $V=(W^\dagger
W)^{1/2}$. We write identically $W=PV$, where $P=WV^{-1}$ is unitary, so we can omit $P$ and use $V$ instead of $W$. We
compute $W^\dagger W=\frac{2\hat p^0}{(mc)^2}\bar\sigma\hat p$, then
\begin{equation}\label{sp6}
V=\frac{1}{mc}\sqrt{\frac{\hat p^0}{\hat p^0+mc}}[(\bar \sigma \hat p)+mc], \quad  V^{-1}=\frac{1}{2\sqrt{\hat p^0(\hat
p^0+mc)}}[mc-\sigma\hat p],
\end{equation}
and final form of the map between canonical and covariant spaces is
\begin{eqnarray}\label{sp4}
V^{-1}: \{\Psi\}\rightarrow\{\psi\}, \quad  \psi=V^{-1}\Psi, ~ \mbox{then} ~ \langle \Psi,\Phi \rangle=\langle V\psi,
V\phi \rangle=(\psi, \phi). \quad
\end{eqnarray}
The transformation between state-vectors induces the map of operators
\begin{equation}\label{cor4}
V^{-1}\hat QV=\hat q,
\end{equation}
acting in spaces $\Psi$ and $\psi$. In the next section we use the covariant formulation of the vector model to
construct manifestly covariant operators which represent our basic variables in the space of $\psi$.  Then we show that
the map (\ref{cor4}) relates operators of canonical and covariant formalism, thus establishing covariant rules for
computation of mean values
\begin{eqnarray}\label{cor5}
\langle \Psi,\hat Q\Phi \rangle=(\psi, \hat q\phi).
\end{eqnarray}

\subsection{Covariant operators of vector model}
\label{ch09:sec9.23} Let us return to the covariant formulation (\ref{mq.1}) and (\ref{mq.11}) of the classical theory.
To take into account the second-class constraints $T_3=p_\mu\omega^\mu=0$ and $T_4=p_\mu\pi^\mu=0$, we construct the
Dirac brackets (\ref{pht.13}) of the variables $x^\mu$, $p^\mu$, $S^{\mu\nu}$  and $s^\mu$. The non vanishing Dirac
brackets are as follows. \par \noindent Spatial sector:
\begin{equation}\label{db3}
\{x^\mu,x^\nu\}=-\frac{1}{2p^2}S^{\mu\nu},\qquad \{x^\mu,p^\nu\}=\eta^{\mu\nu}, \qquad \{p^\mu,p^\nu\}=0.
\end{equation}
\par\noindent Frenkel sector:
\begin{equation}\label{db5}
\{S^{\mu\nu}, S^{\alpha\beta}\}= 2(g^{\mu\alpha} S^{\nu\beta}-g^{\mu\beta} S^{\nu\alpha}-g^{\nu\alpha} J^{\mu\beta}
+g^{\nu\beta} S^{\mu\alpha}),
\end{equation}
\begin{equation}\label{db6}
\{x^\mu, S^{\alpha\beta}\}=\frac{1}{p^2}S^{\mu[\alpha}p^{\beta]}.
\end{equation}
\par \noindent Pauli-Lubanski-sector:
\begin{eqnarray}\label{db7}
\{s^\mu, s^\nu\}=-\frac{1}{\sqrt{-p^2}}\epsilon^{\mu\nu\alpha\beta}p_\alpha s_\beta=\frac{1}{2}S^{\mu\nu},
\end{eqnarray}
\begin{eqnarray}\label{db8}
\{x^\mu, s^\nu\}=-\frac{s^\mu p^\nu}{p^2}=-\frac{1}{4\sqrt{-p^2}}\epsilon^{\mu\nu\alpha\beta}S_{\alpha\beta}-
\frac{s^\nu p^\mu}{p^2}.
\end{eqnarray}
In the equation (\ref{db5}) it has been denoted
$g^\mu{}_\nu\equiv\delta^\mu{}_\nu-\frac{p^{\mu}p_\nu}{p^2}$.

In the covariant scheme, we need to construct operators $\hat{x}^\mu, \hat p^\mu, \hat j^{\mu\nu}, \hat s^\mu$ whose
commutators
\begin{eqnarray}\label{cq23.1}
[\hat q_1 , \hat q_2 ]=i\hbar \left.\{ q_1 , q_2 \}_D\right|_{q_i\rightarrow\hat q_i},
\end{eqnarray}
are defined by the Dirac brackets (\ref{db3})-(\ref{db8}). Inspection of the classical equations
$S^2=\frac{3\hbar^2}{4}$ and $p^2+(mc)^2=0$ suggests that we can look for a realization of operators in the Hilbert space
constructed in Sect.~\ref{ch09:sec9.22}.

With the spin-sector variables we associate the operators
\begin{eqnarray}\label{cq-S-2operator}
s^\mu ~ \rightarrow ~ \hat s_{PL}^\mu=\frac{\hbar}{4\sqrt{-\hat p^2}}\epsilon^{\mu\nu\alpha\beta}\hat p_\nu
\sigma_{\alpha\beta},
\end{eqnarray}
\begin{eqnarray}\label{cq11}
S^{\mu\nu} ~ \rightarrow ~ \hat s^{\mu\nu}\equiv -\frac{2}{\sqrt{-\hat p^2}}\epsilon^{\mu\nu\alpha\beta}\hat
p_\alpha \hat s_{PL \beta}=\hbar\sigma^{\mu\nu}+\hbar\frac{\hat p^\mu(\sigma \hat p)^\nu-\hat p^\nu(\sigma \hat p)^\mu}{\hat p^2}. \qquad
\end{eqnarray}
They obey the desired commutators (\ref{cq23.1}), (\ref{db7}), (\ref{db5}). To find the position operator, we separate
the inner angular momentum $\hat s^{\mu\nu}$ in the expression (\ref{cq9.1}) of Poincar\'{e} generator
\begin{eqnarray}\label{tl-part-of-hatM}
\hat j^{\mu\nu} =\frac12\left[x^\mu+\frac{\hbar(\sigma \hat p)^\mu }{2\hat p^2}\right] \hat p^\nu-\frac12\left[x^\nu+\frac{\hbar(\sigma \hat
p)^\nu }{2\hat p^2}\right]\hat p^\mu+\frac{i}{4}\hat s^{\mu\nu}\,.
\end{eqnarray}
This suggests the operator of ``relativistic position''
\begin{eqnarray}\label{cq13}
x^\mu \quad \rightarrow \quad \hat x_{rp}^\mu=\hat x^\mu+\frac{\hbar}{2\hat p^2}(\sigma \hat p)^\mu\,,
\end{eqnarray}
where $\hat x^\mu\psi=x^\mu\psi$. The operators $\hat p_\mu=-i\hbar\partial_\mu$, (\ref{cq-S-2operator}),
(\ref{cq11}) and (\ref{cq13}) obey the algebra (\ref{cq23.1}),
(\ref{db3})-(\ref{db8}).

\subsection{Proof of relativistic covariance}
\label{ch09:sec9.23} Relativistic invariance of the scalar product (\ref{pha21.2}) has been already shown in
Sect.~\ref{ch09:sec9.22}. Here we show how the covariant formalism can be used to compute mean values of operators of
canonical formulation, thus proving their relativistic covariance. Namely, we confirm \cite{DPM2} the following \par
\noindent {\bf Proposition.} Let
\begin{eqnarray}\label{cor0}
H^+_{can}=\{ ~\Psi(t, \vec x); ~  i\hbar\frac{d\Psi}{dt}=\sqrt{\hat{{\bf p}}^{\,2}+(mc)^2}\Psi, ~ \langle \Psi,\Phi
\rangle=\int d^3x\Psi^\dagger \Phi ~ \}, \quad
\end{eqnarray}
be Hilbert space of canonical formulation and
\begin{eqnarray}\label{cor0.1}
H_{cov}=\{ ~ \psi(x^\mu); ~ (\hat p^2+m^2c^2)\psi=0, ~ (\psi,\phi)=\int\limits_{\Omega} d\Omega_\mu I^\mu,  \cr
I^\mu=\frac{1}{(mc)^2}(\bar\sigma\hat p\psi)^\dagger\sigma^\mu\bar\sigma\hat p\phi- \psi^\dagger\bar\sigma^\mu\phi ~
\}, \qquad
\end{eqnarray}
be Hilbert space of two-component KG equation. With a state-vector $\Psi$ we associate $\psi$ as follows:
\begin{eqnarray}\label{cor0.2}
\psi=V^{-1}\Psi, \qquad V^{-1}=\frac{1}{2\sqrt{\hat p^0(\hat p^0+mc)}}[mc-\sigma\hat p].
\end{eqnarray}
Then $\langle \Psi,\Phi \rangle=(\psi, \phi )$. Besides, mean values of the physical position and spin operators
(\ref{pha.20.2})-(\ref{spin.1}) can be computed as follows
\begin{eqnarray}\label{cor0.3}
\langle \Psi,\hat X^i\Phi \rangle=\mbox{Re}(\psi, \hat x_{rp}^i\phi ), \qquad \langle \Psi,\hat S^{ij}\Phi
\rangle=(\psi, \hat s^{ij}\phi ), \cr \langle \Psi,\hat S^i\Phi \rangle=\frac14\epsilon^{ijk}(\psi, \hat
s^{jk}\phi ), \qquad \qquad \qquad
\end{eqnarray}
where  $\hat x_{rp}^i$ and $\hat s^{ij}$ are spatial components of the manifestly-covariant operators (\ref{cq13}) and (\ref{cq11}).

It will be convenient to work in the momentum representation, $\psi(x^\mu)=\int d^4p \psi(p^\mu){\rm e}^{\frac{i}{\hbar}px}$.
Transition to the momentum representation implies the substitution
\begin{eqnarray}
\hat p_\mu \to p_\mu,\qquad \hat x_\mu \to i\hbar \frac{\partial}{\partial p^\mu},
\end{eqnarray}
in the expressions of covariant operators (\ref{cq-S-2operator}), (\ref{cq11}), (\ref{cq13}) and so on.

An arbitrary solution to the KG equation reads
\begin{eqnarray}
\psi(t, {\bf x})=\int d^3p \left(\psi({\bf p}){\rm e}^{\frac{i}{\hbar}\omega_px^0}+\psi_{-}({\bf p}){\rm e}^{-\frac{i}{\hbar}\omega_p
x^0}\right){\rm e}^{-\frac{i}{\hbar}({\bf p x})}, \cr p^0=-p_0\equiv\omega_p=\sqrt{{\bf p}^2+(mc)^2}, \qquad \qquad
\end{eqnarray}
where $\psi({\bf p})$ and $\psi_{-}({\bf p})$ are arbitrary functions of three-momentum, they correspond to positive
and negative energy solutions. The scalar product can
be written then as follows
\begin{eqnarray}
(\psi,\phi)=2\int \frac{d^3p\, \omega_p}{m^2c^2} \left[\psi^\dagger(\bar\sigma p)\phi-\psi_{-}^\dagger(\sigma
p)\phi_{-}\right],
\end{eqnarray}
We see that this scalar product separates positive and negative energy parts of state vectors. Since our classical
theory contains only positive energies, we restrict our further considerations by the positive-energy solutions. In the
momentum representation the scalar product (\ref{sp1}) reads through the metric $\rho$, or through the operator $V$ as
follows:
\begin{eqnarray}\label{cq-hilber-space-matric}
(\psi,\phi)=\langle W\psi, W\phi \rangle=\langle \psi, W^\dagger W\phi \rangle=\int d^3p\psi^\dagger\rho\phi=\langle V\psi, V\phi \rangle=
\langle \Psi, \Phi \rangle,
\cr \rho=\frac{2\omega_p}{m^2c^2}(\bar\sigma p). \qquad \qquad \qquad \qquad \qquad \qquad
\end{eqnarray}
Now our basic space is composed by arbitrary functions $\psi({\bf p})$. The operators $\hat x^i$, $\hat s^\mu$ and
$\hat s^{\mu\nu}$ act on this space as before, with the only modification, that $\hat p^0\psi({\bf
p})=\omega_p\psi({\bf p})$. The operator $\hat x^0$ and, as a consequence, the operator $\hat x_{rp}^0$, do not act in
this space. Fortunately, they are not necessary to prove the proposition formulated above.

Given operator $\hat A$ we denoted its hermitian conjugated in space $H^+_{can}$ as $\hat A^\dagger$. Hermitian
operators in space $H^+_{can}$ have both real eigenvalues and expectation values. Consider an operator $\hat a$ in
space $H_{cov}$ with  real expectation values $(\psi,\hat a\psi)=(\psi,\hat a\psi)^*$. It should obey $\hat a^\dagger
\rho =\rho \hat a$. That is, such an operator in $H_{cov}$ should be pseudo-Hermitian. We denote pseudo-Hermitian
conjugation in $H_{cov}$ as follows: $\hat a_c=\rho^{-1}\hat a^\dagger \rho$. Then pseudo-Hermitian part of an
arbitrary operator $\hat a$ is given by $\frac12(\hat a+\hat a_c)$.

Let us check the pseudo-Hermicity properties of basic operators. From the following identities:
\begin{eqnarray}
(\sigma^{\mu\nu})^\dagger\rho=\rho\,\left(\sigma^{\mu\nu}+\frac{2i}{p^2} (\sigma p) (p^\mu \bar\sigma^\nu- p^\nu
\bar\sigma^\mu)\right), \quad \cr
(\sigma^{\mu\nu} p_\nu)^\dagger\rho=\rho\,\left(\sigma^{\mu\nu} p_\nu+2i [ p^\mu - (\sigma p)
\bar\sigma^\mu]\right), \qquad \cr
(\hat x_{rp}^j)^\dagger\rho=\rho \left(\hat x_{rp}^j+\frac{i\hbar}{m^2c^2\omega_p}
\left[\frac{m^2c^2}{\omega_p}p^j-p^j(\vec{\sigma}\vec{p})\right] \right)\,,
\end{eqnarray}
we see that operators $\sigma^{\mu\nu}$ and $\hat x^j_{rp}$ are non-pseudo-Hermitian, while operators $\hat p^\mu$, $\hat
s_{PL}^\mu$, $\hat s^{\mu\nu}$ and orbital part of $\hat j^{ij}$ are pseudo-Hermitian.

The transformation between state-vectors induces the map of operators
\begin{equation}\label{cor4.1}
V^{-1}\hat QV=\hat q, \quad\mbox{then}\quad \langle \Psi,\hat Q\Phi \rangle=(\psi, \hat q\phi).
\end{equation}
Due to Hermicity of $V$, $V^\dagger=V$, pseudo-Hermitian operators, $\hat q^\dagger V^2= V^2 \hat q$, transform into
Hermitian operators $\hat Q^\dagger = \hat Q$. For an operator $\hat q$ which commutes with momentum operator,
transformation \eqref{cor4} acquires the following form
\begin{eqnarray}
\hat Q=\frac{1}{2}(\hat q+\hat q^\dagger)-\frac{1}{2(\omega_p+mc)}(\hat q-\hat q^\dagger)(\vec{\sigma}\vec{p}).
\end{eqnarray}
Using this formula, we have checked by direct computations that covariant operators $\hat{\bf{p}}$, $\hat s^{\mu\nu}$
and $\hat s_{PL}^\mu$ transform into canonical operators $\hat{\bf p}$, $\hat S^{\mu\nu}$ and $\hat S_{PL}^\mu$ (recall that
the spatial part of $\hat S^{\mu\nu}$, $\hat S^i=\frac14\epsilon^{ijk}\hat S_{jk}$ represents the classical spin $S^i$).
This result together with Eq. (\ref{cor5}) implies that mean values of these operators of canonical formulation are
relativistic-covariant quantities.

Concerning the position operator, we first apply the inverse to Eq. \eqref{cor4} to our canonical
coordinate $\hat{\tilde X}^i=i\hbar \frac{\partial}{\partial p^i}$ in the momentum representation
\begin{eqnarray}
\hat{\tilde x}_V^i=V^{-1}\hat{\tilde X}^iV=\hat{\tilde X}^i+[V^{-1},\hat{\tilde X}^i]V=
i\hbar \frac{\partial}{\partial p^i}-\frac{i\hbar p^i(\vec{\sigma}\vec{p})}{2mc\omega_p(\omega_p+mc)} \cr
+\frac{i\hbar p^i}{2\omega_p}+
\frac{i\hbar}{2mc}\sigma^i
+\frac{\hbar}{2mc(\omega_p+mc)}\epsilon^{ijk}\sigma_jp_k.\qquad \qquad
\end{eqnarray}
Our position operator (\ref{pha.20}) then can be mapped as follows:
\begin{eqnarray}\label{cov-position-d}
\hat{x}_V^i= V^{-1}\left(i\hbar \frac{\partial}{\partial p^i}+\frac{1}{mc(\omega_p+mc)}\epsilon^{ijk}\hat
S_{PLj}p_k\right)V =i\hbar \frac{\partial}{\partial p^i}+\frac{i\hbar p^i(\vec{\sigma}\vec{p})}{2p^2\omega_p} +\cr \frac{i\hbar
p^i}{2\omega_p}- \frac{i\hbar}{2p^2}\omega_p\sigma^i +\frac{\hbar}{2p^2}\epsilon^{ijk}p_j\sigma_k. \qquad \qquad \qquad
\end{eqnarray}
We note that pseudo-Hermitian part of operator $\hat x_{rp}^i$
coincides with the image $\hat{x}_V^i$,
\begin{eqnarray}
\hat{x}_V^i=\frac{1}{2}\left(\hat x_{rp}^i+\left[\hat x_{rp}^i\right]_c\right).
\end{eqnarray}
Since $\hat x_{rp}^\mu$ has explicitly covariant form, this also proves covariant
character of position operator $\hat{X}^i$. Indeed, \eqref{cor4} means that
matrix elements of $\hat{X}^i$ are expressed through the real part of manifestly covariant matrix elements
\begin{eqnarray}
\langle \Psi,\hat{X}^i \Phi \rangle=(\psi, \hat x_V^i \phi)={\rm Re}(\psi, \hat{x}_{rp}^i \phi).
\end{eqnarray}
In summary, we have proved the proposition formulated above. It could be formulated also as follows. The operators
$\hat s^{\mu\nu}$ and $\hat{x}_{rp}^\mu$, which act on the space of two-component KG equation, represent
manifestly-covariant form of the Pryce (d)-operators.

Table \ref{tabular:non-covariantFW-covariant-quantization} summarizes manifest form of operators of canonical formalism
and their images in covariant formalism.
\begin{center}
\begin{table}
\caption{Operators of canonical and manifestly covariant formulations in momentum representation}
\label{tabular:non-covariantFW-covariant-quantization}
\begin{center}
\begin{tabular}{c|c|c}
{}  & Canonical formalism $\Psi({\bf p})$         & Covariant formalism $\psi({\bf p})$\\
\hline \hline
$\hat p_j \to \hat p_j $   & $p_j$            & $p_j$
\\
$\hat S^i \to \hat s^i $   & $\frac{\hbar}{2mc}\left(\omega_p
\sigma^i-\frac{1}{(\omega_p+mc)}(\vec{p}\vec{\sigma})p^i\right)$            &
$\frac{\hbar\omega_p}{2(mc)^2}\left(\omega_p \sigma^i-(\vec{p}\vec{\sigma})p^i-i\epsilon_{imn}p^m\sigma^n\right)$
\\
$\hat X^i \to \hat x^i_V $  & $i\hbar \frac{\partial}{\partial
p^i}-\frac{\hbar}{2mc(\omega_p+mc)}\epsilon^{ijk}p_j\sigma_k$  &    $i\hbar \frac{\partial}{\partial p^i}+\frac{i\hbar
p^i(\vec{\sigma}\vec{p})}{2p^2\omega_p} +\frac{i\hbar p^i}{2\omega_p}- \frac{i\hbar}{2p^2}\omega_p\sigma^i
+\frac{\hbar}{2p^2}\epsilon^{ijk}p_j\sigma_k$
\\
$\hat S^{ij}\to \hat s^{ij} $& $\frac{\hbar}{mc}\epsilon^{ijk}\left(\omega_p
\sigma_k-\frac{1}{(\omega_p+mc)}(\vec{p}\vec{\sigma})p_k\right)$  &
$\frac{\hbar\omega_p}{m^2c^2}\epsilon^{ijk}\left(\omega_p
\sigma_k-(\vec{p}\vec{\sigma})p_k-i\epsilon_{kmn}p^m\sigma^n\right)$
\\
$\hat S^{0i}\to \hat s^{0i} $& $-\frac{\hbar}{mc}\epsilon^{ijk}p_j \sigma_k$  & $-\frac{\hbar}{m^2c^2}\epsilon^{ijk}
\left(\omega_p \sigma_k- i \epsilon_{kml}p^m\sigma^l\right)p_j$
\\
$\hat S_{PL}^0 \to \hat s_{PL}^0 $   & $\frac{\hbar}{2mc}(\vec{p}\vec{\sigma})$            &
$\frac{\hbar}{2mc}(\vec{p}\vec{\sigma})$
\\
$\hat S_{PL}^i \to \hat s_{PL}^i$   &
$\frac{\hbar}{2}\left(\sigma^i+\frac{1}{mc(\omega_p+mc)}(\vec{p}\vec{\sigma})p^i\right)$ &
$\frac{\hbar}{2mc}(\omega_p\sigma^i +i\epsilon^{ijk}p_j\sigma_k )$
\\
\end{tabular}
\end{center}
\end{table}
\end{center}

\subsection{Vector model and Dirac equation}
\label{ch09:sec9.23} Here we demonstrate the equivalence of quantum mechanics of two-component Klein-Gordon and Dirac
equations. As a consequence, probabilities and mean values of canonical operators (\ref{pha.20.2})-(\ref{spin.1}) can
be computed, using an appropriately constructed covariant operators on the space of Dirac spinors.

Let us replace two
equations of second order, (\ref{cq16}), by an equivalent system of four equations of the first order. To achieve this,
with the aid of the identity $\hat p^\mu \hat p_\mu=\sigma^\mu \hat p_\mu\bar\sigma^\nu\hat p_\nu$, we represent
(\ref{cq16}) in the form
\begin{eqnarray}\label{cq18}
\sigma^\mu \hat p_\mu\bar\sigma^\nu\hat p_\nu \psi+m^2c^2\psi=0.
\end{eqnarray}
Consider an auxiliary two-component function $\bar\xi$ (Weyl spinor of opposite chirality), and define evolution of
$\psi$ and $\bar\xi$ according the equations
\begin{eqnarray}\label{cq19}
\sigma^\mu \hat p_\mu(\bar\sigma^\nu\hat p_\nu) \psi+m^2c^2\psi=0, \\
(\bar\sigma^\nu\hat p_\nu)\psi-mc\bar\xi=0.\label{cq19-2}
\end{eqnarray}
That is dynamics of $\psi$ is determined by (\ref{cq18}), while $\bar\xi$ accompanies $\psi$: $\bar\xi$ is determined
from the known $\psi$ taking its derivative, $\bar\xi=\frac{1}{mc}(\bar\sigma\hat p)\psi$. Evidently, the systems
(\ref{cq16}) and (\ref{cq19}), (\ref{cq19-2}) are equivalent.  Rewriting the system (\ref{cq19}), (\ref{cq19-2}) in a
more symmetric form, we recognize the Dirac equation
\begin{eqnarray}\label{cq20}
\left(
\begin{array}{cc}
0& \sigma^\mu\hat p_\mu\\
-\bar\sigma^\nu\hat p_\nu & 0
\end{array}
\right)\left(
\begin{array}{c}
\psi\\
\bar\xi
\end{array}
\right)+mc\left(
\begin{array}{c}
\psi\\
\bar\xi
\end{array}
\right)=0, \quad  {\mbox or} \quad (\gamma_W^\mu\hat p_\mu+mc)\Psi_{DW}=0, \qquad
\end{eqnarray}
for the Dirac spinor $\Psi_{DW}=\left(\psi, \bar\xi\right)$
in the Weyl representation of $\gamma$\,-matrices
\begin{eqnarray}\label{cq22}
\gamma_W^0=\left(
\begin{array}{cc}
0 & {\bf 1}\\
{\bf 1} & 0
\end{array}
\right), \qquad \gamma_W^i=\left(
\begin{array}{cc}
0& \sigma^i\\
-\sigma^i& 0
\end{array}
\right).
\end{eqnarray}
This gives one-to-one correspondence among two spaces. With each solution $\psi$ to KG equation we associate the solution
\begin{eqnarray}\label{cq21}
\Psi_{DW}[\psi]=\left(
\begin{array}{c}
\psi\\
\frac{1}{mc}(\bar\sigma\hat p)\psi
\end{array}
\right), \nonumber
\end{eqnarray}
to the Dirac equation. Below we also use the Dirac representation of $\gamma$\,-matrices
\begin{eqnarray}\label{cq22}
\gamma^0=\left(
\begin{array}{cc}
{\bf 1} & 0\\
0&    -{\bf 1}
\end{array}
\right), \qquad \gamma^i=\left(
\begin{array}{cc}
0& \sigma^i\\
- \sigma^i& 0
\end{array}
\right).
\end{eqnarray}
In this representation, the Dirac spinor corresponding to $\psi$ reads
\begin{eqnarray}\label{cq22.1}
\Psi_D[\psi]=\frac{1}{\sqrt{2}}\left(
\begin{array}{cc}
\,\,\,\,{\bf 1} & {\bf 1}\\
-{\bf 1} & {\bf 1}
\end{array}
\right)\left(
\begin{array}{c}
\psi\\
\frac{1}{mc}(\bar\sigma\hat p)\psi
\end{array}
\right)=\frac{1}{\sqrt{2}mc}\left(
\begin{array}{c}
{[(\bar\sigma\hat p)+mc]}\psi\\
{[(\bar\sigma\hat p)-mc]}\psi
\end{array}
\right).
\end{eqnarray}
The conserved current (\ref{cq17}) of Klein-Gordon equation (\ref{cq16}), being rewritten in terms of Dirac spinor,
coincides with the Dirac current (\ref{1.5}). Therefore, the scalar product (\ref{cq-inv-scalar-product}) coincides
with that of Dirac, $(\Psi_D, \Phi_D)_D=\int d^3x\bar\Psi_D\gamma^0\Psi_D$
\begin{eqnarray}\label{cq23}
I^\mu[\psi_1,\psi_2]=\bar\Psi_{D}[\psi_1]\gamma^\mu\Psi_{D}[\psi_2], \quad\mbox{then}\quad (\psi, \phi)=(\Psi_{D}[\psi], \Psi_{D}[\phi])_D.
\end{eqnarray}

This allows us to find manifestly-covariant operators in the Dirac theory which have the same expectation values as
$\hat{s}^{\mu\nu}$ and $\hat{x}_{rp}^\mu$ . Consider the following analog of $\hat s^{\mu\nu}$ on the space of
4-component Dirac spinors
\begin{equation}\label{cq24}
\hat{s}_D^{\mu\nu}=\hbar\gamma^{\mu\nu}+\hbar\frac{\hat p^\mu\gamma^{\nu\alpha} \hat p_\alpha-\hat p^\nu\gamma^{\mu\alpha} \hat
p_\alpha}{\hat p^2} =\hbar\gamma^{\mu\nu}+\frac{i\hbar}{\hat p^2}\left(\hat p^\mu\gamma^{\nu} -\hat
p^\nu\gamma^{\mu}\right)(\gamma\hat p),
\end{equation}
where $\gamma^{\mu\nu}=\frac{i}{2}(\gamma^\mu\gamma^\nu-\gamma^\nu\gamma^\mu)$. This definition is independent
from a particular representation of $\gamma$-matrices. In the representation \eqref{cq22} this reads
\begin{eqnarray}
\gamma^{\mu\nu}=\left(
\begin{array}{cc}
\sigma^{\mu\nu} & 0\\
0 & (\sigma^{\mu\nu})^\dagger
\end{array}
\right)\,,
\end{eqnarray}
and can be used to prove the equality of matrix elements
\begin{eqnarray}\label{cq25}
(\Psi_{D}[\psi]\hat{s}_D^{\mu\nu}\Phi_{D}[\phi])_D=(\psi, \hat{s}^{\mu\nu} \phi),
\end{eqnarray}
for arbitrary solutions $\psi$, $\phi$ of two-component Klein-Gordon equation. The covariant position operator can be defined
as follows:
\begin{equation}\label{cq26}
\hat x_D^{\mu}=x^{\mu}+\frac{\hbar\gamma^{\mu\alpha}\hat p_\alpha}{2\hat p^2}+\frac{i\hbar(\gamma^5-1)\hat p^\mu}{2\hat
p^2}= x^{\mu}+\frac{i\hbar\gamma^{\mu}}{2\hat p^2}(\gamma\hat p)+\frac{i\hbar\gamma^5\hat p^\mu}{2\hat p^2},
\end{equation}
where $\gamma_5=-i\gamma^0\gamma^1\gamma^2\gamma^3$. Again, one can check that matrix elements in two theories coincide
\begin{eqnarray}
(\Psi_D[\psi]\hat{x}_D^{\mu}\Phi_D[\phi])_D=(\psi, \hat{x}_{rp}^{\mu} \phi).
\end{eqnarray}
As a result, the manifestly-covariant operators $\hat s_D^{\mu\nu}$ and $\hat x_D^{\mu}$ of the Dirac equation represent
position ${\bf x}$ and spin ${\bf S}$ of the spinning particle .  Their mean values can be computed as follows
\begin{eqnarray}\label{cq27}
\langle \Psi,\hat X^i\Phi \rangle=\frac12\mbox{Re}(\Psi_D[\psi], [\hat x_{D}^i+\hat x_{D}^{i\dagger}]\Phi_D[\phi] )_D,
\cr \langle \Psi,\hat S^i\Phi \rangle=\frac14\epsilon^{ijk}(\Psi_D[\psi], \hat s_{D}^{jk}\Phi_D[\phi] )_D. \qquad
\end{eqnarray}
We emphasize that the observables of vector model have an expected behavior both on classical and quantum level. In
particular, the position operator $\hat X^i$ does not experiences \textit{Zitterbewegung}, contrary to some other
classical models \cite{Bruno14}. Note also that the covariant operator (\ref{cq26}), that represents the position of
spinning particle in the Dirac theory, is different from the naive expression used in equation (\ref{4.2}).

{\bf The map $V$ and Foldy-Wouthuysen transformation.} Our map $V$, that relates canonical and two-component
Klein-Gordon spaces, turns out to be in close relation with the Foldy-Wouthuysen transformation. The latter is given by
unitary operator
\begin{eqnarray}\label{cq28}
U_{FW}=\frac{\omega_p+mc+(\vec{\gamma}\vec{p})}{\sqrt{2(\omega_p+mc)\omega_p}},
\end{eqnarray}
and relates the Dirac and four-component Klein-Gordon equations. Applying it to the  Dirac spinor $\Psi_D[\psi]$, we
obtain
\begin{eqnarray}
U_{FW}\Psi_D[\psi]=
\left(
\begin{array}{c}
V\psi\\
0
\end{array}
\right)=\left(
\begin{array}{c}
\Psi\\
0
\end{array}
\right),
\end{eqnarray}
That is the operator $V$ is a restriction of $U_{FW}$ to the space of positive-energy right Weyl spinors $\psi$.

\section{Conclusion}

In non relativistic theory, spin can be described on the base of semiclassical Lagrangian (\ref{nr2}) which is
invariant under spin-plane local symmetry. The symmetry yields two first-class constraints (\ref{intr12.1}) on
spin-sector variables in Hamiltonian formalism. The resulting theory has an expected number of physical degrees of
freedom, in particular, the only observable quantities of spin-sector turn out to be the components of spin-vector
(\ref{intr.9.1}).

To treat spin in a manifestly Lorentz-covariant way, we extended phase space with two auxiliary degrees of freedom,
adding null-components to the basic variables ${\boldsymbol{\omega}}$ and ${\boldsymbol{\pi}}$. To supply their
auxiliary character, we used two second-class constraints, see (\ref{f1.0}). This implies drastic modification of the
formalism: the constraints induce noncommutative geometry in the phase space even in a free relativistic theory. In
particular, spin induces the noncommutative classical brackets (\ref{pha.15}) of position variables. This must be taken
into account in construction of quantum mechanics of a spinning particle. For a spinning electron in Coulomb electric
and constant magnetic field, our model yields the non relativistic Hamiltonian (\ref{18}) with correct factors in front
of spin-orbit and Pauli terms. Hence the spin-induced noncommutativity explains the famous one-half factor in the Pauli
equation on the classical level, without appeal to the Thomas precession, Dirac equation or Foldy-Wouthuysen
transformation. Besides, for a spinning body in gravitational field, the spin-induced noncommutativity clarifies the
discrepancy in expressions for three-acceleration obtained by different methods, see \cite{DPW2}.

Lagrangian of the vector model admits interaction with an arbitrary gravitational and electromagnetic fields and has
reasonable properties both on classical and on quantum level. Dealing with the variational problem, we were able to
determine both Hamiltonian and classical brackets of the theory in unambiguous way.

Regarding the interaction with electromagnetic field, equations (\ref{FF.6})-(\ref{FF.8}) of our particle with a
magnetic moment generalize the approximate equations of Frenkel and BMT to the case of an arbitrary field.
Straightforward canonical quantization of the model yields quantum mechanics in Hilbert space of two-component Weyl
spinors. All solutions to the Schr\"{o}dinger equation (\ref{pha.21.1}) have positive energy. Since our basic variables
obey non-canonical brackets, the operators which represent them have a non standard form, see (\ref{pha.20.2}) and
(\ref{spin.1}). To establish the relativistic covariance of obtained quantum mechanics, we first developed manifestly
relativistic quantum mechanics of two-component Klein-Gordon equation. Then we related states and operators of two
formalisms and demonstrated on this base the relativistic invariance of the scalar product (\ref{pha21.2}), and
relativistic covariance of mean values of operators of canonical formalism, see Eqs. (\ref{sp4}) and (\ref{cor0.3}).
Using the relationship (\ref{cq22.1}) between Klein-Gordon and Dirac formalisms, we also formulated the rules
(\ref{cq27}) for computation the mean values in the framework of Dirac formalism. Here we emphasize once again, that we
have not tried to find an interpretation of negative-energy states presented in the covariant KG and Dirac formalisms.
The formalisms were considered as an auxiliary constructions that allow us to prove relativistic covariance of the
quantum mechanics formulated in Sect.~\ref{ch09:sec9.20}.

Regarding the interaction with gravity, the minimal coupling gives the equations (\ref{m003})-(\ref{m005}) equivalent
to MPTD equations of a rotating body (\ref{r003})-(\ref{r005}). Spin-gravity interaction induces an effective metric
(\ref{gmm16}) in the Lagrangian equation (\ref{gml1}) for a trajectory. To study it in ultra-relativistic limit, we
used the Landau-Lifshitz approach to define the three-dimensional geometry, and defined on this base the
three-dimensional acceleration (\ref{La.8.5}) which guarantees the impossibility for a particle in geodesic motion to
overcome the speed of light. The effective metric causes unsatisfactory behavior of MPTD particle in the
ultra-relativistic regime. In particular, its acceleration grows with velocity and becomes infinite in the limit. To
improve this, we constructed a non minimal spin-gravity interaction (\ref{acc-11}) through a gravimagnetic moment, and
showed that a fast-moving body with unit gravimagnetic moment has a satisfactory behavior.

\section*{Acknowledgments}
The work of AAD has been supported by the Brazilian foundations CNPq (Conselho Nacional de Desenvolvimento Científico e
Tecnológico - Brasil) and FAPEMIG (Fundação de Amparo à Pesquisa do Estado de Minas Gerais - Brasil). WGR thanks CAPES
for the financial support (Programm PNPD/2011).

\end{document}